\documentclass[]{article}
\usepackage{jheppub}
\pdfoutput=1

\usepackage{amsmath}
\usepackage{amssymb}
\usepackage{latexsym}
\usepackage{amsfonts}
\usepackage{graphicx}
\usepackage{epstopdf}

\usepackage{color}%

\def\hhref#1{\href{http://arxiv.org/abs/#1}{arXiv:#1}} 

\newcommand{\ft}{\mathfrak t}
\newcommand{\K}{\mathbb{K}}
\newcommand{\E}{\mathbb{E}}
\newcommand{\PP}{{\cal P}}

\newcommand{\be}{\begin{equation}}
\newcommand{\ee}{\end{equation}} 
\def\bea{\begin{align}}
\def\ena{\end{align}}
\def\beqa{\begin{eqnarray}}
\def\enqa{\end{eqnarray}}

\begin{document}

\title{Quantum Geometry of Resurgent Perturbative/Nonperturbative Relations}

\author[1]{G\"ok\c ce Ba\c sar,}
\author[2]{Gerald~V.~Dunne,}
\author[3]{Mithat~\"Unsal}
\affiliation[1]{Maryland Center for Fundamental Physics, University of Maryland, College Park, MD, 20742, USA}
\affiliation[2]{Department of Physics, University of Connecticut, Storrs CT 06269-3046, USA}
\affiliation[3]{Department of Physics, North Carolina State University, Raleigh, NC 27695-8202, USA}
\emailAdd{gbasar@umd.edu}
\emailAdd{gerald.dunne@uconn.edu}
\emailAdd{munsal@ncsu.edu}

\abstract{
For a wide variety of quantum potentials, including the textbook
`instanton' examples of the periodic cosine and symmetric double-well
potentials, the perturbative data coming from fluctuations about the
vacuum saddle encodes all non-perturbative data in all higher
non-perturbative sectors. Here we unify these examples in geometric terms,
arguing that the all-orders quantum action determines the all-orders
quantum dual action for quantum spectral problems associated with a
classical genus one elliptic curve.  Furthermore, for a special class of
genus one potentials this relation is particularly simple: this class
includes the cubic oscillator, symmetric double-well, symmetric degenerate
triple-well, and periodic cosine potential. These are related to the
Chebyshev potentials, which are in turn related to certain ${\mathcal
N}=2$ supersymmetric quantum field theories, to mirror maps for
hypersurfaces in projective spaces, and also to  topological $c=3$
Landau-Ginzburg models and `special geometry'. These systems inherit a
natural modular structure corresponding to Ramanujan's theory of elliptic
functions in alternative bases, which is especially important for the
quantization. Insights from supersymmetric quantum field theory suggest
similar structures for more complicated potentials, corresponding to
higher genus.
Our approach is very elementary, using basic classical geometry combined
with all-orders WKB.
}
 
 \date{\today}

\maketitle

\section{Introduction} 

One of the most intriguing aspects of resurgent asymptotics \cite{Dingle:1973,Ecalle:1981,delabaere,sauzin,Costin:2009}, as applied to quantum theories,  is that the characteristic divergence of fluctuations about certain saddle points, such as the perturbative vacuum, may encode detailed information about the global non-perturbative structure of the system. Two distinct types of resurgent behavior have been identified in quantum spectral problems. The first is a generic type of  ``large order/low order'' form of resurgence \cite{ddp,kawai-takei,ZinnJustin:2004ib,Jentschura:2004jg}, whereby the large order growth of the perturbative coefficients of fluctuations about a given non-perturbative sector is related to the low-order perturbative coefficients of fluctuations about other non-perturbative sectors. This resurgent structure encodes an intricate network of relations between different non-perturbative sectors, and reflects to a surprising degree the generic resurgent structure of the all-orders steepest descents analysis of ordinary exponential integrals \cite{BerryHowls,delabaere-howls}, and indeed the general resurgent structure of real trans-series \cite{Aniceto:2013fka}.
A second type of resurgent behavior is less generic, yielding a  ``low order/low order'' form of resurgence, in which the fluctuations about all non-perturbative sectors are explicitly encoded in the perturbative expansion about the vacuum sector.  This `constructive' form of resurgence appears to have first been noticed in formulas for the ionization rate for hydrogenic atoms \cite{hoe}, a result that motivated a systematic study by \'Alvarez and Casares in the context of one dimensional oscillators  \cite{alvarez-cubic,alvarez,alvarez-howls-silverstone}, in which such explicit perturbative/non-perturbative (P/NP) relations were found in the cubic and quartic oscillator systems. Later studies found further examples of such P/NP relations in the periodic cosine (Mathieu), supersymmetric double-well, radial anharmonic oscillator, and supersymmetric Mathieu potentials \cite{Dunne:2013ada,Dunne:2014bca,Dunne:2016qix,Misumi:2015dua,Dunne:2016jsr}, and more recently in quasi-exactly soluble models \cite{Kozcaz:2016wvy}. In a recent paper, Codesido and Mari\~no have demonstrated the precise connection of these P/NP relations for some 1d quantum oscillator systems with the refined holomorphic anomaly equation of topological string theory   \cite{Codesido:2016dld}.

This constructive type of resurgence has the following form: consider the Schr\"odinger spectral problem
\begin{eqnarray}
-\frac{\hbar^2}{2} \frac{d^2}{dx^2} \psi + V(x) \psi = u\, \psi
\label{eq:schrodinger}
\end{eqnarray}
with energy $u$. Then, given the perturbative series $u_{\rm pert}(\hbar, N)$, where $\hbar$ is the coupling and $N$ labels the unperturbed harmonic energy level,  it is possible to write an explicit and constructive expression for the fluctuations about any higher non-perturbative sector, directly in terms of the perturbative data, $u_{\rm pert}(\hbar, N)$. 
For example, for the cosine (Mathieu) potential, $V(x)=\cos^2 x$,
the edges of the $N^{th}$ band, for $N\hbar\ll 1$, are given by a trans-series expression:
\begin{eqnarray}
u_\pm(\hbar, N)=u_{\rm pert}(\hbar, N) \pm   \sqrt{2 \over\pi} \frac{1}{N!} \left(\frac{2^{7/2}}{\hbar}\right)^{N+\frac{1}{2}} \exp\left[-\frac{2\sqrt{2}}{\hbar}\right]\,{\mathcal P}_{\rm inst}(\hbar, N)\,  +\dots
\label{eq:one}
\end{eqnarray}
Both the perturbative series, $u_{\rm pert}(\hbar, N)$, and the fluctuations about the one-instanton sector, ${\mathcal P}_{\rm inst}(\hbar, N)$, are formal divergent series:
\begin{eqnarray}
u_{\rm pert}(\hbar, N)=\sum_{n=0}^\infty \hbar^{n} u_{n}(N)\qquad, \qquad
{\mathcal P}_{\rm inst}(\hbar, N)=\sum_{n=0}^\infty \hbar^{n} p_{n}(N)
\label{eq:perturbation}
\end{eqnarray}
where $u_{n}(N)$ and $p_{n}(N)$ are polynomials in $N$.
For the Mathieu system, the perturbative/non-perturbative relation result is that the exponentially suppressed one-instanton term in the trans-series, including the all orders fluctuation factor ${\mathcal P}_{\rm inst}(\hbar, N)$ is expressed entirely in terms of the perturbative expansion $u_{\rm pert}(\hbar, N)$ \cite{Dunne:2013ada,Dunne:2014bca,Dunne:2016qix}
\footnote{In refs \cite{Dunne:2013ada,Dunne:2014bca,Dunne:2016qix}, the Mathieu potential was written as  $V(x)=\cos x$. In this paper we use $V(x)=\cos^2 x$, to emphasize the connection to other potentials, normalizing the wells to have the common energy range $u\in [0, 1]$, as shown in Figure \ref{fig:T3}. These normalizations can be translated by: $\hbar_{there}=2\sqrt{2}\,\hbar_{here}$ and $u_{there}=2\,u_{here}-1$.}:
\begin{eqnarray}
{\mathcal P}_{\rm inst}(\hbar, N)=\frac{\partial u_{\rm pert}(\hbar, N)}{\partial N}\,  \exp\left[\frac{S_{\cal I}}{\omega_{c}} \int_0^{\hbar} \frac{d\hbar}{\hbar^3}\left( \frac{\partial u_{\rm pert}(\hbar, N)}{\partial N} -\hbar \omega_{c} +\frac{\hbar^2 \omega_{c} \left(N+\frac{1}{2}\right)}{S_{\mathcal I}}\right)\right] 
\label{eq:prefactor-sg}
\end{eqnarray}
Here $S_{\cal I}=2\sqrt{2}$ is the one-instanton action, and $\omega_{c}=\sqrt{2}$ is the classical frequency of harmonic motion at the bottom of the potential well for the Mathieu potential $V(x)=\cos^2 x$. Note that this expression shows that {\it all} the $\hbar$ dependent
prefactors in (\ref{eq:one}) are encoded in the perturbative series
$u_{\rm pert}(\hbar, N)$. We stress that this manifestation of resurgence is completely {\it constructive}: given a certain number of terms of the $\hbar$ expansion of $u_{\rm pert}(\hbar, N)$, the expression (\ref{eq:prefactor-sg}) generates a similar number of terms in the fluctuations about the one-instanton sector, ${\mathcal P}_{\rm inst}(\hbar, N)$. Furthermore, these relations propagate throughout the entire trans-series, so that perturbation theory encodes the fluctuations about each non-perturbative sector \cite{alvarez-cubic,alvarez,alvarez-howls-silverstone,Dunne:2013ada,Dunne:2014bca,Dunne:2016qix,Misumi:2015dua,Dunne:2016jsr,Codesido:2016dld}. 

These results are particularly interesting when interpreted not only in terms of differential equations, but in terms of a formal saddle point (Lefschetz thimble) decomposition of the associated path integral. 
The extent to which these resurgent structures are inherited from the basic resurgent structure of ordinary exponential integrals \cite{BerryHowls,delabaere-howls} is still not fully understood, even though much of our physical intuition about the non-perturbative physics of path integrals is based on analogies drawn from saddle point analysis of ordinary integrals. It is also quite surprising that such powerful P/NP relations exist in such a disparate set of spectral problems, in various dimensions, with and without tunneling, and with and without supersymmetry. Further, the interpretation of these P/NP relations between perturbative and instanton sectors is quite mysterious in conventional Feynman diagrammatic language, where they have been explicitly confirmed at three loop order of fluctuations about the one-instanton sector, for the symmetric double-well potential and the periodic (Mathieu) potential \cite{Escobar-Ruiz:2015rfa}. These diagrammatic computations  require a complicated summation of many multi-loop Feynman diagrams, each of which involves propagators in an instanton background. Moreover, remarkable cancellations of irrational terms occur between different Feynman diagrams, producing the final rational coefficients that come naturally from the P/NP relation. These cancellations are reminiscent of behavior in multi-loop QFT \cite{Broadhurst:1995dq}. 

The goal of this paper is to understand in very elementary terms the origin, and generality, of this constructive form of resurgence. To address this question,  we first translate the previous results for these \'Alvarez-Casares-type perturbative/non-perturbative (P/NP)  relations \cite{alvarez-cubic,alvarez,alvarez-howls-silverstone,Dunne:2013ada,Dunne:2014bca,Dunne:2016qix,Dunne:2016jsr,Kozcaz:2016wvy} into a more geometric language, better suited for a path integral formulation. We adopt the semiclassical path integral approach of Balian and Bloch \cite{Balian:1974ah,Balian:1978et} and the geometric Stokes diagram and monodromy picture \cite{ddp,kawai-takei,voros}, to express the P/NP relations in terms of all-orders WKB (``exact WKB'') actions and dual actions. In so doing, it proves useful to adopt some of the language and ideas from supersymmetric gauge theories, integrability, conformal field theory and wall-crossing \cite{Gorsky:1995zq,Klemm:1994qs,Sonnenschein:1995hv,Klemm:1997gg,Nekrasov:2002qd,Nekrasov:2009rc,Nekrasov:2003rj,agt,fateev,neitzke,Teschner:2016yzf}, given the close connection of such theories with exact WKB \cite{Mironov:2009uv,Mironov:2009ib,He:2010xa,Huang:2011qx,KashaniPoor:2012wb,Krefl:2013bsa,Gorsky:2014lia,piatek,Basar:2015xna,kpt,Ashok:2016yxz}.  This is similar to the philosophy of  \cite{Codesido:2016dld}, where the refined holomorphic anomaly equation of topological string theory is shown to describe these P/NP relations for some 1d quantum oscillator systems.  Further motivation along these lines comes from the ODE/Integrable Model correspondence \cite{Dorey:1998pt,Bazhanov:1998wj,Bazhanov:2003ni}, which provides explicit mappings between monodromy operators in certain Schr\"odinger systems and Yang-Baxter operators in integrable models. We are also strongly motivated by the geometric relation between supersymmetric gauge theories, matrix models and topological strings \cite{Dijkgraaf:2002fc,Aganagic:2003qj,Aganagic:2011mi}, for which a rich web of resurgent structures has been comprehensively established both analytically and numerically \cite{Marino:2007te,Marino:2008vx,Pasquetti:2009jg,Aniceto:2011nu,Marino:2012zq,Couso-Santamaria:2014iia,Aniceto:2014hoa,Couso-Santamaria:2015wga,marcos-book,Grassi:2016vkw,Couso-Santamaria:2016vwq}. There are surprisingly close parallels between the resurgent structures found in such theories  for the partition function (or free energy) as a function of (at least) two parameters, $g_s$ and $N$, and the resurgent structure of the Schr\"odinger energy eigenvalue $u(\hbar, N)$, as a function of $\hbar$ and the perturbative level number $N$.

 In this language, the explicit relation between the perturbative and non-perturbative all-orders WKB actions can be expressed as a ``quantum Matone relation'', relating the energy eigenvalue $u$ to the all-orders WKB action $a(u, \hbar)$ and dual action $a^D(u, \hbar)$. For the Mathieu system, which is associated with $\mathcal N=2$ supersymmetric gauge theory \cite{Mironov:2009uv,Mironov:2009ib,He:2010xa,Huang:2011qx,KashaniPoor:2012wb,Krefl:2013bsa,Gorsky:2014lia,piatek,Basar:2015xna,kpt,Ashok:2016yxz}, the P/NP relation is:
\begin{eqnarray}
\frac{\partial u(a, \hbar)}{\partial a}=\frac{i}{8\pi}\left(a^D(a, \hbar)-a \frac{\partial a^D(a, \hbar)}{\partial a} -\hbar  \frac{\partial a^D(a, \hbar)}{\partial \hbar}\right)
\label{eq:quantum-matone}
\end{eqnarray}
In the gauge theory formalism, this ``quantum Matone relation'' generalizes the classical Matone relation  \cite{Matone:1995rx} with the inclusion of gravitational couplings \cite{Klemm:2002pa,Flume:2004rp,Poghossian:2010pn,Maruyoshi:2010iu}, and it also has a natural interpretation in all-orders WKB \cite{Gorsky:2014lia,Basar:2015xna,Codesido:2016dld}. At the classical level, $\hbar\to 0$, the relation (\ref{eq:quantum-matone}) is a simple consequence of the associated classical Picard-Fuchs equation, which characterizes the energy dependence of the classical action variables (see Section \ref{sec:aw-chebyshev} below). The quantum version has only a simple modification, to all orders in $\hbar$, as shown in (\ref{eq:quantum-matone}). A practical consequence of the result (\ref{eq:quantum-matone}) is that given the all-orders $\hbar$ expansion of the quantum action $a(u, \hbar)$, one can immediately deduce, term-by-term, the corresponding  all-orders $\hbar$ expansion of the quantum dual action $a^D(u, \hbar)$. In the language of the quantum spectral problem, the left-hand-side of (\ref{eq:quantum-matone}) is proportional to the derivative of the energy with respect to the level number, which is inherently perturbative information; while the right-hand-side describes the inherently non-perturbative information of the tunneling action $a^D(u, \hbar)$. Expression (\ref{eq:quantum-matone}) says that they are constructively related, to all orders in $\hbar$.
It is in this sense that all the non-perturbative physics of the problem (that is, $a^D(u, \hbar)$) is encoded in the perturbative physics (that is, $a(u, \hbar)$). 

It is also useful to view the perturbative/non-perturbative relation between $a(u, \hbar)$ and $a^D(u, \hbar)$ as a ``quantum corrected Wronskian condition''. This is the inverse function version of the quantum Matone relation in (\ref{eq:quantum-matone}). For the Mathieu system,  the Wronskian condition of  the classical Picard-Fuchs equation becomes a ``quantum Wronskian condition'' \cite{Basar:2015xna}:
\begin{eqnarray}
\left(a(u, \hbar)- \hbar  \frac{\partial a(u, \hbar)}{\partial \hbar} \right)\frac{\partial a^D(u, \hbar)}{\partial u}
-\left(a^D(u, \hbar)-\hbar \frac{\partial a^D(u, \hbar)}{\partial \hbar} \right) \frac{\partial a(u, \hbar)}{\partial u}=8\pi i
\label{eq:quantum-wronskian}
\end{eqnarray}
At the classical level, (\ref{eq:quantum-wronskian}) reduces to the Wronskian of the second order Picard-Fuchs differential equation for the Mathieu system.
In this paper we find a class of quantum systems for which the all-orders quantum Wronskian condition takes exactly the same form as in (\ref{eq:quantum-wronskian}), just with a different constant. 
It is remarkable that for a class of quantum problems, when the actions become non-trivial functions of $\hbar$, the only modification to the classical Wronskian condition, to all-orders in $\hbar$, is the extra $\hbar$ derivative terms on the left-hand-side of (\ref{eq:quantum-wronskian}). Ultimately this fact can be traced to the ${\mathcal N}=2$ supersymmetry of the related quantum field theories \cite{Nekrasov:2002qd,Nekrasov:2009rc,Nekrasov:2003rj,Klemm:2002pa,Flume:2004rp,Poghossian:2010pn,Maruyoshi:2010iu}. 

Concerning the generality of these results, we first note that the common feature of the examples studied in  \cite{hoe,alvarez-cubic,alvarez,alvarez-howls-silverstone,Dunne:2013ada,Dunne:2014bca,Dunne:2016qix} is that their classical energy-momentum relation $p^2=2(u-V(x))$ defines a genus 1 elliptic curve: a torus. In Section \ref{sec:proof}  we formulate a simple geometric argument which shows that for all potentials  corresponding to such  a genus 1 elliptic curve, the all-orders (``non-perturbative'') dual action $a^D(u, \hbar)$ is constructively encoded in the (``perturbative'')  action $a(u, \hbar)$. Furthermore, we show that there is a subclass of special genus 1 systems for which this encoding takes {\it exactly} the same form as the Mathieu system perturbative/non-perturbative relations in (\ref{eq:quantum-matone}) and (\ref{eq:quantum-wronskian}), just with different numerical constants. Section \ref{sec:chebyshev} introduces the Chebyshev class of potentials and their relevant geometric, modular and number theoretic properties. 
Their quantization is presented in Section \ref{sec:higher}. Section \ref{sec:more-general} discusses more general genus 1 models, and in the conclusions we discuss higher genus, where we conjecture that there are similar relations, but of a more general structure whose direct computation involves hyperelliptic functions. 

Another  aspect of our approach is that we find a deep connection between the all-orders P/NP relations (\ref{eq:quantum-matone}) and (\ref{eq:quantum-wronskian}), and Ramanujan's theory of elliptic functions with respect to alternative bases \cite{ramanujan,fricke,borwein,berndt,cooper,shen,shen-egs}, and extensions to modular functions \cite{zagier}. These number theoretic functions are also associated with topological $c=3$ Landau-Ginzburg models \cite{Verlinde:1991ci,Klemm:1991vw} and certain superconformal quantum field theories \cite{Ashok:2016oyh}. In addition, the special class of Chebyshev models also has an interesting geometric interpretation in terms of mirror maps in weighted projective spaces \cite{Klemm:1994wn,Lian:1994zv,Lian:1995js,Lian:1999rq,Brandhuber:1996ng}. These  correspond to the normal forms of functions with unimodular  singularities \cite{arnold}. We hope that our gauge theory-motivated approach may provide new hints towards generalizing some of the results on resurgence in quantum mechanical systems to more general differential equations, and ultimately to implementation for more general quantum field theories, with less supersymmetry. 
Conversely, the results and techniques of SUSY QFT provide new insights into the resurgent spectral properties of higher genus Schr\"odinger spectral problems. We stress that despite these wide-ranging motivations from both physics and mathematics, our approach is extremely elementary, relying solely on some basic classical geometry and all-orders WKB.

\section{General Proof of the Perturbative/Non-perturbative Relation for Genus 1 Systems}
\label{sec:proof}

To see where these novel perturbative/non-perturbative relations   are coming from, we first consider potentials for which the corresponding classical mechanics is genus 1, in the sense that the classical relation between energy $u$ and momentum $p$
\begin{eqnarray}
p^2=2(u-V(x))
\label{eq:curve1}
\end{eqnarray}
defines a genus 1 elliptic curve. There is a systematic procedure in classical geometry for reduction to standard forms, determining the geometric invariants from the elliptic curve \cite{bateman,byrd}, summarized in Appendix A in Section \ref{app:uniformization}.  The examples in \cite{hoe,alvarez-cubic,alvarez,alvarez-howls-silverstone,Dunne:2013ada,Dunne:2014bca,Dunne:2016qix,Misumi:2015dua,Dunne:2016jsr,Codesido:2016dld} are all genus 1 (recall that the Hydrogenic Stark problem separates in parabolic coordinates to anharmonic oscillator problems \cite{alvarez-stark}).

The general proof of the perturbative/non-perturbative relation for genus 1 systems  proceeds in three simple steps:
\begin{enumerate}
\item
For systems where the classical mechanics defines a genus 1 elliptic curve (\ref{eq:curve1}),  a classical action and period is associated with each of the two cycles, $\alpha$ and $\beta$, on the torus:
\begin{eqnarray}
a_0(u)&=&\sqrt{2}\oint_\alpha dx\, \sqrt{u-V(x)} \qquad, \qquad \omega_0(u):= \frac{d}{du} a_0(u) =\frac{1}{\sqrt{2}}\oint_\alpha\frac{dx}{\sqrt{u-V(x)}} \\
a_0^D(u)&=&\sqrt{2} \oint_\beta dx\, \sqrt{u-V(x)} \qquad, \qquad \omega_0^D(u) := \frac{d}{du} a_0^D(u) =\frac{1}{\sqrt{2}}\oint_\beta\frac{dx}{\sqrt{u-V(x)}}
\label{eq:action-period}
\end{eqnarray}
Physically, the cycle $\alpha$ defines a closed path in phase space and characterizes the oscillatory motion around a local minimum of the potential.  At a given energy  $u$, the classical action, $a_0(u)$, is the phase space volume enclosed by this closed path, and the classical period, $\omega_0(u)$, is the period of the oscillatory motion. The orthogonal cycle,  $\beta$, is associated with classical motion under a barrier of the potential. 
At leading semi-classical order, $a_0(u)$ is related to bound state energy eigenvalues, and 
the dual action $a_0^D(u)$ is related to tunneling. 

It is a standard result of the classical geometry of a torus that $a_0(u)$ and $a^D_0(u)$ are explicitly related to one another \cite{bateman,byrd}. They are independent solutions to a linear third-order differential equation known as the Picard-Fuchs equation. (There is a special class of genus 1 systems, to which the Mathieu and symmetric-double-well potentials belong, for which this third-order Picard-Fuchs equation reduces to a second order equation. In this case, further simplifications occur, as discussed in detail in Section \ref{sec:chebyshev}). For every genus 1 system, while the classical actions $a_0(u)$ and $a^D_0(u)$  generically satisfy a third-order equation, the periods $\omega_0(u)$ and $\omega_0^D(u)$  satisfy a second-order equation. Thus the third linearly independent classical action solution is just the trivial constant action solution, and the two non-trivial independent solutions $a_0(u)$ and $a^D_0(u)$ are related to one another by a Wronskian relation. Furthermore, as discussed below there is a natural modular structure underlying the Picard-Fuchs equation, and the two actions $a_0(u)$ and $a^D_0(u)$ can be related by modular transformations. This modular structure means we can easily extend this analysis into the complex plane with well-controlled analytic continuations and monodromies. Thus, there is only one independent classical action function on the torus. We give explicit expressions in a number of concrete  examples in Section  \ref{sec:chebyshev}.

These are classical geometric facts, and our goal is to see how they are modified by quantization. At leading semiclassical order, the classical actions have the following quantum consequences. The classical action $a_0(u)$ enters the quantum description of the corresponding Schr\"odinger equation (\ref{eq:schrodinger})
via the leading WKB Bohr-Sommerfeld quantization condition:
\begin{eqnarray}
a_0(u)=2\pi \hbar \left(N+\frac{1}{2}\right)\qquad, \quad N=0, 1, 2, \dots
\label{eq:bs0}
\end{eqnarray}
This quantization condition is of course modified at higher orders in $\hbar$: see Eq (\ref{eq:bs1}). The classical dual action $a_0^D(u)$ characterizes the leading non-perturbative splitting of the energy levels due to tunneling between wells  \cite{dykhne,connor,keller}:
\begin{eqnarray}
\Delta u(\hbar, N) \sim \frac{2}{\pi}\frac{\partial u}{\partial N}\exp\left[-\frac{1}{2\hbar} {\rm Im}\, a_0^D(u)\right] 
\label{eq:splitting}
\end{eqnarray}
Noting that the classical actions $a_0(u)$ and $a_0^D(u)$ are related to one another, in these expressions we already see (at this level of leading-order WKB)  the germ of the full all-orders result (\ref{eq:one}, \ref{eq:prefactor-sg}). We next consider how these leading WKB expressions can be extended to all orders in $\hbar$, with the full quantum action and dual action also being related.

\item
The next step is to consider all-orders WKB \cite{ddp,kawai-takei,dunham,carl-book} via the formal expansions:
\begin{eqnarray}
a(u, \hbar)&=&\sqrt{2}\Big( \oint_\alpha \sqrt{u-V}dx -\frac{\hbar^2}{2^6}\oint_\alpha \frac{(V^\prime)^2}{(u-V)^{5/2}} dx -\frac{\hbar^4}{2^{13}} \oint_\alpha \left(\frac{49 (V^\prime)^4}{(u-V)^{11/2}} -\frac{16 V^\prime V^{\prime\prime\prime}}{(u-V)^{7/2}} \right) dx 
\nonumber\\
&&  \hskip0.8cm-\dots\Big)
\label{eq:dunham1}
\\
a^D(u, \hbar)&=&\sqrt{2}\Big( \oint_\beta \sqrt{u-V}dx -\frac{\hbar^2}{2^6}\oint_\beta \frac{(V^\prime)^2}{(u-V)^{5/2}} dx -\frac{\hbar^4}{2^{13}} \oint_\beta \left(\frac{49 (V^\prime)^4}{(u-V)^{11/2}} -\frac{16 V^\prime V^{\prime\prime\prime}}{(u-V)^{7/2}} \right) dx
\nonumber\\
&&  \hskip0.8cm-\dots\Big)
\label{eq:dunham2}
\end{eqnarray}
Here the contour integrals encircle the appropriate turning points for the cycles $\alpha$ and $\beta$. Note that the integrands are identical; only the integration cycles differ. This fact will be important below. Also note the homogeneity properties of each term in this expansion under rescalings of $V$, $u$ and $\hbar$, as follows from the Schr\"odinger equation (\ref{eq:schrodinger}).

We write these all-orders WKB expansions for the full quantum action and dual action, and associated periods, as formal expansions
\begin{eqnarray}
a(u, \hbar)&:=& \sum_{n=0}^\infty \hbar^{2n} \, a_n(u) \qquad, \qquad \omega(u, \hbar):= \frac{\partial}{\partial u}a(u, \hbar) := \sum_{n=0}^\infty \hbar^{2n} \, \omega_n(u) 
\label{eq:fullactions1}
\\
a^D(u, \hbar)&:=& \sum_{n=0}^\infty \hbar^{2n} \, a_n^D(u)  \qquad, \qquad \omega^D(u, \hbar):= \frac{\partial}{\partial u}a^D(u, \hbar) := \sum_{n=0}^\infty \hbar^{2n} \, \omega_n^D(u) 
\label{eq:fullactions2}
\end{eqnarray}

For genus 1 systems, all terms $a_n(u)$ and $a_n^D(u)$ in the expansion (\ref{eq:dunham1}, \ref{eq:dunham2}) can be reduced, by suitable changes of variable, to integrals that give the three basic elliptic functions: $\mathbb K$, $\mathbb E$, and $\Pi$. Furthermore, these functions are `closed' under differentiation with repect to the energy $u$ \cite{bateman,byrd}. These facts follow from results  concerning the reduction of genus 1 integrals to standard elliptic forms, and are described in more detail in the language of uniformization in Appendix A in Section \ref{app:uniformization}. Concrete examples are given in Section \ref{sec:chebyshev}.

Furthermore, these reductions of $a_n(u)$ and $a_n^D(u)$ to elliptic function form can be expressed as differential operators $ {\mathcal D}_u^{(n)}$, with respect to the energy $u$, acting on the integrands in (\ref{eq:dunham1}, \ref{eq:dunham2}). And since these are deformable contour integrals, these differential operators may be taken outside the integral, with the consequence that:
\begin{eqnarray}
a_n(u)&=& {\mathcal D}_u^{(n)}\, a_0(u)
\label{eq:same1} \\
a_n^D(u)&=& {\mathcal D}_u^{(n)}\, a_0^D(u) 
\label{eq:same2}
\end{eqnarray}
Here ${\mathcal D}_u^{(n)}$ is a differential operator in $u$.
Again, we present explicit examples in Section \ref{sec:chebyshev}, but the construction applies to all genus 1 systems.

The most important fact is that since these expressions arise from manipulations of the integrands, and the integrands for $a(u, \hbar)$ and $a^D(u, \hbar)$ are the same, the differential operators  ${\mathcal D}_u^{(n)}$ in (\ref{eq:same1}, \ref{eq:same2}) are  the same for $a_n(u)$ and $a_n^D(u)$. Also, since $a_0(u)$ and $a_0^D(u)$ satisfy the same third-order classical Picard-Fuchs equation, these differential operators can be reduced in order so that $a_n(u)$ and $a_n^D(u)$ are expressed as  linear combinations of $a_0(u)$, $a_0^\prime (u)$ and $a_0^{\prime\prime}(u)$:
\begin{eqnarray}
a_n(u)&=& f_n^{(0)}(u) a_0(u)+f_n^{(1)}(u) \frac{d a_0(u)}{d u}+f_n^{(2)}(u) \frac{d^2 a_0(u)}{d u^2} 
\label{eq:ans1}\\
a_n^D(u)&=& f_n^{(0)}(u) a_0^D(u)+f_n^{(1)}(u) \frac{d a_0^D(u)}{d u}+f_n^{(2)}(u) \frac{d^2 a_0^D(u)}{d u^2}
\label{eq:ans2}
\end{eqnarray}
We stress that the coefficient functions $f_n^{(0)}(u)$, $f_n^{(1)}(u)$ and $f_n^{(2)}(u)$ in (\ref{eq:ans1}, \ref{eq:ans2}) are the same for $a_n(u)$ and  $a_n^D(u)$. See for example, equations (\ref{eq:s2areduced}, \ref{eq:s2adreduced}) and (\ref{eq:Lame_higherorder1}, \ref{eq:Lame_higherorder2}) below.
For the special class of genus 1 systems (mentioned above), for which the third-order Picard-Fuchs equation reduces to a second order equation, the  $a_0^{\prime\prime}(u)$ and $(a_0^D)^{\prime\prime}(u)$ terms in (\ref{eq:ans1}, \ref{eq:ans2}) can be expressed in terms of the lower order derivatives, so only two functions, $f_n^{(0)}(u)$ and  $f_n^{(1)}(u)$, are needed. We discuss this special class in detail in Section \ref{sec:chebyshev}. For explicit formulae of this type, see the Appendix B in Section \ref{app:results}.

\item
The general result now follows because the energy spectrum is determined by an exact quantization condition which is a monodromy condition expressed in terms of $a(u, \hbar)$ and $a^D(u, \hbar)$ \cite{voros,ddp,kawai-takei}. Physically, we can think  of the cycle $\alpha$ as describing the usual Bohr-Sommerfeld perturbative physics, while the other cycle $\beta$ describes the non-perturbative tunneling physics. Thus  perturbative  and non-perturbative physics are explicitly and quantitatively related if $a(u, \hbar)$ and $a^D(u, \hbar)$ are explicitly and quantitatively related. 

As a consequence  of (\ref{eq:ans1}, \ref{eq:ans2}),  knowledge of the action $a(u, \hbar)$ to some order $n_c$ in the $\hbar^2$ expansion  (\ref{eq:fullactions1}, \ref{eq:fullactions2}) involves knowing the differential operators ${\mathcal D}_u^{(n)}$ for $n=1, ..., n_c$, and because $a_0^D(u)$ is directly related to $a_0(u)$, we see that  knowledge of $a(u, \hbar)$ to order $n_c$ in the $\hbar^2$ expansion determines $a^D(u, \hbar)$ to the same order $n_c$.
Thus, the non-perturbative information in $a^D(u, \hbar)$, to some order in $\hbar^2$, is encoded in the knowledge of the perturbative information in $a(u, \hbar)$, to the same order in $\hbar^2$. 

For example, standard Rayleigh-Schr\"odinger perturbation theory is obtained by expanding the all-orders Bohr-Sommerfeld quantization condition (generalizing the leading order (\ref{eq:bs0}))
\begin{eqnarray}
a(u, \hbar)=2\pi \hbar \left(N+\frac{1}{2}\right) \qquad, \quad N=0, 1, 2, \dots
\label{eq:bs1}
\end{eqnarray}
at low energies, and then inverting to find $u=u_{\rm pert}(N, \hbar)$. Recall that $N$ labels the perturbative level number. The non-perturbative splitting of energy levels is characterized by the dual action $a^D(u, \hbar)$. The full non-perturbative spectrum is generated from an exact quantization condition (specific to the particular potential under consideration) which requires as input \textit{both} the all-orders WKB action, $a(u, \hbar)$, and the all-orders dual action $a^D(u, \hbar)$, obtained from the two different cycles.  But since $a^D(u, \hbar)$ is actually encoded in $a(u, \hbar)$, this shows that all non-perturbative information is encoded in the perturbative action $a(u, \hbar)$.

Furthermore, this perturbative/non-perturbative relation holds throughout the entire energy spectrum, not just at low energies below the potential barrier. Indeed, it can be analytically continued into the complex $u$ plane. This is because for these genus 1 cases, all expressions for the coefficient functions $a_n(u)$ and $a_n^D(u)$ in (\ref{eq:fullactions1}, \ref{eq:fullactions2}) reduce to elliptic functions, and we have complete control over the analytic continuations and modular transformations connecting different spectral regions.

\end{enumerate}

\section{Chebyshev  Potentials}
\label{sec:chebyshev}

In the previous Section we showed that for general genus 1 systems, the all-orders non-perturbative action $a^D(u, \hbar)$ is encoded in the all-orders perturbative action $a(u, \hbar)$. In this Section we discuss a special subclass of genus 1 potentials for which all these relations and inversions can be made particularly explicit, resulting in simple expressions of the form of a quantum-Matone relation (\ref{eq:quantum-matone}),  or (equivalently) a quantum Wronskian relation (\ref{eq:quantum-wronskian}). 

\subsection{Classical Properties of Chebyshev Potentials}
\label{sec:2nd_pf}

We first specialize to a class of potentials for which  the classical Picard-Fuchs equation for the classical action and dual action reduces to a simpler second-order equation. This reduction of order has important consequences at both the classical and quantum level.

\begin{figure}[htb]
\includegraphics[scale=.4]{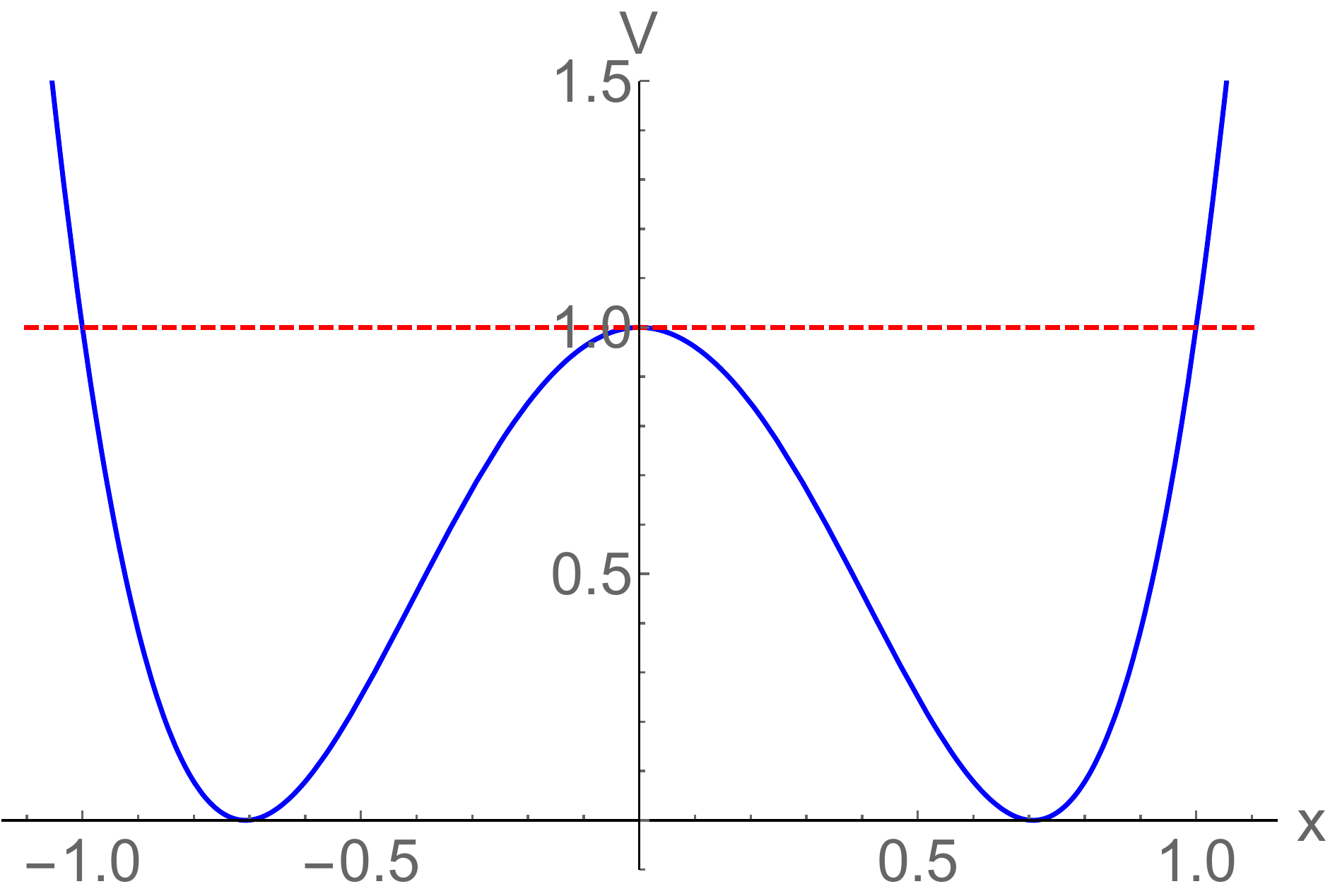}\qquad
\includegraphics[scale=.4]{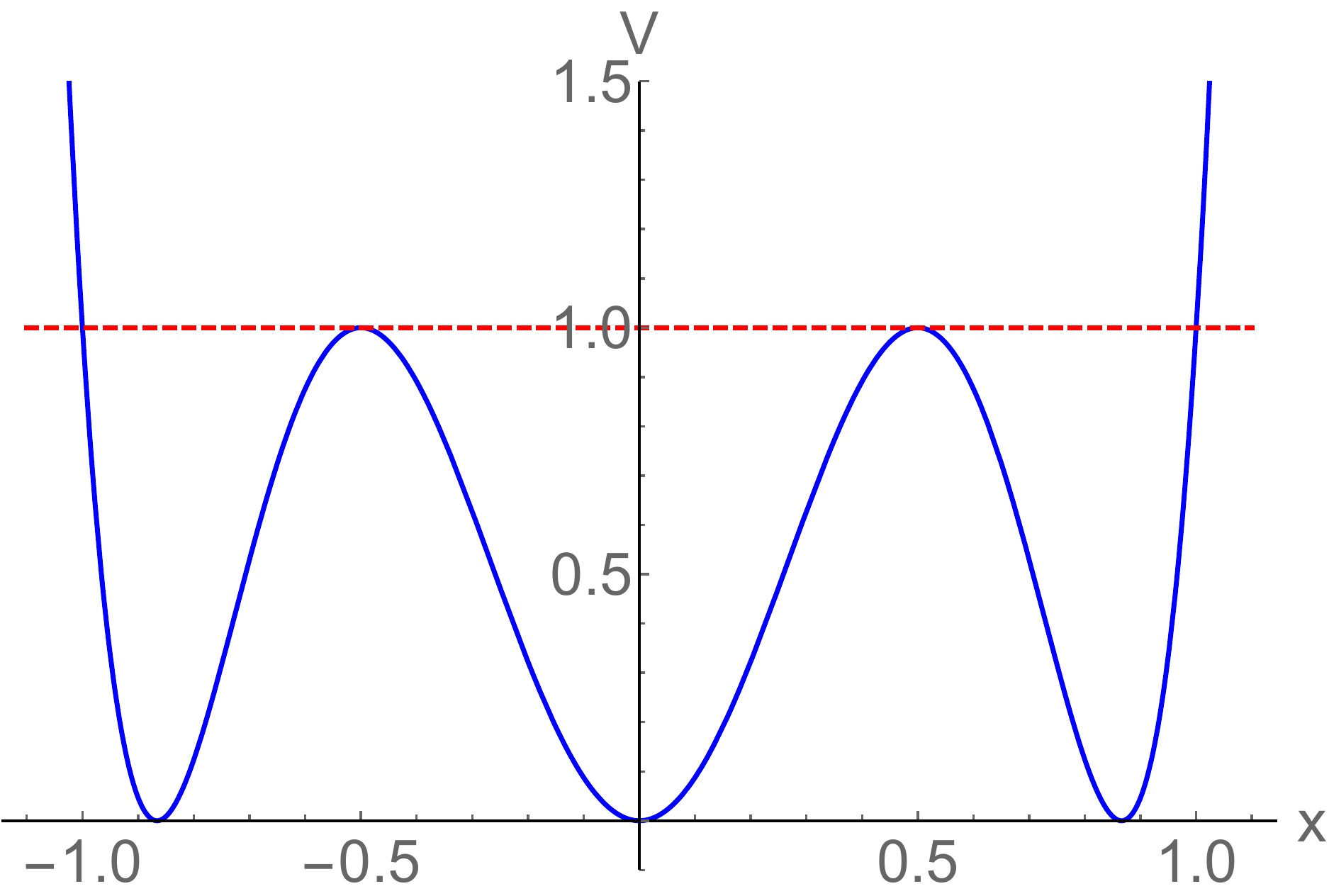}\\
\includegraphics[scale=.4]{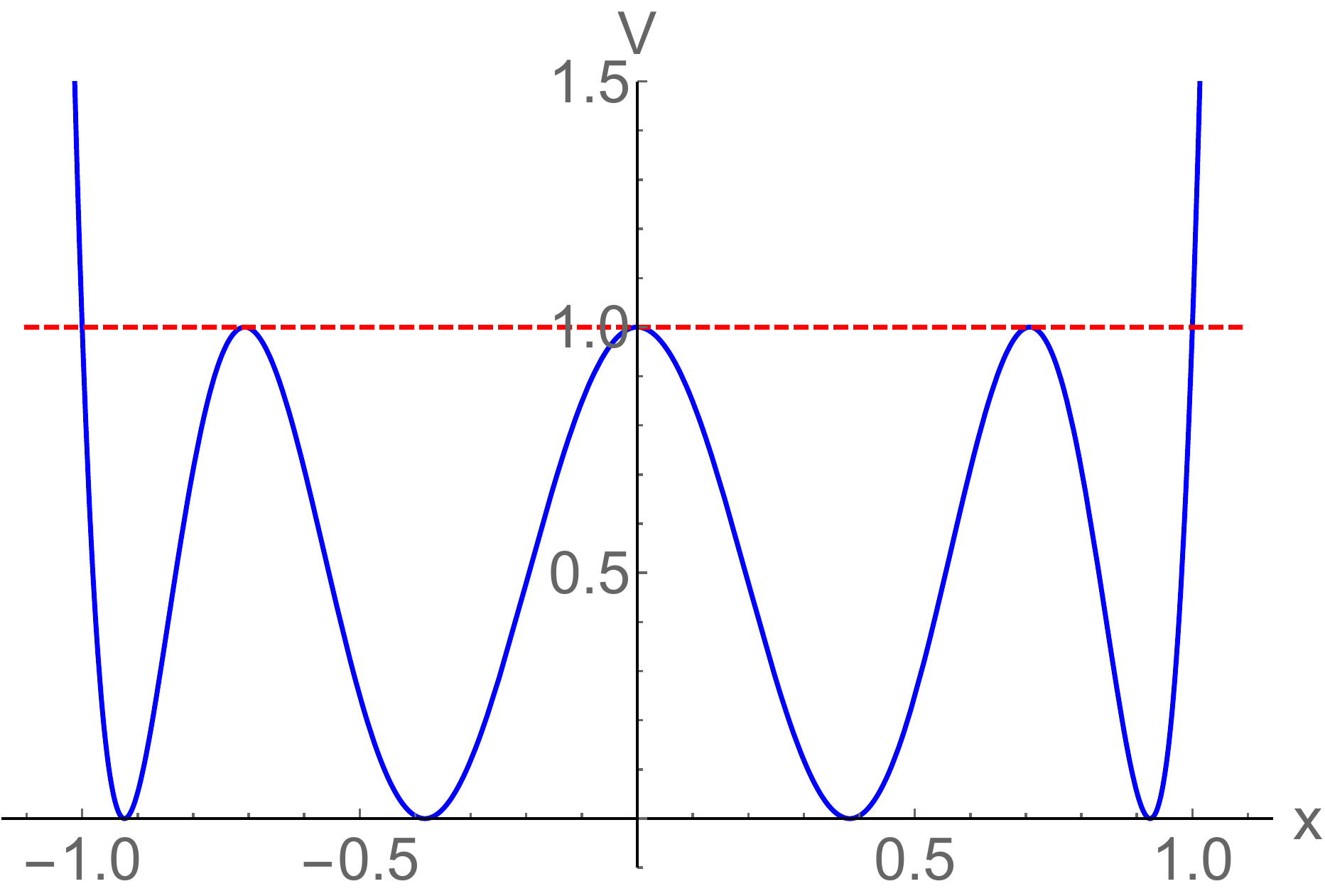}\qquad
\includegraphics[scale=.4]{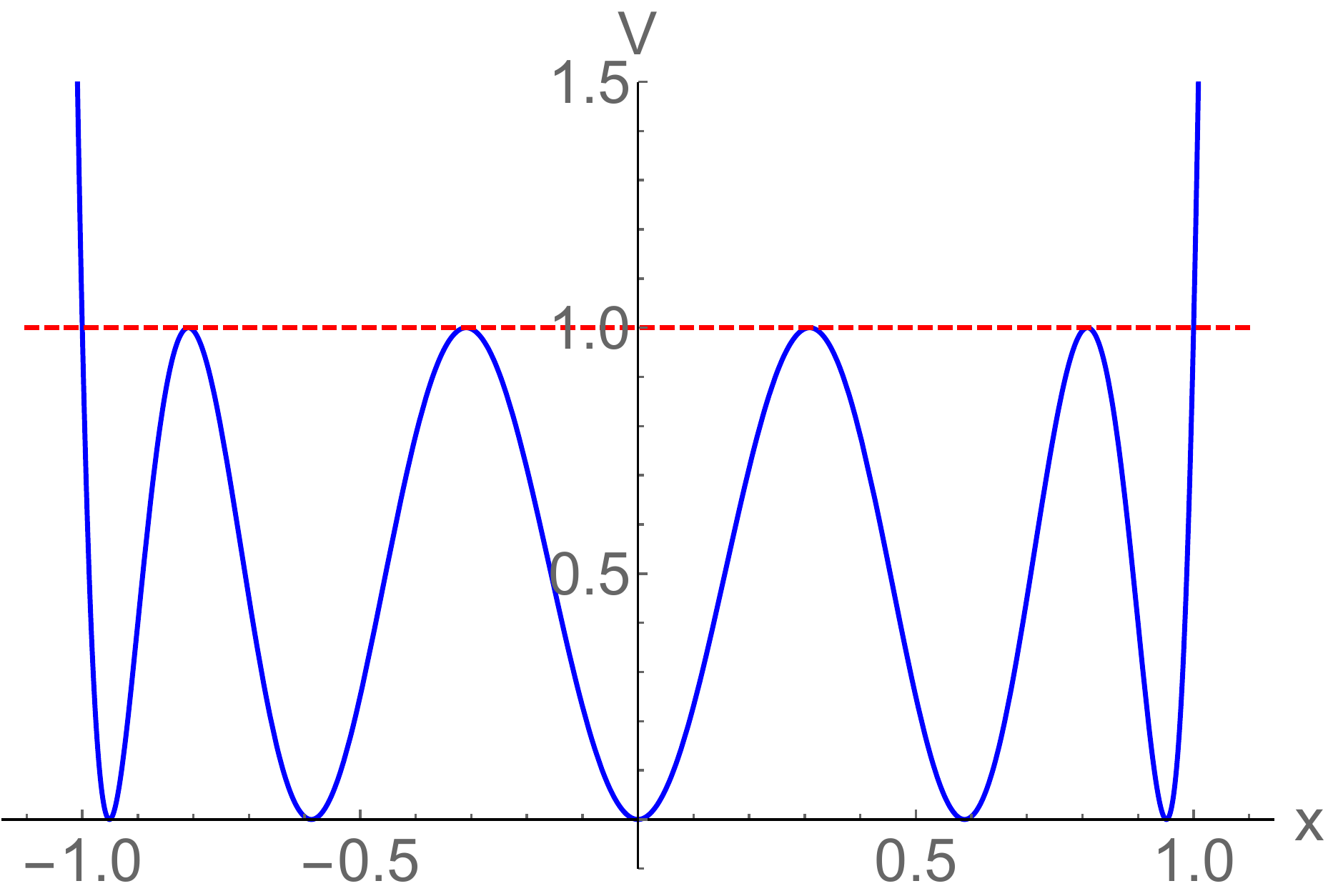}
\caption{The Chebyshev potentials $V(x)=T_m^2(x)$ in (\ref{eq:chebyshev}), for $m= 2, 3, 4, 5$. [The case $m=1$ corresponds to the soluble quadratic potential, i.e. harmonic oscillator.] Note that all maxima are at energy $u=1$, and all minima are at energy $u=0$. This class generalizes the symmetric double well potential  $V(x)=T_2^2(x)$ }
\label{fig:T1}
\end{figure}

\begin{figure}[htb]
\includegraphics[scale=.4]{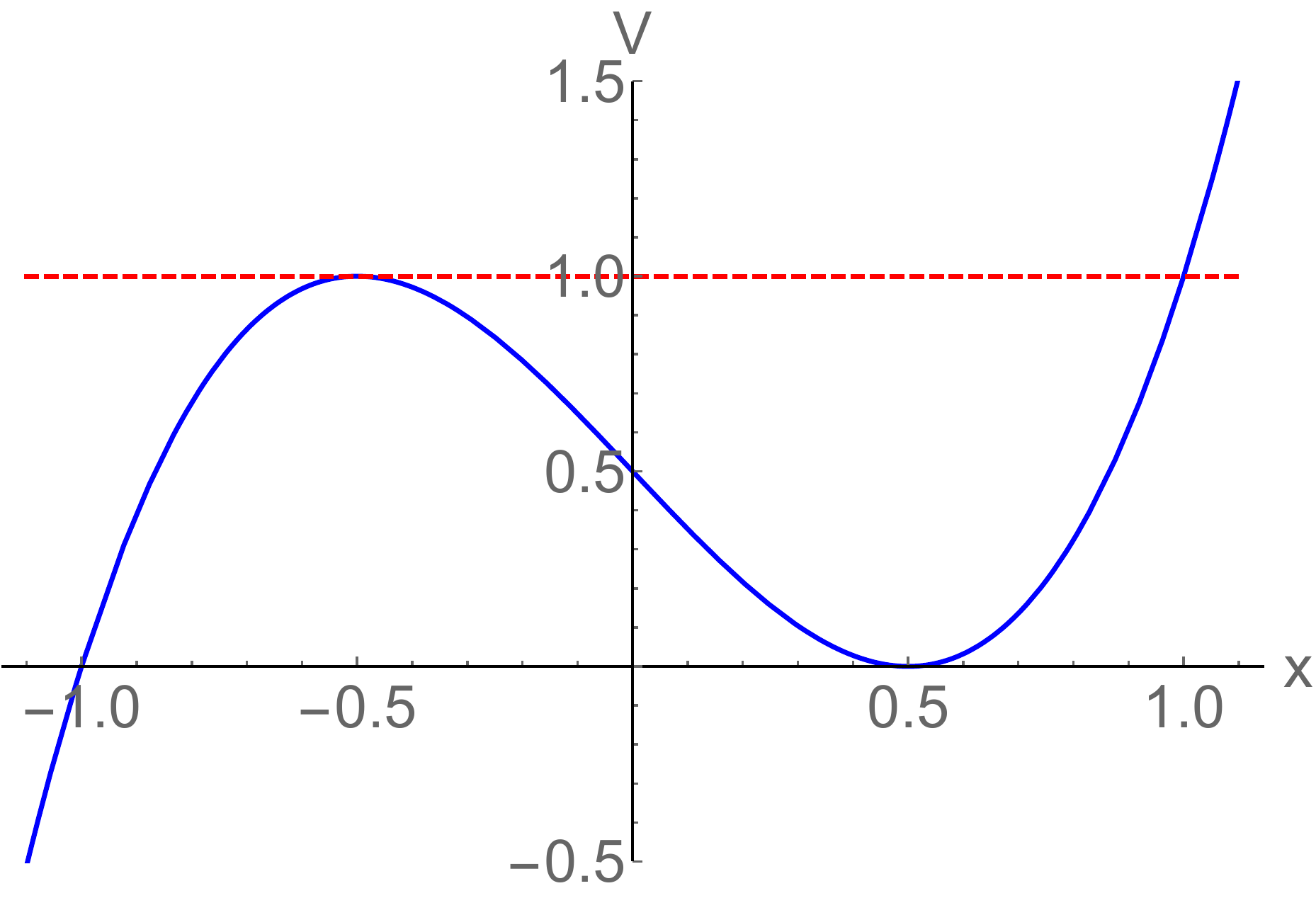}\qquad
\includegraphics[scale=.4]{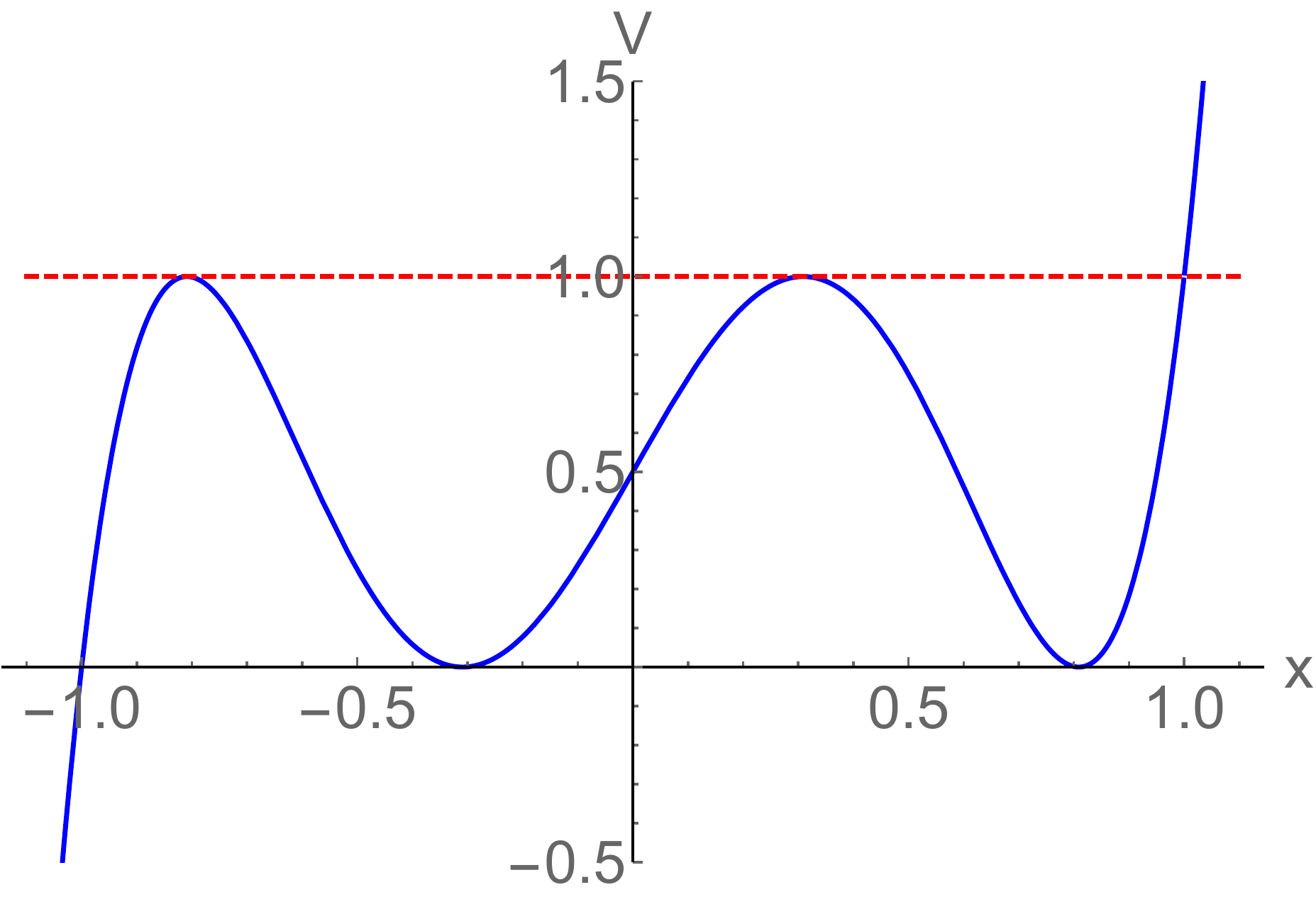}\\
\includegraphics[scale=.4]{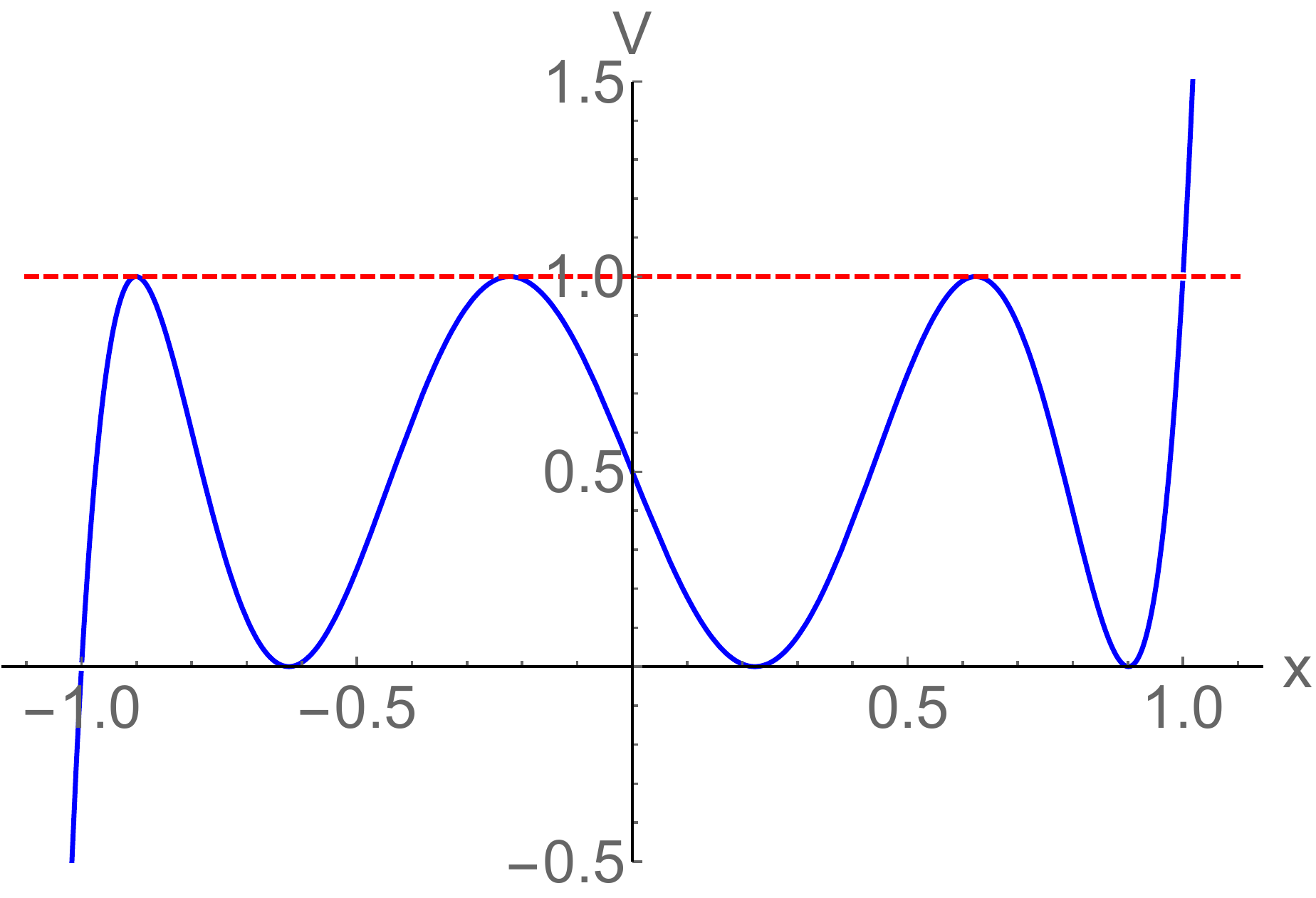}\qquad
\includegraphics[scale=.4]{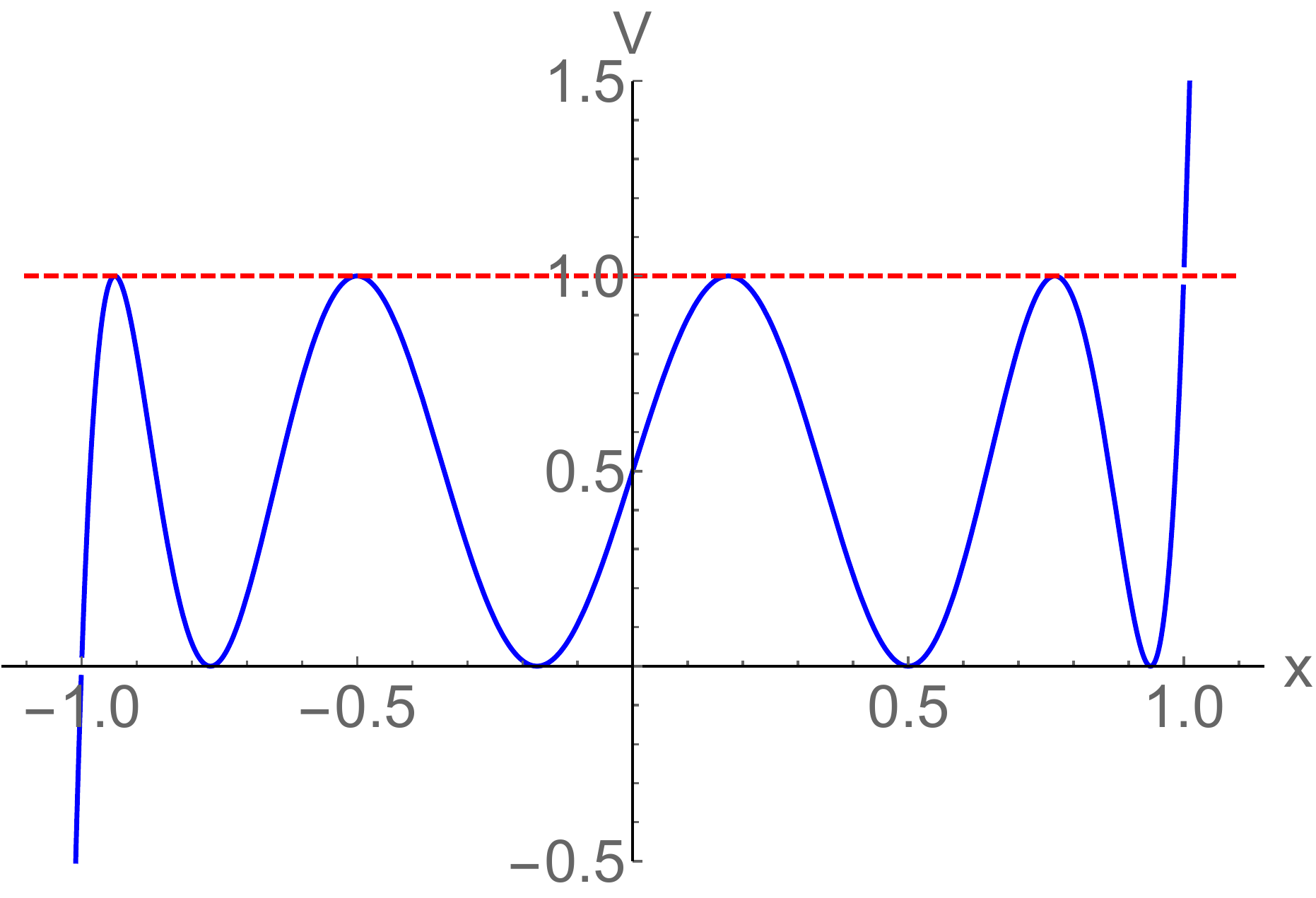}
\caption{The Chebyshev potentials $V(x)=T_m^2(x)$ in (\ref{eq:chebyshev}), for $m= 3/2, 5/2, 7/2, 9/2$. [The case $m=1/2$ corresponds to the  soluble linear potential.] Note that all maxima are at energy $u=1$, and all minima are at energy $u=0$. This class generalizes the cubic oscillator well potential  $V(x)=T_{3/2}^2(x)$.}
\label{fig:T2}
\end{figure}

Consider the class of potentials with
\begin{eqnarray}
V(x)=T_m^2(x)
\label{eq:chebyshev}
\end{eqnarray}
where $T_m(x)$ is the Chebyshev polynomial of the first kind. We choose $m$ to be a half-integer, resulting in smooth potentials for $x\in {\mathbb R}$.
Some plots are shown in Figures \ref{fig:T1} and \ref{fig:T2}. For integer valued $m$ the potentials are even functions,  generalizing the familiar symmetric-double-well potential to a symmetric-$m$-well potential (see Figure \ref{fig:T1}). For half-odd-integer values of $m$, the potentials are odd functions (shifted up by 1/2), generalizing the familiar cubic-oscillator potential (see Figure \ref{fig:T2}). Apart from the trivially soluble cases of $m=1/2$ and $m=1$, each potential has a number of harmonic wells and barriers, with different characteristic frequencies. These Chebyshev potentials are all normalized to have minima at energy $u=0$, and maxima  at energy $u=1$. 

\subsubsection{Classical Actions and Periods for Chebyshev Potentials}
\label{sec:aw-chebyshev}

We consider the energy region $0\leq u\leq 1$ below the potential barrier, and associate a classical period $\omega_0(u)$ and action $a_0(u)$ with each well, and a classical dual period $\omega_0^D(u)$ and dual action $a_0^D(u)$ with each barrier. All periods and actions can be expressed in terms of hypergeometric functions, so the extension to other regions of $u$ is achieved by straightforward analytic continuation.

These Chebyshev potentials (\ref{eq:chebyshev}) have the following remarkable property at the classical level: even though the different wells and barriers have different curvatures, the classical  action $a_0(u)$ in each well is a common function of energy, just with a  different overall multiplicative factor (associated with the harmonic frequency of the well), and the classical dual  action $a_0^D(u)$ in each barrier is another common function of energy, just with a  different overall multiplicative factor (associated with the harmonic frequency of the barrier):
\begin{eqnarray}
\oint_{\alpha_1} dx\, \sqrt{2\big(u-V(x)\big)}&\propto&\oint_{\alpha_2} dx\,\sqrt{2\big(u-V(x)\big)}\propto\dots\propto\oint_{\alpha_{\rm outermost}} dx\, \sqrt{2\big(u-V(x)\big)}=a_0(u) \hskip1cm
\\
\oint_{\beta_1} dx\, \sqrt{2\big(u-V(x)\big)}&\propto&\oint_{\beta_2} dx\,\sqrt{2\big(u-V(x)\big)}\propto\dots\propto\oint_{\beta_{\rm outermost}} dx\,\sqrt{2\big(u-V(x)\big)}=a_0^D(u)  \hskip1cm
\end{eqnarray}
 where $\alpha_i$ and $\beta_i$ denote closed contours around the turning points for the $i^{th}$ well and $i^{th}$  barrier  respectively. Therefore to compute the classical actions and periods we can concentrate on just one well and one barrier: for example, the outermost well and the neighboring outermost barrier.
Computation of the classical action can be done by the basic property of Chebychev polynomials: $T_m(\cos\theta)=\cos(m \theta)$. Then using the identity \cite{ramanujan3} 
\begin{eqnarray}
~_2 F_1\left(\frac{1}{p}, 1-\frac{1}{p}, 1; z\right)=\frac{2}{\pi} \int_0^{{\rm arcsin} \sqrt{z}}d\theta\,  \frac{\cos\left(\left(\frac{2}{p}-1\right)\theta\right)}{\sqrt{z-\sin^2 \theta}} 
\label{eq:ramanujan}
\end{eqnarray}
and its variants, leads to the following basic classical period and classical dual period for the Chebyshev potentials in (\ref{eq:chebyshev}) for the outermost well and the neighboring barrier:
\begin{eqnarray}
\omega_0(u)&=& {\sqrt{2} \,\pi\over m}\sin\left({\pi\over2m}\right) ~_2F_1\left(\frac{1}{2}-\frac{1}{2m}, \frac{1}{2}+\frac{1}{2m}, 1; u\right) 
\label{eq:hypermagic-w} \\
\omega_0^D(u)&=& i\, {\sqrt{2} \,\pi\over m}\sin\left({\pi\over m}\right) \, ~_2F_1\left(\frac{1}{2}-\frac{1}{2m}, \frac{1}{2}+\frac{1}{2m}, 1; 1-u\right) 
\label{eq:hypermagic-wd}
\end{eqnarray}
(Note the different normalization factors of $\omega_0$ and $\omega_0^D$).
 These periods satisfy a  hypergeometric equation:
\begin{eqnarray}
u(1-u) \omega_0^{\prime\prime} +(1-2u) \omega_0^\prime-\frac{1}{4}\left(1-\frac{1}{m^2}\right) \omega_0=0 
\label{eq:hyper}
\end{eqnarray}
This can be integrated once,  implying that the corresponding classical action and dual action are also hypergeometric:
\begin{eqnarray}
a_0(u)&=& {\sqrt{2} \,\pi\over m}\sin\left({\pi\over2m}\right) \, u ~_2F_1\left(\frac{1}{2}-\frac{1}{2m}, \frac{1}{2}+\frac{1}{2m}, 2; u\right)  
\label{eq:hypermagic-a}
\\
a_0^D(u)&=&- i\, {\sqrt{2} \,\pi\over m}\sin\left({\pi\over m}\right) \, (1-u)~_2F_1\left(\frac{1}{2}-\frac{1}{2m}, \frac{1}{2}+\frac{1}{2m}, 2; 1-u\right)
\label{eq:hypermagic-ad}
\end{eqnarray}
The classical action $a_0(u)$ and classical dual action $a_0^D(u)$ are two independent solutions to the {\it second-order} Picard-Fuchs equation:
\begin{eqnarray}
u(1-u)\frac{d^2 a_0}{du^2}=\frac{1}{4}\left(1-\frac{1}{m^2}\right)\, a_0(u)
\label{eq:pfp}
\end{eqnarray}
 The normalized actions and periods are shown in Table \ref{table:ramanujan}, for four special cases to be studied further below. (The normalization for the Mathieu potential is treated separately as it has an infinite number of wells and barriers.)
 
 \begin{table}[htb]
\centerline{\begin{tabular}{|c|c|c|c|c|}  
\hline 
\qquad & \qquad \qquad & \qquad  \qquad & \qquad  \qquad & \qquad  \\
$m$ & potential & $V(x)$& classical period $\omega_0(u)$ & classical action $a_0(u)$  \\ &&&& \\
\hline
&&&& \\
$\infty$ & Mathieu & $\cos^2 x$ \qquad & \qquad  $\sqrt{2} \pi~_2 F_1\left(\frac{1}{2}, \frac{1}{2}, 1; u\right)$  \qquad & \qquad $\sqrt{2} \pi\,u~_2 F_1\left(\frac{1}{2}, \frac{1}{2}, 2; u\right)$  \qquad  \\ 
&&&& \\
\hline
&&&& \\
$3$ & symmetric triple-well &$x^2(3-4x^2)^2$ &  \qquad  $ {\pi\over 3\sqrt{2}}~_2 F_1\left(\frac{1}{3}, \frac{2}{3}, 1; u\right)$  \qquad & \qquad $ {\pi\over 3\sqrt{2}}\,u~_2 F_1\left(\frac{1}{3}, \frac{2}{3}, 2; u\right)$  \qquad  \\
&&&& \\
\hline
&&&& \\
$2$ & symmetric double-well & $(1-2x^2)^2$ &  \qquad  ${\pi\over 2}~_2 F_1\left(\frac{1}{4}, \frac{3}{4}, 1; u\right)$  \qquad & \qquad ${\pi\over 2}\,u~_2 F_1\left(\frac{1}{4}, \frac{3}{4}, 2; u\right)$  \qquad  \\
&&&& \\
\hline
&&&& \\
$3/2$ & cubic oscillator & $\frac{1}{2}(1+x)(1-2x)^2$ &  \qquad  $\pi \sqrt{2\over3} ~_2 F_1\left(\frac{1}{6}, \frac{5}{6}, 1; u\right)$  \qquad & \qquad $\pi \sqrt{2\over3} \,u~_2 F_1\left(\frac{1}{6}, \frac{5}{6}, 2; u\right)$  \qquad  \\
&&&& \\
\hline
\end{tabular}}
\caption{The classical periods and actions for a special class of Chebyshev potentials, which will be studied in more detail in Sections \ref{sec:ramanujan} and \ref{sec:higher}. This is the class of genus 1 Chebyshev potentials  for which the quantum Matone and quantum Wronskian conditions take the special all-orders form in (\ref{eq:quantum-matone}) and (\ref{eq:quantum-wronskian}). For plots of these potentials, see Figure \ref{fig:T3}.}
\label{table:ramanujan}
\end{table}

The fact that the classical actions $a_0(u)$ and $a_0^D(u)$ satisfy the second-order Picard-Fuchs equation (\ref{eq:pfp}) implies that their Wronskian is constant (here normalized to their values in the outermost well and barrier, as described above):
\begin{eqnarray}
a_0(u) \omega_0^D(u)- a_0^D(u) \omega_0(u)=i \,2 \,S_{\cal I} \,T={4\pi i\over m^2-1} \sin^2\left({\pi \over m}\right)
 \label{eq:classical-wronskian}
\end{eqnarray}
  The constants on the right hand side of (\ref{eq:classical-wronskian}) can be understood as follows. The Wronskian \eqref{eq:classical-wronskian} defines a flux on the genus 1 Riemann surface via the Riemann bilinear identity. This flux is a constant and equal to twice the instanton action of the associated barrier, multiplied by the period of the harmonic oscillator around the minimum of the associated well:
  \begin{eqnarray}
  S_{\cal I}&:=&= -i {1\over 2} a_0^D(0)={\sqrt{2}\, m \over m^2-1}4\cos^2\left({\pi\over 2m}\right)\sin\left({\pi\over 2m}\right)\quad\text{(instanton action for outermost barrier)}
  \\
   T&:=&\omega_0(0)= \sqrt{2} \,{\pi \over m}\sin\left({\pi\over 2m}\right)\quad \text{(period of the harmonic oscillator at the outermost well)}\hskip1.5cm
   \label{eq:inst-period}
  \end{eqnarray}
The classical Wronskian identity (\ref{eq:classical-wronskian}) is the classical limit of the quantum Wronskian identity (\ref{eq:quantum-wronskian}). From a geometrical perspective, at the quantum level the flux remains unchanged.

Another interesting fact about these Chebyshev potentials is that the classical action can be expressed in terms of the classical period in an especially simple way. In general, the classical action is an {\it integral} of the classical period with respect to the energy $u$, but for these Chebyshev cases the action can be obtained by {\it differentiation} operations, as follows. Consider the ratio, $t_0$, of the classical dual period and period\footnote{The normalization factor $r$ is not important at this stage, but will be important in the next section when we discuss modular properties. }:
\begin{eqnarray}
t_0(u) :={1\over r}  \frac{\omega_0^D(u)}{\omega_0(u)}\qquad \text{where} \quad r:=4 \cos^2\left({\pi \over 2m}\right)
\label{eq:modtau}
\end{eqnarray}
Then the Picard-Fuchs equation \eqref{eq:pfp}, combined with the classical Wronskian identity \eqref{eq:classical-wronskian} for normalization, implies that
\begin{eqnarray}
\frac{d t_0(u)}{du}=-{i  \over 2\pi} \frac{1}{u(1-u)}{T^2\over\omega_0^2(u)} 
\label{eq:dtdu}
\end{eqnarray}
This has the consequence that the classical actions can be obtained by differentiating the corresponding inverse classical periods:
\begin{eqnarray}
a_0(u)= i\, c(m) \frac{d}{dt_0}\left(\frac{1}{\omega_0}\right)  \quad, \quad 
a_0^D(u)= -i\, c(m) \frac{d}{dt^D_0}\left(\frac{1}{\omega_0^D}\right) 
\label{eq:a0}
\end{eqnarray}
where the constant $c(m)$ is given by
\begin{equation}
c(m):={4\,\pi\over m^2-1}\,\sin^2\left( {\pi \over 2m} \right)={2S_{\cal I} T \over r} 
\end{equation}
 and 
\begin{eqnarray}
t_0^D := \frac{\omega_0(u)}{\omega_0^D(u)}=-{1\over r}\frac{1}{t_0(u)}
\label{eq:modtaud}
\end{eqnarray}
Given that $\omega_0(u)$ and $\omega_0^D(u)$ are the same function (up to a factor of $\sqrt{r}$), with the replacement $u\leftrightarrow 1-u$, we see that the relation between the classical action and the classical dual action is particularly explicit for these Chebyshev potentials. We can express the classical Wronskian condition as:
\begin{eqnarray}
a_0^D(u)=r\,t_0(u) \, a_0(u) -\frac{i\, 2S_{\cal I} T}{\omega_0(u)}\qquad, \qquad a_0(u)=t_0^D(u) \, a_0^D(u) +\frac{i\, 2S_{\cal I} T}{\omega_0^D(u)}
\label{eq:simple-dual}
\end{eqnarray}
Thus, in these cases, given $a_0(u)$, we immediately know $\omega_0(u)$, and hence $\omega_0^D(u)$, and therefore $a_0^D(u)$. This is an explicit example of the general fact that the classical dual action, $a_0^D(u)$, is determined by the classical action, $a_0(u)$, and vice versa. 

Additionally, in terms of the ratio of the periods, the hypergeometric equation for the classical periods \eqref{eq:hyper} translates into a Schwarzian type differential equation 
\begin{eqnarray}
\{t_0,u\}-2Q(u)=0
\quad {\text{where}}\quad \{t_0,u\}:={t_0^{\prime\prime\prime}\over t_0^\prime}-{3\over2}\left({t_0^{\prime\prime}\over t_0^\prime}\right)^2\quad,\quad Q(u)=\frac{(1-m^{-2})(u-1) u+1}{4 (u-1)^2 u^2}\qquad
\label{eq:schwarz}
\end{eqnarray}
This hints at the possibility of an underlying modular structure. We show in the next Section that there is indeed an underlying modular structure, but only for \textit{four} special cases $m= 2, 3, 3/2$ along with the Mathieu equation ($m=\infty$), corresponding to the classical actions and periods shown in Table \ref{table:ramanujan}, with corresponding potentials plotted in Figure \ref{fig:T3}. The reason for this has an interesting  number theoretic explanation, discussed in the next section. 

The form of the Picard-Fuchs equation also leads to a natural definition of a classical prepotential \cite{Matone:1995rx}. To see this, simply invert the Picard-Fuchs equation (\ref{eq:pfp}) by writing the energy $u$ as a function of the classical action $a_0$:
\begin{eqnarray}
u:= {\mathcal G}_0(a_0)
\label{eq:G0}
\end{eqnarray}
Then the fact that both $a_0(u)$ and $a_0^D(u)$ satisfy the Picard-Fuchs equation (\ref{eq:pfp}) implies that ${\mathcal G}_0(a_0)$ satisfies the nonlinear equations
\begin{eqnarray}
{\mathcal G}_0(1-{\mathcal G}_0) \frac{d^2 {\mathcal G}_0}{d a_0^2} &=& - \frac{r^2}{r-1}\, a_0 \,\left(\frac{d {\mathcal G}_0}{d a_0} \right)^3
\\
{\mathcal G}_0(1-{\mathcal G}_0)\left( \frac{d {\mathcal G}_0}{d a_0}\frac{d^2 a_0^D}{d a_0^2} -\frac{d^2 {\mathcal G}_0}{d a_0^2} \frac{d a_0^D}{d a_0} \right) &=& \frac{r^2}{r-1}\, a_0^D \, \left(\frac{d {\mathcal G}_0}{d a_0} \right)^3
\label{eq:nonlinear-G0}
\end{eqnarray}
Combining these two equations, we see that
\begin{eqnarray}
\frac{d {\mathcal G}_0}{d a_0}  \equiv \frac{d u}{d a_0}=\text{constant}\times \left( a_0^D-a_0\, \frac{d a_0^D}{d a_0} \right)
\label{eq:classical-matone1}
\end{eqnarray}
This expression is the classical ($\hbar\to 0$) limit of the all-orders quantum Matone relation\footnote{In the context of gauge theory, the Mathieu system describes the $\Omega$-deformed ($\epsilon_1=\hbar,\epsilon_2=0$) ${\cal N}=2$ $SU(2)$ SUSY gauge theory \cite{Nekrasov:2002qd,Nekrasov:2009rc,Nekrasov:2003rj}. In the $\hbar=0$ limit one obtains the undeformed gauge theory (Seiberg-Witten theory). Thus, equation (\ref{eq:classical-matone1}) is the classical Matone relation associated with the Seiberg-Witten theory.} (\ref{eq:quantum-matone}), found previously for the Mathieu system. This can be integrated in terms of a classical prepotential ${\mathcal F}_0(a_0)$ as
\begin{eqnarray}
{\mathcal G}_0 := u(a_0)= \text{constant}\times \left( a_0\, a_0^D(a_0) -2 {\mathcal F}_0(a_0)\right)
\label{eq:classical-matone2}
\end{eqnarray}
where ${\mathcal F}_0(a_0)$ is defined as
\begin{eqnarray}
a_0^D:= \frac{d {\mathcal F}_0(a_0)}{d a_0}
\label{eq:classical-F0}
\end{eqnarray}
An immediate consequence of the classical Matone relations (\ref{eq:classical-matone1}, \ref{eq:classical-matone2}) is that if we know the expansion of $u(a_0)$ we can immediately deduce the corresponding expansion of $a_0^D(a_0)$ as a function of the classical action $a_0$, and therefore the expansion of the classical prepotential ${\mathcal F}_0(a_0)$.

\subsection{Ramanujan's Theory of Elliptic Functions in Alternative Bases and Hecke Groups}
\label{sec:ramanujan}

The classical mechanics of these Chebyshev potentials carries a strict modular interpretation only for the cases $m= 3, 2, 3/2$,  and $m=\infty$ (with appropriate scaling). This modular structure can be formulated within Ramanujan's theory of elliptic functions in alternative bases \cite{ramanujan,fricke,borwein,berndt,zagier,cooper,shen,shen-egs} and it plays an important role in the corresponding {\it quantum} theories, as we discuss below in Section \ref{sec:higher}. But here we first review some number theoretic results concerning the {\it classical} theories. We start with a familiar example: the Mathieu equation ($m=\infty$), whose classical limit (the simple pendulum) is described by elliptic functions. Then we will discuss the generalizations to the three Chebyshev potentials with $m= 3, 2, 3/2$. 

\subsubsection{Classical Modular Structure of the Mathieu System}

As is well known, the classical mechanics of the Mathieu potential (suitably normalized with $V(x)=\cos^2(x)$) naturally leads to elliptic function expressions for the classical actions and periods in terms of $\mathbb K(u)$ and $\mathbb E(u)$:
\begin{eqnarray}
\omega_0(u)&=&  \pi\,\sqrt{2} ~_2F_1\left(\frac{1}{2}, \frac{1}{2}, 1; u\right)=2\sqrt{2}\,  \mathbb K(u)
\label{eq:elliptic1} \\
a_0(u) &=& \pi\, \sqrt{2} u~_2F_1\left(\frac{1}{2}, \frac{1}{2}, 2; u\right) =4\sqrt{2} \,\left(-(1-u) \mathbb K(u)+\mathbb E(u)\right)
 \label{eq:elliptic2} \\
\omega_0^D(u)&=& i \, \pi\, \sqrt{2} ~_2F_1\left(\frac{1}{2}, \frac{1}{2}, 1; 1-u\right)=i\, 2\sqrt{2} \, \mathbb K(1-u)
\label{eq:elliptic3}  \\
a_0^D(u) &=&-i \, \pi\, \sqrt{2} (1-u)~_2F_1\left(\frac{1}{2}, \frac{1}{2}, 2; 1-u\right) =-i \,4\sqrt{2} \left(-u\, \mathbb K(1-u)+\mathbb E(1-u)\right) 
\label{eq:elliptic4}
\end{eqnarray}
The classical actions satisfy the classical Picard-Fuchs equation (\ref{eq:pfp}) with $m=\infty$. Because of the infinite number of wells and barriers, the normalization factors  different from those in 
(\ref{eq:hypermagic-w}, \ref{eq:hypermagic-wd}, \ref{eq:hypermagic-a}, \ref{eq:hypermagic-ad}). The normalized Mathieu classical actions and periods satisfy the classical Wronskian relation
\begin{eqnarray}
a_0(u) \omega_0^D(u)- a_0^D(u) \omega_0(u)=i \,2 \,S_{\cal I} \,T= 8\pi i 
\label{eq:classical-wronskian-Mathieu}
\end{eqnarray}
which is the classical limit of (\ref{eq:quantum-wronskian}).
This classical Wronskian condition reduces to the familiar Legendre identity for these elliptic functions:
\begin{eqnarray}
\E(u)\K(1-u)+ \K(u) \E(1-u) -\K(u) \K(1-u)=\frac{\pi}{2}  
\label{eq:legendre-identity}
\end{eqnarray}

Furthermore, Jacobi's inversion formula leads to the following inversion, 
\begin{eqnarray}
u(\tau_0) = \frac{\vartheta_2^4(2\tau_0)}{\vartheta_3^4(2\tau_0)} 
\label{eq:jacobi}
\end{eqnarray}
where $\vartheta_i$ are Jacobi theta functions, expressing $u$ in terms of the classical periods via a classical modular parameter\footnote{ As we will see below, the factor of 2 in the denominator stems from the fact that the relevant modular group is the congruence group $\Gamma_0(4)$ and this choice fixes the T transformation to $\tau_0\to\tau_0+1$. In some references the modular parameter is defined without this factor. The relevant T transformation in that case is  $\tau_0\to\tau_0+2$.  } 
\begin{eqnarray}
\tau_0\equiv \frac{\omega^D_0(u)}{2\, \omega_0(u)}=\frac{i\, \mathbb  K(1-u)}{2\, \mathbb K(u)}
\label{eq:mathieu-tau0}
\end{eqnarray}
This is the analogue of the Mirror map that appears in the study of algebraic K3 surfaces, as we discuss further in Section \ref{sec:mirror}.
The classical period and classical action for the Mathieu system can also be expressed directly in terms of the modular parameter $\tau_0$:
\begin{eqnarray}
\omega_0(\tau_0) &=& \vartheta_3^2(2\tau_0) \\
a_0(\tau_0) &=& {2 \sqrt{2} \pi\over 3 \vartheta_3^2(2\tau_0)} \left(E_2 (2\tau_0)+\vartheta_2^4(2\tau_0)-\vartheta_4^4(2\tau_0)\right)
\label{eq:mathieu-inversion}
\end{eqnarray}
where $E_2$ is the classical Eisenstein series. These classical periods and actions  have simple transformation properties under the modular (congruence) group, $\Gamma_0(4)$, generated by:
\begin{eqnarray}
{\text S}:\tau_0\to-{1\over4\tau_0}\quad,\quad  {\text T}:\tau_0\to\tau_0+1
\end{eqnarray}
Using the transformation properties of the Jacobi theta functions, the energy transforms as:
\begin{eqnarray}
{\text S}: u(\tau_0)\to u(-1/(4\tau_0))=1-u(\tau_0)\qquad, \qquad {\text T}: u(\tau_0)\to u(\tau_0+1)=u(\tau_0)
\label{eq:energy-modular}
\end{eqnarray}
Notice that the locations of the degenerate points of the moduli space of the tori, $u=0,1, \infty$, (corresponding to the singular points of the associated Picard-Fuchs equation) either stay invariant or are exchanged under modular transformations, but no new singularity is introduced. 

The classical period, $\omega_0(\tau_0)$, is a modular form of weight 1 in the sense that
\begin{equation}
{\text S}:\omega_0(\tau_0)\to \omega_0(-1/(4\tau_0))=-2i\tau_0\,\omega_0(\tau_0)=-i\, \omega^D_0(\tau_0)\qquad, \qquad
{\text T}:\omega_0(\tau_0)\to \omega_0(\tau_0+1)=\omega_0(\tau_0)
\label{eq:period-modular}
\end{equation}
In other words,  the S transformation interchanges the two periods of the torus, as expected. 
This also follows from the transformations of the energy $u$ \eqref{eq:energy-modular}, and the hypergeometric expressions for the classical periods (\ref{eq:elliptic1}-\ref{eq:elliptic4}).
However, the modular properties of the classical actions, $a_0$ and $a_0^D$, are more interesting. Notice that $a_0$ in (\ref{eq:mathieu-inversion}) includes the second Eisenstein series, $E_2$, which is a \textit{quasi}-modular form, with the S-transformation property
\begin{equation}
{\text S}: E_2(2\tau_0)\to E_2(-1/(2\tau_0))=(2\tau_0)^2 \, E_2(2\tau_0)- {6 i\over \pi}\, (2\tau_0)
\label{eq:E2-quasi}
\end{equation}
This implies that $a_0$ is also a quasi-modular form with S-transformation property:
\begin{eqnarray}
{\text S}: a_0(\tau_0)\to a_0(-1/(4\tau_0))=2i\,\tau_0\, a_0(\tau_0)+{2S_{\cal I} T \over \omega_0(\tau_0)}= i\, a^D_0(\tau_0) 
\label{eq:mathieu_action_S}
\end{eqnarray}
Note that with the Mathieu potential normalized as $V(x)=\cos^2(x)$,  the instanton action is $S_{\cal I}=2\sqrt{2}$, and the period of the harmonic oscillator near the minimum of the well is $T=\sqrt{2}\pi$.  Expression (\ref{eq:mathieu_action_S}) is the classical Wronskian identity \eqref{eq:simple-dual} that we derived earlier, which relates the classical dual action $a_0^D$ to the classical action $a_0$.
It is consistent with the modular transformations (\ref{eq:energy-modular}) of the energy $u$, and with the hypergeometric expressions for the classical actions.
Thus, since the classical actions, $a_0$ and $a_0^D$, enter the leading semiclassical quantization of the  Mathieu system, the connection between the perturbative and non-perturbative physics is directly related to the quasi-modular nature of the actions, and the deviation from modularity is proportional to the instanton action. A similar pattern exists in the all-orders quantum theory, and ultimately is related to the holomorphic anomaly where the non-modular actions can be traded off with modular but non-holomorphic actions  \cite{Huang:2006si,Huang:2011qx,KashaniPoor:2012wb,Huang:2013eja,Billo:2013jba,Billo:2014bja,Ashok:2015cba,Ashok:2016oyh,Codesido:2016dld}.

\begin{figure}[htb]
\includegraphics[scale=.4]{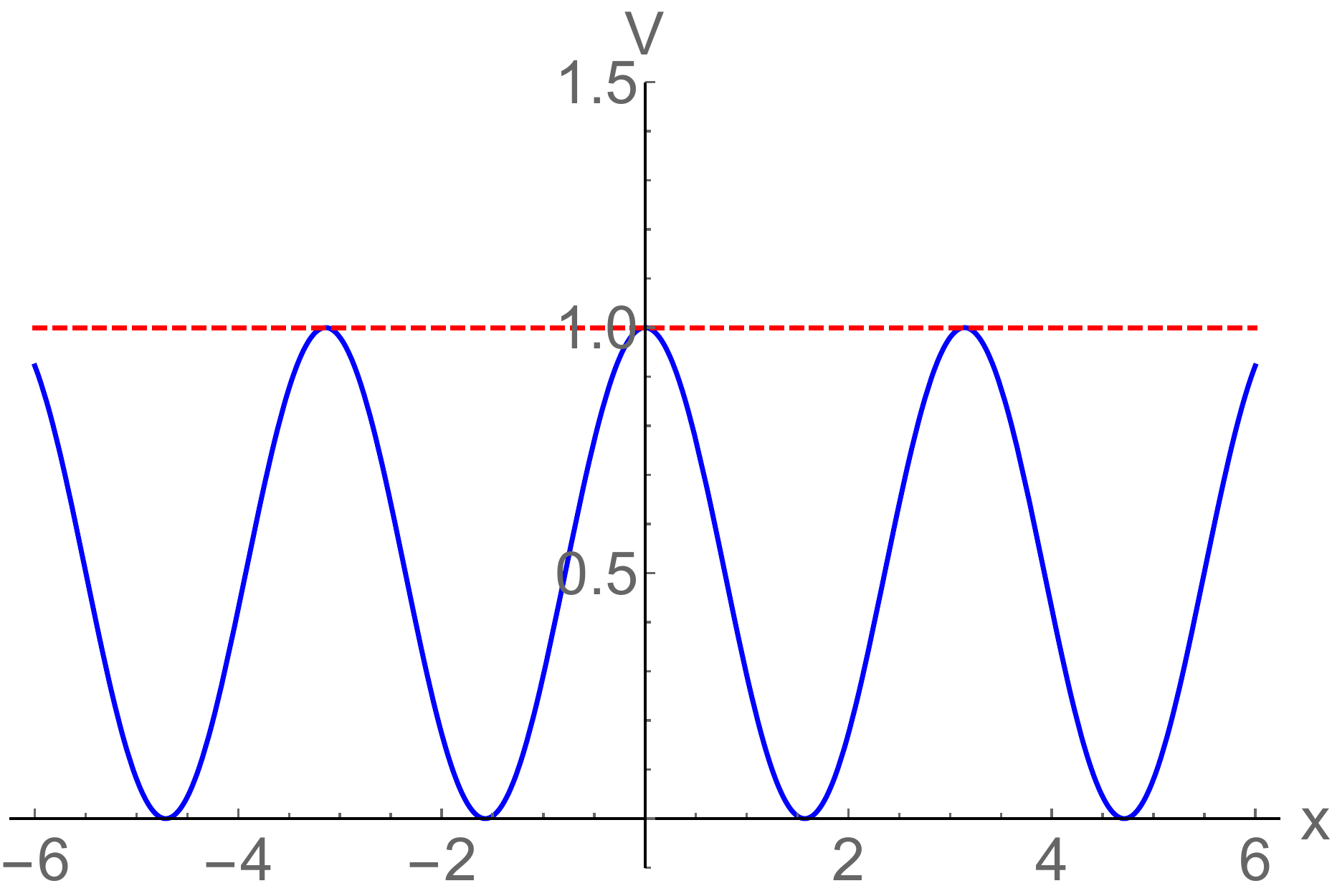}\qquad
\includegraphics[scale=.4]{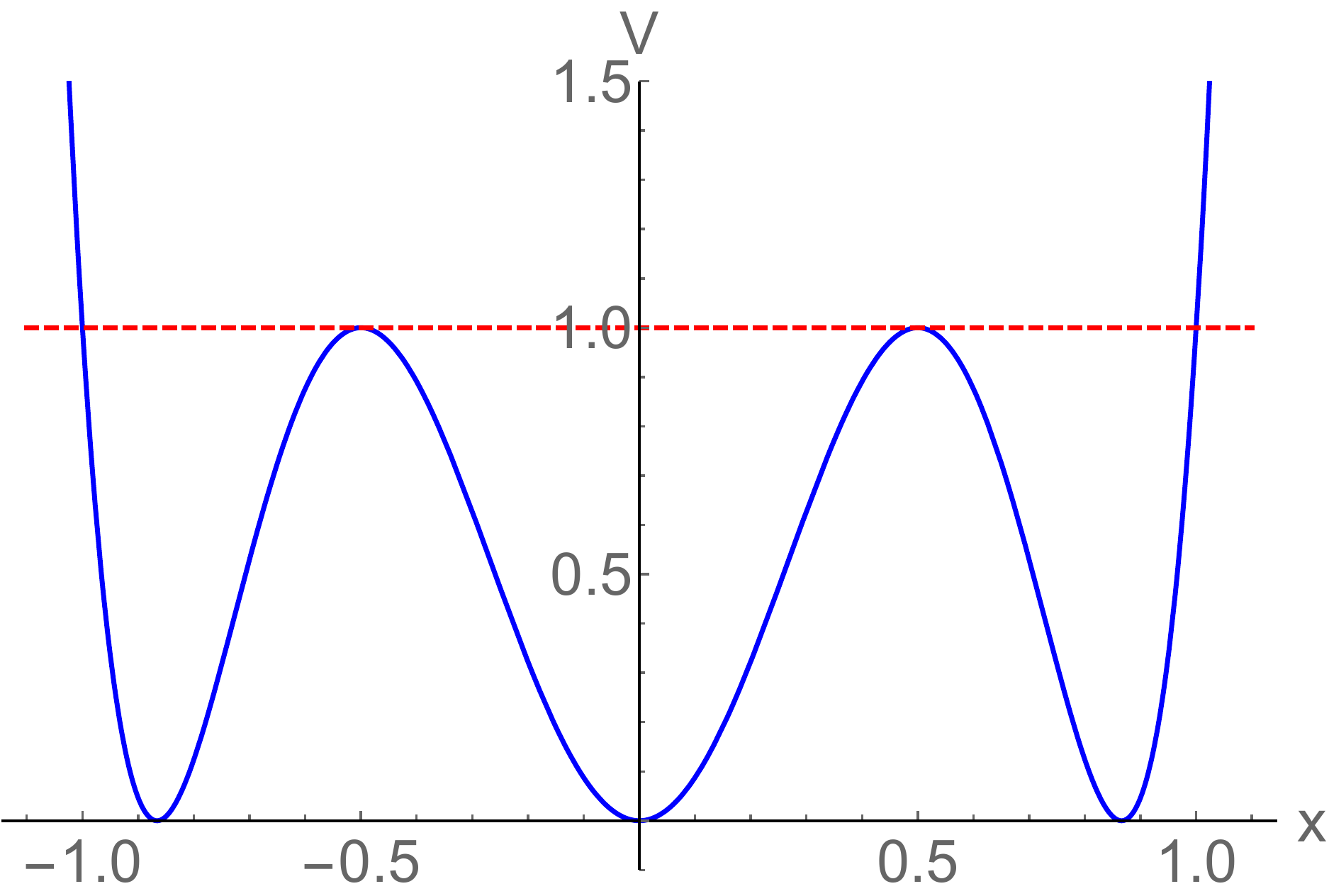}\\
\includegraphics[scale=.4]{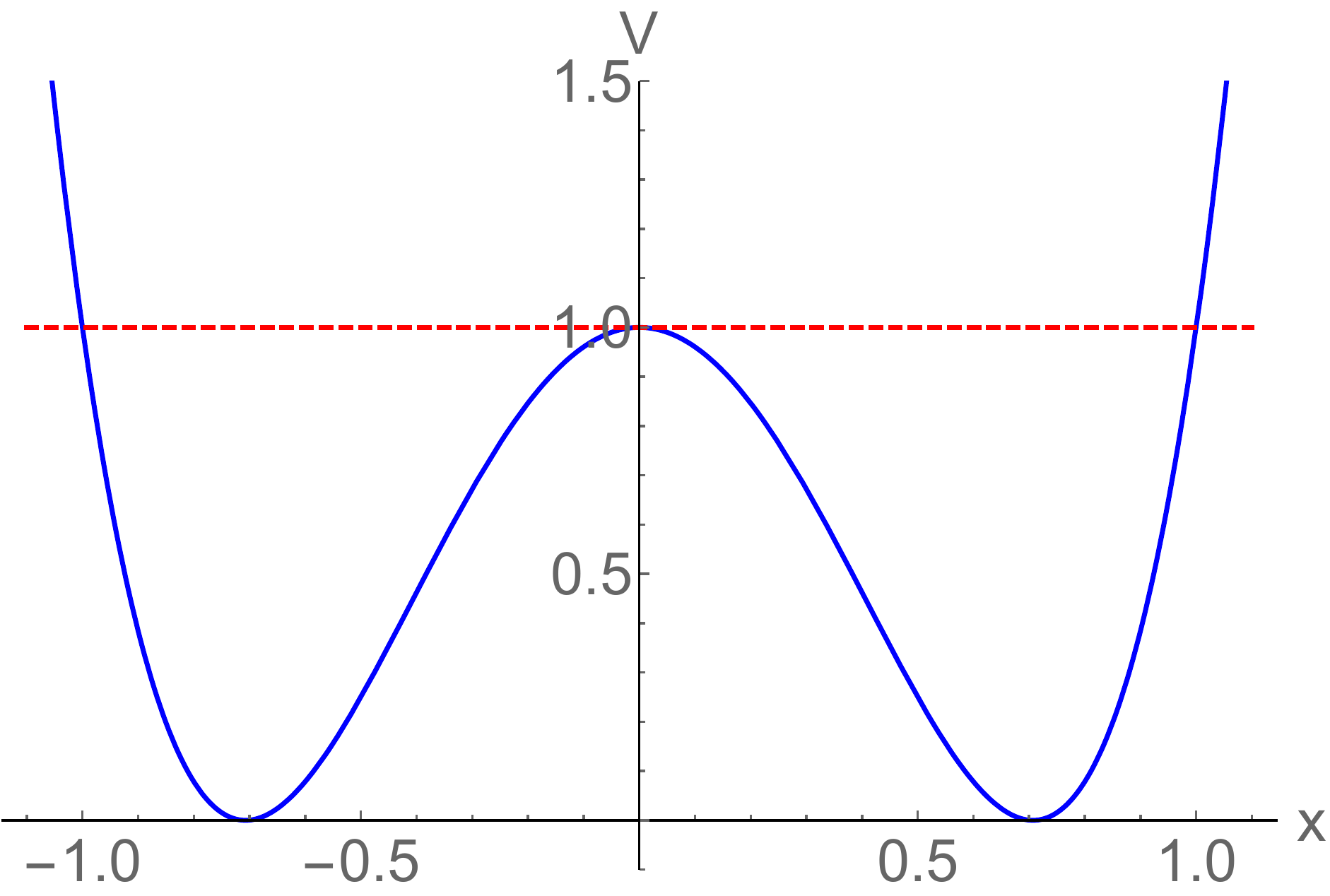}\qquad
\includegraphics[scale=.4]{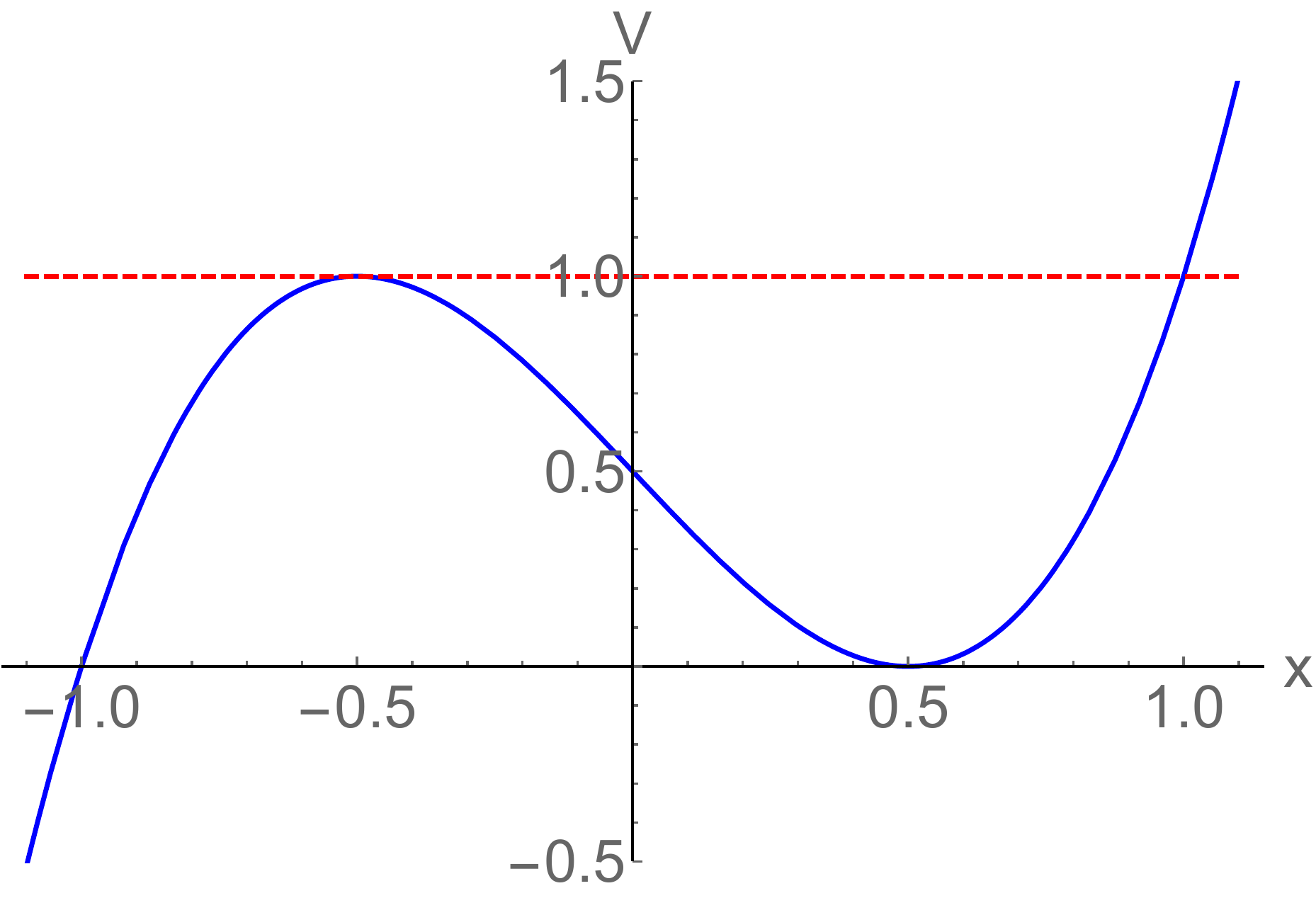}
\caption{Plots of the special class of Chebyshev potentials (respectively, Mathieu, symmetric degenerate triple well, symmetric double well, and cubic oscillator) whose classical actions and periods are associated with Ramanujan's theory of elliptic functions with respect to alternative bases, as featured in Sections \ref{sec:ramanujan}. At the quantum level, these are the potentials for which the quantum Matone and quantum Wronskian conditions take the special all-orders form in (\ref{eq:quantum-matone}) and (\ref{eq:quantum-wronskian}).}
\label{fig:T3}
\end{figure}

\begin{table}
\centerline{\begin{tabular}{|c|c|c|c|c|c|}  
\hline 
\qquad & \qquad \qquad & \qquad  \qquad & \qquad  \qquad & \qquad  & \qquad  \\
 potential & Chebyshev index $m$ &  modular signature $M$  & modular level $r$ & modular group & $\tau_0/( {\omega^D_0\over\omega_0})$   \\ &&&&& \\
\hline
&&&&& \\
Mathieu &$\infty$ &  $2$ \qquad & \qquad  $4$  \qquad & \qquad $\Gamma_0(4)$   \qquad  &  1/2   \qquad  \\ 
&&&&& \\
\hline
&&&&& \\
symmetric triple-well & $3$ &  $3$ &  \qquad  $3$  \qquad & \qquad  $\Gamma_0(3)$  \qquad &  1/3\qquad  \\
&&&&& \\
\hline
&&&&&\\
 symmetric double-well & $2$ & $4$ &  \qquad  $2$  \qquad & \qquad  $\Gamma_0(2)$ \qquad &  1/2  \qquad  \\
&&&&& \\
\hline
&&&&& \\
 cubic oscillator & $3/2$ & $6$ &  \qquad  $1$  \qquad & \qquad  $\Gamma_0(1)$ \qquad &  1  \qquad  \\
&&&&& \\
\hline
\end{tabular}}
\caption{The  modular signature $M$ and level $r$, and their associated QM Chebyshev potential and modular group. This is the the special class of Chebyshev potentials (respectively, Mathieu, symmetric triple well, symmetric double well, and cubic oscillator) associated with Ramanujan's theory of elliptic functions with respect to alternative bases,  featured in Sections \ref{sec:ramanujan} and \ref{sec:higher}, and for which the quantum Matone and quantum Wronskian conditions take the special all-orders form in (\ref{eq:quantum-matone}) and (\ref{eq:quantum-wronskian}).}
\label{table:hecke}
\end{table}

\subsubsection{Ramanujan's Generalized Classical Modular Structure}

There is a generalization of this familiar modular structure of the Mathieu system that includes a special subclass of the Chebyshev potentials introduced  in Section \ref{sec:2nd_pf}. In the process of studying rapidly converging series for $1/\pi$, Ramanujan defined elliptic functions with respect to alternative bases (or `signatures' $M$) as \cite{ramanujan,berndt}
\begin{eqnarray}
\mathbb K_M(u)=  {\pi \over 2}\,~_2F_1\left(\frac{1}{M}, 1-\frac{1}{M}, 1; u\right) \qquad, \qquad \mathbb E_M(u)= {\pi \over 2}\,~_2F_1\left(-1+\frac{1}{M}, 1-\frac{1}{M}, 1; u\right) 
\label{eq:elliptic-ramanujan}
\end{eqnarray}
and developed a corresponding  theory of modular functions. The modular signature $M$ is identified with the Chebyshev index $m$ in (\ref{eq:chebyshev}) as:
\begin{eqnarray}
M:= \frac{2m}{m-1} 
\label{eq:M}
\end{eqnarray}
Compare, for example, the Ramanujan elliptic functions (\ref{eq:elliptic-ramanujan}) with the classical periods (\ref{eq:hypermagic-w}, \ref{eq:hypermagic-wd}) and actions (\ref{eq:hypermagic-a}, \ref{eq:hypermagic-ad}) for the Chebyshev potentials (\ref{eq:chebyshev}).
For $M=2$, corresponding to $m=\infty$, this is the classical theory of modular functions associated with the Mathieu system, reviewed in the previous subsection. For $M=3, 4, 6$ ({\it i.e.}, $m=3, 2, 3/2$) Ramanujan found that much of this classical modular structure generalizes in a straightforward manner. Fricke independently developed the $M=6$ ($m=3/2$) case \cite{fricke}. The special modular signature values, $M= 2, 3, 4, 6$, correspond to the potentials shown in Figure \ref{fig:T3}, and listed in Tables \ref{table:ramanujan} and \ref{table:hecke}. 

For example, the generalized elliptic functions in (\ref{eq:elliptic-ramanujan}) satisfy a generalized Legendre identity:
\begin{eqnarray}
\E_M(u)\K_M(1-u)+ \K_M(u) \E_M(1-u) -\K_M(u) \K_M(1-u)=\frac{\pi}{4}  {M\over M-1} \sin\left({\pi\over M}\right)
\label{eq:gen-legendre-identity}
\end{eqnarray}
The classical Wronskian expression for the  classical Picard-Fuchs equation (\ref{eq:pfp}) of the Chebyshev potentials reduces to this generalized Legendre identity (\ref{eq:gen-legendre-identity}).
It is also useful to characterize these cases by the modular {\it level} parameter $r$, defined earlier in (\ref{eq:modtau}), which is related to $M$ and $m$ as:
\begin{eqnarray}
r=4\cos^2\left(\frac{\pi}{2m}\right)= 4\sin^2\left(\frac{\pi}{M}\right) 
\label{eq:r}
\end{eqnarray}
Table \ref{table:hecke} shows the relations between the three equivalent ways to parametrize the special Chebyshev potentials, using the Chebyshev index $m$, or the modular signature $M$, or the modular level $r$. Note that these special cases are the only smooth Chebyshev  potentials for which $M$ and $r$ are integers.
The significance of the modular level $r$ is discussed below.

Very similar to the Mathieu case, the ratio of the periods is identified with the modular parameter of the torus\footnote{One distinction is that for the Mathieu case, $\tau_0$ is normalized as: $\tau_0= {\omega^D_0\over 2\omega_0}$.}:
\begin{equation}
\tau_0={1\over r} {\omega^D_0 \over\omega_0}={i\over \sqrt{r}} {\,_2F_1\left(\frac{1}{2}-\frac{1}{2m}, \frac{1}{2}+\frac{1}{2m}, 1; 1-u\right) \over \,_2F_1\left(\frac{1}{2}-\frac{1}{2m}, \frac{1}{2}+\frac{1}{2m}, 1; u\right) }
\label{eq:r-tau}
\end{equation}
The related modular group is generated by the transformations
 \begin{eqnarray}
{\text S}:\tau_0\to-{1\over r\tau_0}\quad,\quad  {\text T}:\tau_0\to\tau_0+1
\end{eqnarray}
This is the so-called \textit{Hecke group}\footnote{The Hecke group sometimes is defined via the generators $\tau\to\tau+\sqrt{r}$ and $\tau\to-1/\tau$, which is related to our definition by a rescaling of $\tau$ by a factor of $\sqrt{r}$.}.
From the number theory perspective, the Hecke group coincides with the congruence subgroup $\Gamma_0(r)$, of the modular group $SL(2,\mathbb Z)$ only when $r$ is an integer \cite{cooper,shen}. This is also the condition for Ramanujan's special generalizations to occur. This only happens in the four cases listed in Table \ref{table:hecke}, which we associate with four special Chebyshev potentials. In the number theory literature, these values of $M$ (or $r$) for which the Hecke group is commensurable with the modular group are referred to as the ``arithmetic cases'' \cite{shen}. 

For these arithmetic cases, Jacobi's inversion formula also generalizes \cite{borwein,fricke,berndt,cooper,shen,shen-egs}. Physically, this means that one can invert and express the energy $u$, and also the classical period $\omega_0(u)$ and classical action $a_0(u)$, as explicit number-theoretic functions of the modular parameter $\tau_0$ defined in  \eqref{eq:r-tau}
\begin{eqnarray}
u&=& \frac{1}{2}\left(1-\frac{E_6(\tau_0)}{(E_4(\tau_0))^{3/2}}\right) \qquad, \quad 
r=1
\label{eq:u-r1}\\
\frac{u}{1-u}&=& \left(\frac{r^{\frac{1}{4}}\, \eta(r\, \tau_0)}{\eta(\tau_0)}\right)^{\frac{24}{r-1}}\qquad, \quad 
 r=2, 3, 4
\label{eq:u-r2}
\end{eqnarray}
Here $\eta$ is the Dedekind eta function, and $E_4$ and $E_6$ are the Eisenstein series of weight $4$ and $6$.  A particular modular invariant combination is the Klein $J$-invariant, which is the basic building block of modular invariant functions:  any holomorphic, modular function can be written as a rational function of $J$. In terms of $u$, $J$ is proportional to $1/(u(1-u))$, which is a modular invariant (recall (\ref{eq:energy-modular})). This generalizes to all the arithmetic cases: 
\begin{eqnarray}
J=J_1&=& \frac{432}{u(1-u)} \qquad, \quad 
r=1\\
J_r&=& \frac{r^{6\over r-1}}{u(1-u)} \qquad, \quad 
 r=2, 3, 4
\label{eq:j_r}
\end{eqnarray}
Upon the identification of the modular $q_0$ as:
\begin{eqnarray}
 q_0:= \exp\left(2\pi i \tau_0\right)
 \label{eq:q0}
 \end{eqnarray}
 with $\tau_0$ defined now as in (\ref{eq:r-tau}), 
 we obtain the following $q$ expansions for the $J$ invariants:
 \begin{eqnarray}
 J_1(q_0)&=&{1\over q_0}+ 744 +196884\,q_0+21493760\,q_0^2+864299970\,q_0^3+20245856256\,q_0^4+\dots \nonumber\\
 J_2(q_0)&=&{1\over q_0}+ 104 +4372\,q_0+96256\,q_0^2+1240002\,q_0^3+10698752\,q_0^4+\dots \nonumber\\
  J_3(q_0)&=&{1\over q_0}+ 42 +783\,q_0+8672\,q_0^2+65367\,q_0^3+371520\,q_0^4\dots \nonumber\\
  J_4(q_0)&=&{1\over q_0}+ 24 +276\,q_0+2048\,q_0^2+11202\,q_0^3+49152\,q_0^4+\dots
  \label{eq:j_q_expansions}
 \end{eqnarray} 
Here $J_1$ is the original Klein $J$-invariant, and the other $J_r$ are the generalizations to the associated Hecke groups. The normalization is chosen such that the coefficient of the $q_0^{-1}$ term is unity. 

From these expansions, and the relations (\ref{eq:j_r}), we obtain expansions of the energy in terms of $q_0$:
\begin{eqnarray}
r=1: \frac{1}{432} u(q_0)&=& q_0-312q_0^2+87084 q_0^3-23067968 q_0^4+5930898126 q_0^5+\dots
\label{eq:energy-inversions1}\\
r=2: \frac{1}{64} u(q_0)&=& q_0-40q_0^2+1324 q_0^3-39872 q_0^4+1136334 q_0^5 +\dots
\label{eq:energy-inversions2}\\
r=3: \frac{1}{27} u(q_0)&=& q_0-15q_0^2+171 q_0^3-1679 q_0^4+15054 q_0^5+\dots
\label{eq:energy-inversions3}\\
r=4:  \frac{1}{16} u(q_0)&=& q_0-8 q_0^2+44 q_0^3-192 q_0^4+718 q_0^5 + \dots
\label{eq:energy-inversions4}
\end{eqnarray}
These are consistent with the direct expansions of the expressions in (\ref{eq:u-r1}-\ref{eq:u-r2}). These  Ramanujan modular functions also appear in the study of superconformal SUSY ${\mathcal N}=2$ QFT, for $SU(M)$ with $N_f=2M$ massless hypermultiplets \cite{Argyres:1994xh,Argyres:1995wt,Argyres:1995fw,Douglas:1995nw,Hanany:1995na,Minahan:1995er,Ashok:2006du,Ashok:2015cba,Ashok:2016oyh}. For $M=2, 3, 4, 6$, these were called "arithmetic" cases in \cite{Ashok:2015cba,Ashok:2016oyh},  and these same inversions appear in the expressions for the $q_0$ in terms of the couplings.

The classical periods are modular forms and they can be expressed explicitly in terms of the modular parameter $\tau_0$:
\begin{eqnarray}
\omega_0(\tau_0)&=& \sqrt{2\over3}\pi\left(E_4(\tau_0)\right)^{1/4} \qquad, 
\quad r=1
\\
\omega_0(\tau_0)&=&{\sqrt{2}\,\pi\over m}\sin\left({\pi\over 2m}\right) \sqrt{\frac{r E_2(r\,\tau_0)-E_2(\tau_0)}{ (r-1)}}\qquad, \quad 
r=2, 3
\\
\omega_0(\tau_0)&=&\sqrt{2}\,\pi\ \sqrt{\frac{r E_2(r\,\tau_0)-E_2(\tau_0)}{ (r-1)}}\qquad, \quad 
r=4
\label{eq:w0_r}
\end{eqnarray}
Even though the appearance of $E_2$ might suggest otherwise,  for each $r=1,2,3,4$, the period $\omega_0(\tau_0)$ is  in fact  modular forms of weight 1,  transforming under the related arithmetic Hecke group. 

The classical actions are 
\begin{eqnarray}
a_0(\tau_0)&=& \frac{\sqrt{6}}{5} \pi\frac{\left(E_2(\tau_0)E_4(\tau_0)-E_6(\tau_0)\right)}{\left(E_4(\tau_0)\right)^{5/4}} \qquad,  \qquad r=1
\\
a_0(\tau_0)&=&\frac{\sqrt{2}\,m \,\pi\,\sin\left({\pi\over 2m}\right)\sqrt{r-1}}{6\,(m^2-1)}\,\frac{r^2\left(E_2^2(r\, \tau_0)-E_4(r\,\tau_0) \right)-E_2^2( \tau_0)+E_4(\tau_0)}{\left(r E_2(r\, \tau_0)-E_2(\tau_0) \right)^{3/2}}\qquad, \quad 
r=2, 3\hskip1.5cm
\\
a_0(\tau_0)&=&{\sqrt{2}\,\pi\over 6}\sqrt{r-1}\, \frac{r^2\left(E_2^2(r\, \tau_0)-E_4(r\,\tau_0) \right)-E_2^2( \tau_0)+E_4(\tau_0)}{\left(r E_2(r\, \tau_0)-E_2(\tau_0) \right)^{3/2}}\qquad, \quad 
r=4
\label{eq:a0-r}
\end{eqnarray}
These expressions for $a_0(\tau_0)$ are most easily derived from (\ref{eq:a0}) in terms of the $\tau_0$ derivative of the inverse of the classical period. The dual expressions for $\omega_0^D$ and $a_0^D$ are obtained by the classical modular transformation, $\tau_0\to -1/(r\tau_0)$:
\begin{eqnarray}
a^D_0(\tau_0)&=&-i a_0(-1/(r\tau_0))=r \tau_0 a_0+i { S_{\cal I} T/r \over \omega_0(\tau_0)}\quad r=1,2,3 \\
a^D_0(\tau_0)&=&-i a_0(-1/(r\tau_0))=2 \tau_0 a_0+i {8\pi \over \omega_0(\tau_0)}\quad r=4
\end{eqnarray}
 where the instanton action, $S_{\cal I}$,  and the period, $T$, are defined in \eqref{eq:inst-period}. These expressions reflect the fact that the classical actions are {\it quasi-modular} forms (in contrast to the periods, which are true modular forms of weight 1).  
 
These are all classical results which can be understood in terms of the classical Picard-Fuchs
equations (\ref{eq:pfp}). The special values $M=2, 3, 4, 6$, corresponding to the special potentials listed in Table \ref{table:hecke}, are singled out for number theoretic reasons, namely the overlap between the modular and Hecke groups. 
In Section \ref{sec:higher} we show that this same special set is precisely the class of potentials for which the classical modular structure generalizes to the all-orders quantum Matone  (\ref{eq:quantum-matone}) and quantum Wronskian (\ref{eq:quantum-wronskian}) expressions, found previously for the Mathieu system.

\subsection{Chebyshev Potentials and Mirror Curves}
\label{sec:mirror}

In addition to having an interesting number theoretic interpretation, as discussed in the previous section, the classical ``arithmetic'' Chebyshev systems also have an interesting geometric interpretation in terms of mirror symmetry of hypersurfaces in weighted projective space.
 This can be traced directly to the hypergeometric form of the classical Picard-Fuchs equations (\ref{eq:pfp}), and the associated Schwarzian form discussed in Section \ref{sec:aw-chebyshev}. Indeed, it is well known \cite{nehari}  that ratios of solutions to the hypergeometric equation describe conformal maps of spherical triangles. These form the simplest nontrivial example of a mirror map \cite{Klemm:1994wn,Lian:1994zv,Lian:1995js,Lian:1999rq}:
\begin{eqnarray}
q=e^{2\pi i t}\qquad, \qquad t\equiv \frac{i ~_2F_1(a, b, 1, 1-z)}{~_2F_1(a, b, 1, z)}
\label{eq:mirror}
\end{eqnarray}
As noted in \cite{Klemm:1994wn}, the one-parameter family of cubic curves in $\mathbb P^2$
\begin{eqnarray} 
X_s:\quad x_1^3+x_2^3+x_3^3-s\, x_1\, x_2\,x_3=0
\label{eq:constraint}
\end{eqnarray}
can be transformed to an elliptic curve in Weierstrass form, with invariants: $g_2=3s(8+s^3)$ and $g_3=8+20s^3-s^6$. Then, identifying $z=1/s^3$, the mirror map is given by the modular $J$ function $J(t)$, which has an integer coefficient expansion in powers of $q$:
\begin{eqnarray}
1728\, J(q)=\frac{1}{q}+744+196884 q +21493760 q^2+\dots 
\label{eq:j1}
\end{eqnarray}
which we recognize from the equation for $J_1(q_0)$ in (\ref{eq:j_q_expansions}). 

Furthermore, Reference \cite{Klemm:1994wn} identifies three realizations of elliptic curves as hypersurfaces in weighted projective spaces, as shown in Table \ref{table:klemm} (see the Table below equation (3.10) in  \cite{Klemm:1994wn}).
 These examples correspond to the unique normal forms of functions with unimodular  parabolic singularities \cite{arnold}, which appear in the description of $c=3$ topological Landau-Ginzburg models \cite{Verlinde:1991ci,Klemm:1991vw}. Inverting the mirror map,  $z$ is expanded in terms of $q$ as:
\begin{eqnarray}
P_8\, : \quad z(q)&=& q-15q^2+171 q^3-1679 q^4+15054 q^5 -126981 q^6+\dots 
\label{eq:mirror-inversions1}\\
X_9\, : \quad z(q)&=& q-40q^2+1324 q^3-39872 q^4+1136334 q^5 -31239904 q^6+\dots 
\label{eq:mirror-inversions2}\\
J_{10}\, : \quad z(q)&=& q-312q^2+87084 q^3-23067968 q^4+5930898126 q^5 -1495818530208 q^6+\dots
\label{eq:mirror-inversions3}
\end{eqnarray}

\begin{table}
\centerline{\begin{tabular}{|c|c|c|c|c|}  
\hline 
\qquad & \qquad \qquad & \qquad  \qquad & \qquad  \qquad & \qquad  \\
~ & projective space & constraint & diff. operator & $1728\times J(z)$ \\ &&&& \\
\hline
&&&& \\
$P_8$ &  $\mathbb P^2(1,1,1)$  & $x_1^3+x_2^3+x_3^3-\frac{1}{z^{1/3}} x_1\, x_2\,x_3=0$  & $\theta^2-3z(3\theta+2)(3\theta+1)$  & $\frac{(1+216 z)^3}{z(1-27 z)^3}$  \\ 
&&&& \\
\hline
&&&& \\
$X_9$ & $\mathbb P^2(1,1, 2)$ & $x_1^4+x_2^4+x_3^2-\frac{1}{z^{1/4}} x_1\, x_2\,x_3=0$ & $\theta^2-4z(4\theta+3)(4\theta+1)$ & $\frac{(1+192 z)^3}{z(1-64 z)^2}$ \\
&&&& \\
\hline
&&&& \\
$J_{10}$ & $\mathbb P^2(1, 2, 3)$ & $x_1^6+x_2^3+x_3^2-\frac{1}{z^{1/6}} x_1\, x_2\,x_3=0$ & $\theta^2-12z(6\theta+5)(6\theta+1)$ & $\frac{1}{z(1-432 z)}$ \\
&&&& \\
\hline
\end{tabular}}
\caption{List, from \cite{Klemm:1994wn}, of three realizations of elliptic curves as hypersurfaces in weighted projective spaces. Here $\theta$ is the differential operator $\theta\equiv z\partial_z$. These same examples appear in Arnold's classification of normal forms of functions with unimodular parabolic singularities \cite{arnold}.}
\label{table:klemm}
\end{table}

\begin{table}
\centerline{\begin{tabular}{|c|c|c|c|c|}  
\hline 
\qquad & \qquad \qquad & \qquad  \qquad & \qquad  \qquad & \qquad  \\
potential (in Chebyshev form) & Chebyshev $m$ & hypergeometric operator  & $J(u)$ & identification \\ &&&& \\
\hline
&&&& \\
triple-well: $V(x)=T_3(x)^2$ &  $3$  & $u(1-u)\frac{d^2}{du^2}+(1-2u)\frac{d}{du}-\frac{2}{9}$  & $\frac{1}{64}\ \frac{(9-8 u)^3}{u^3 (1-u)}$  &  $u\leftrightarrow 1-27 z$ \\ 
&&&& \\
\hline
&&&& \\
double-well: $V(x)=T_2(x)^2$ & $2$  & $u(1-u)\frac{d^2}{du^2}+(1-2u)\frac{d}{du}-\frac{3}{16}$  & $\frac{1}{27}\,\frac{(4-3u)^3}{u^2(1-u)}$  &  $u\leftrightarrow 1-64 z$ \\
&&&& \\
\hline
&&&& \\
cubic oscillator: $V(x)=T_{3/2}(x)^2$ &  $\frac{3}{2}$  & $u(1-u)\frac{d^2}{du^2}+(1-2u)\frac{d}{du}-\frac{5}{36}$  & $\frac{1}{4 u (1-u)}$  &  $u\leftrightarrow 1-432 z$ \\
&&&& \\
\hline
\end{tabular}}
\caption{Arithmetic Chebyshev potentials, with their associated hypergeometric differential operator for the periods, and the corresponding modular $J(u)$ derived from the elliptic curve [see Appendix A in Section \ref{app:uniformization}]. The last column shows the conversion needed for the comparison with Table \ref{table:klemm}.}
\label{table:p2}
\end{table}

Now compare this with the modular $J(u)$ function obtained from the uniformization of the elliptic curves for the Chebyshev potentials, as described in Appendix A in Section \ref{app:uniformization}. The relevant data is reproduced here in Table \ref{table:p2}. There is a one-to-one mapping between the elliptic curves from the $\mathbb P^2$ hypersurfaces in Table \ref{table:klemm}, and those coming from the new ``arithmetic" Chebyshev potentials. 
\begin{eqnarray}
P_8 :  \mathbb P^2(1,1,1) &\longleftrightarrow& \text{symmetric degenerate triple-well potential}\\
X_9 :  \mathbb P^2(1,1,2) &\longleftrightarrow& \text{symmetric double-well potential}\\
J_{10} :  \mathbb P^2(1, 2, 3) &\longleftrightarrow& \text{cubic oscillator potential}
\end{eqnarray}
For example, with the simple (linear) identifications between $z$ and the Chebyshev energy $u$ shown in Table \ref{table:p2}, the differential operators associated with the mirror maps reduce to the hypergeometric operators acting on the Chebyshev periods, as in (\ref{eq:hyper}), with the appropriate choice of Chebyshev index $m$. Furthermore, the $J(u)$ invariants agree precisely with the $J(z)$ invariants, with the overall normalization factor of $1728$.
And note that the inversions listed in (\ref{eq:mirror-inversions1} - \ref{eq:mirror-inversions3}) agree precisely with the energy inversions in terms of $q_0$ for the Chebyshev systems in (\ref{eq:energy-inversions1} - \ref{eq:energy-inversions3}).

We  can also identify Chebyshev potentials with specific ${\mathcal N}=2$ SUSY quantum field theories. For example, the Mathieu system is well known to be associated with the pure $SU(2)$ ${\mathcal N}=2$ SUSY quantum field theory \cite{Mironov:2009uv,Mironov:2009ib,He:2010xa,Huang:2011qx,KashaniPoor:2012wb,Krefl:2013bsa,Gorsky:2014lia,Basar:2015xna,kpt,Ashok:2016yxz}. This correspondence can be seen already at the classical level by comparing the elliptic curve data. For example, the elliptic curves for the $SU(2)$ ${\mathcal N}=2$ SUSY quantum field theories with $N_f$ massless flavors of matter fields are shown in Table \ref{table:curves}, along with the corresponding modular $J$ function obtained by uniformization (see, for example, \cite{Brandhuber:1996ng}, and the procedure summarized in Appendix A in Section \ref{app:uniformization}). Table \ref{table:curves} also shows the identifications with the quantum mechanical energy $u$, and the converted expression for the $J$ invariant. This leads to the following correspondences:
\begin{table}
\centerline{\begin{tabular}{|c|c|c|c|c|}  
\hline 
 &  &  &  &  \\
$N_f$ & elliptic curve & $J(v, \Lambda)$ & identification & $J(u)$  \\ &&&& \\
\hline
&&&& \\
0 &  $y^2=(x^2-\Lambda^2)(x-v)$  & $\frac{1}{27} \frac{(v^2+3\Lambda^4)^3}{\Lambda^4(v^2-\Lambda^4)^2}$  & $u=\frac{1}{2}+\frac{v}{2\Lambda^2}$  &  $\frac{4}{27}\frac{(1-u+u^2)^3}{u^2(1-u)^2}$ \\ 
&&&& \\
\hline
&&&& \\
1 & $y^2=x^2(x-v)-\frac{\Lambda^6}{64}$  & $-\frac{16384}{27} \frac{v^6}{\Lambda^6(27\Lambda^6+256v^3)}$ & $u=-\frac{27\Lambda^6}{256 v^3}$ & $\frac{1}{4 u (1-u)}$   \\
&&&& \\
\hline
&&&& \\
2 &  $y^2=\left(x^2-\frac{\Lambda^4}{64}\right)(x-v)$  & $\frac{1}{27} \frac{(64v^2+3\Lambda^4)^3}{\Lambda^4(64v^2-\Lambda^4)^2}$  & $u=1-\frac{\Lambda^4}{64 v^2}$  &  $\frac{1}{27}\,\frac{(4-3u)^3}{u^2(1-u)}$ \\
&&&& \\
\hline
&&&& \\
3& $y^2=x^2(x-v)-\frac{\Lambda^4}{64}(x-v)^2$ & $\frac{1}{27\times 2^{22}}  \frac{(\Lambda^4-256 v \Lambda^2 +4096 v^2)^3}{v^4 \Lambda^2(\Lambda^2-256 v)}$ & $u=1-\frac{256 v}{\Lambda^2} $ & $\frac{1}{108} \frac{(1+14 u+u^2)^3}{u(1-u)^4}$ \\
&&&& \\
\hline
\end{tabular}}
\caption{Elliptic curves for the $SU(2)$ ${\mathcal N}=2$ SUSY quantum field theories with $N_f$ massless flavors of matter fields, and the associated modular $J$ functions. Comparisons with the $J$ functions in Tables \ref{table:p2} and \ref{table:uni} from the uniformization of the elliptic curves  for the arithmetic Chebyshev potentials leads to the identifications in (\ref{eq:nf0}, \ref{eq:nf1}, \ref{eq:nf2}).}
\label{table:curves}
\end{table}
\begin{eqnarray}
N_f=0  &\longleftrightarrow& \text{Mathieu potential}
\label{eq:nf0}\\
N_f=1 &\longleftrightarrow&  \text{cubic oscillator potential} 
\label{eq:nf1}\\
N_f=2  &\longleftrightarrow& \text{symmetric double-well potential}
\label{eq:nf2}
\end{eqnarray}
The $N_f=3$ case corrsponds to the projective space ${\mathbb P}^3(1, 1, 1, 1)$ discussed in Section 5 of \cite{Lian:1994zv}.

\section{Quantum Properties of Chebyshev Systems: All Orders WKB}
\label{sec:higher}

So far the discussion of the Chebyshev potentials (\ref{eq:chebyshev}) has been classical. Now we consider their quantization. We took the detour of the previous Section to describe the classical number theoretic, modular and geometric properties  of the special ``arithmetic'' subclass of Chebyshev potentials, because this class is precisely the one for which classical Wronskian condition (\ref{eq:classical-wronskian})  and the  classical Matone relation (\ref{eq:classical-matone1}) generalize upon quantization to the same form as those for the Mathieu system, the  quantum Matone relation (\ref{eq:quantum-matone}) and 
quantum Wronskian condition (\ref{eq:quantum-wronskian}), but simply with different constants. In Section \ref{sec:more-general}, we discuss a more general genus 1 system, for which there is also a simple all-orders perturbative/non-perturbative relation, but of a different form.

We adopt the all-orders WKB approach outlined in Section \ref{sec:proof}, and compute the higher order WKB coefficients $a_n(u)$ and $a_n^D(u)$ appearing in the $\hbar^2$ expansions (\ref{eq:fullactions1}, \ref{eq:fullactions2}) of the full quantum action, $a(u, \hbar)$, and full quantum dual action, $a^D(u, \hbar)$.
For the Chebyshev potentials, all higher order terms in the quantum actions and periods can also be expressed in terms of hypergeometric functions, and we can investigate in  detail the relation between the quantum action $a(u, \hbar)$ and  the quantum dual action $a^D(u, \hbar)$. We show in this section that precisely for the special ``arithmetic'' Ramanujan cases, with modular signature $M=2, 3, 4, 6$  (equivalently, modular level $r=4, 3, 2, 1$),  corresponding to the Mathieu, symmetric degenerate triple-well, symmetric double-well,  and cubic oscillator potentials, the perturbative/non-perturbative relations (\ref{eq:quantum-matone}) and (\ref{eq:quantum-wronskian}) found previously for the Mathieu system, apply also with the same form, just with different constants.
For other members of the Chebyshev potential class, the perturbative/non-perturbative relation takes a more general form. This is because these are the only Chebyshev potentials that are genus 1.  Note that it also appears to be true (although we do not have a simple proof) that only for the special "arithmetic" class does the classical proportionality of all the classical actions and classical periods in each well (or each barrier) persist at higher WKB orders.

We concentrate on the quantum Wronskian condition (\ref{eq:quantum-wronskian}), which we repeat here for easy reference,
\begin{eqnarray}
\left(a(u, \hbar)- \hbar  \frac{\partial a(u, \hbar)}{\partial \hbar} \right)\frac{\partial a^D(u, \hbar)}{\partial u}
-\left(a^D(u, \hbar)-\hbar \frac{\partial a^D(u, \hbar)}{\partial \hbar} \right) \frac{\partial a(u, \hbar)}{\partial u}=2 i \mathcal S_I \mathcal T
\label{eq:quantum-wronskian2}
\end{eqnarray}
with the quantum Matone relation (\ref{eq:quantum-matone}) being obtained by inversion. Expanding the quantum acton and dual action in formal series in powers of $\hbar^2$, as in (\ref{eq:fullactions1}, \ref{eq:fullactions2}), we see that (\ref{eq:quantum-wronskian2}) becomes a set of recursion formulas relating the different coefficients $a_k(u)$ and $a_k^D(u)$ at $k^{th}$ order of the WKB expansion \cite{Basar:2015xna}:
\begin{eqnarray}
a_0(u) \frac{d a^D_0(u)}{du} - a_0^D(u) \frac{d a_0(u)}{du} &=& 2 i \mathcal S_I \mathcal T
\label{eq:ladder0}\\
\sum_{k=1}^{n-1} (1-2k)\left(a_k(u) \frac{d a^D_{n-k}(u)}{du} - a_k^D(u) \frac{d a_{n-k}(u)}{du}\right)&=&0\quad, \quad n\geq 1
\label{eq:laddern}
\end{eqnarray}
The first condition (\ref{eq:ladder0}) is just the classical Wronskian identity (\ref{eq:classical-wronskian}), but the higher-order conditions (\ref{eq:laddern}) encode non-trivial constraints on the quantum corrections.

The arithmetic Chebyshev class is special because they define genus 1 systems, and so fall within the discussion of Section \ref{sec:proof}. Therefore, for each WKB order $n$, the $a_n(u)$ can be expressed in the form (\ref{eq:same1}, \ref{eq:same2}), as a differential operator acting  on $a_0(u)$, and $a_n^D(u)$ is given by exactly the same differential operator acting on $a_0^D(u)$. A further simplification occurs because the 
classical Picard-Fuchs equation (\ref{eq:pfp}) for $a_0(u)$ and $a_0^D(u)$ is second order (rather than third-order, as for the general genus 1 case), and therefore the second derivative terms in (\ref{eq:ans1}, \ref{eq:ans2}) can be further reduced to zeroth and first order derivatives:
\begin{eqnarray}
a_n(u)&=& f_n^{(0)}(u) a_0(u)+f_n^{(1)}(u) \frac{d a_0(u)}{d u}
\label{eq:ans1_special}\\
a_n^D(u)&=& f_n^{(0)}(u) a_0^D(u)+f_n^{(1)}(u) \frac{d a_0^D(u)}{d u}
\label{eq:ans2_special}
\end{eqnarray}
The functions $f_n^{(0)}(u)$ and $f_n^{(1)}(u) $ are  sums of poles located at $u=0$ and $u=1$:
\begin{eqnarray}
f_n^{(0)}(u)= \sum_{k=1}^{2n-1} \left({c^{(0)}_{n,k}\over (1-u)^k}+{d^{(0)}_{n,k}\over u^k} \right)
\quad,\quad
f_n^{(1)}(u)= \sum_{k=1}^{2n-2} \left({c^{(1)}_{n,k}\over (1-u)^k}+{d^{(1)}_{n,k}\over u^k} \right)
\label{eq:fs_special}
\end{eqnarray}
where $c^{(i)}_k$ and $d^{(i)}_k$ are rational numbers that depend on the potential. We have verified this structure up to WKB order $\hbar^{10}$. For the arithmetic cases, the higher order terms are straightforward to produce, and we list the first 2 nontrivial WKB orders of the functions $f_n^{(i)}(u)$ in the following subsections, and a further two orders in  Appendix B in Section \ref{app:results}. The location of the poles are precisely the points in moduli space where the torus degenerates. Quantization does not create extra singularities in addition to these classical degenerate points. This structure is related to the quantum modification of the Picard-Fuchs equation as we detail below in Section \ref{sec:qpf}.

The $n=1$ condition in (\ref{eq:laddern}) can be expressed as
\begin{eqnarray}
\frac{d}{du}\left(a_0(u) a^D_1(u) - a_0^D(u) a_{1}(u)\right)=0
\label{eq:ladder1}
\end{eqnarray}
In fact, for the Chebyshev potentials, $a_1(u)$ and  $a_1^D(u)$ can be written as simple linear combinations:
\begin{eqnarray}
a_1(u)&=& f_1^{(0)}(u) a_0(u)+f_1^{(1)} \frac{d a_0(u)}{d u}
\label{eq:a1}
\\
a_1^D(u)&=& f_1^{(0)}(u) a_0^D(u)+f_1^{(1)}\frac{d a_0^D(u)}{d u}
\label{eq:a1d}
\end{eqnarray}
where $f_1^{(0)}(u)$ is a rational function of $u$, and $f_1^{(1)}$ is actually a constant. Further, note that  $f_1^{(0)}$ and $f_1^{(1)}$ are the same in (\ref{eq:a1}) and (\ref{eq:a1d}). Therefore, the first-order WKB condition (\ref{eq:ladder1}) is  indeed satisfied.

At the next order of the WKB expansion, the condition  (\ref{eq:laddern}) places a non-trivial condition on $a_0(u)$, $a_1(u)$, $a_2(u)$ and the corresponding duals:
\begin{eqnarray}
\left(a_0(u) \frac{d a^D_{2}(u)}{du} - a_0^D(u) \frac{d a_{2}(u)}{du}\right)-\left(a_1(u) \frac{d a^D_{1}(u)}{du} - a_1^D(u) \frac{d a_{1}(u)}{du}\right)
\nonumber\\
-3\left(a_2(u) \frac{d a^D_{0}(u)}{du} - a_2^D(u) \frac{d a_{0}(u)}{du}\right)=0
\label{eq:ladder2}
\end{eqnarray}
In the following subsections, for each of the arithmetic Chebyshev potentials we present explicit expressions for $a_0(u)$, $a_1(u)$, $a_2(u)$, and the associated dual actions $a_0^D(u)$, $a_1^D(u)$, $a_2^D(u)$, and verify that the WKB condition (\ref{eq:ladder2}) is satisfied. It is straightforward to generate higher order terms but we do not write the increasingly lengthy expressions here. The next two orders are listed in Appendix B in Section \ref{app:results}.

The quantum Wronskian identity \eqref{eq:laddern} can also be viewed as relating the functions $f_n^{(i)}(u)$ in a highly nontrivial way. Plugging the expressions (\ref{eq:ans1_special} - \ref{eq:ans2_special}) in \eqref{eq:laddern} leads to the conditions on the coefficient functions $f_n^{(0)}(u)$ and $f_n^{(1)}(u)$:
\begin{eqnarray}
\sum_{k=0}^{n}(1-2k)  \left( f_k^{(0)}(u) f_{n-k}^{(0)}(u)- p_m(u) f_{k}^{(1)}(u) f_{n-k}^{(1)}(u)
     + f_k^{(0)}(u) {d f_{n-k}^{(1)}\over du} -  f_{n-k}^{(0)}(u) {d f_{k}^{(1)}\over du}\right)=0\quad\forall n>0
     \nonumber\\
     \label{eq:f-conditions}
\end{eqnarray}
where $p_m(u)=\frac{1}{4u (1-u)}\left(1-\frac{1}{m^2}\right)$ is the function appearing in the classical Picard-Fuchs equation \eqref{eq:pfp}. The conditions (\ref{eq:f-conditions}) generate an infinite tower of equations which we have verified up to order $\hbar^{10}$. 

In the next subsections we list the results for the first few orders of the WKB expansion, for the Mathieu, symmetric-degenerate-triple-well, symmetric-double-well, and cubic oscillator potentials, which classically correspond to the ``arithmetic" cases of Ramanujan's theory of elliptic functions in alternative bases. In Appendix B in Section \ref{app:results} we list two further orders. 

\subsection{Mathieu potential: $V(x)=\cos^2(x)$}
\label{sec:q-mathieu}

The normalized classical actions and periods are:
\begin{eqnarray}
a_0(u)= \sqrt{2} \pi\, u\,_2F_1\left(\frac{1}{2}, \frac{1}{2}, 2; u\right)\ \quad,\quad a^D_0(u)=- i \sqrt{2} \pi\, (1-u)\,_2F_1\left(\frac{1}{2}, \frac{1}{2}, 2; 1-u\right)\\
\omega_0(u)=  \sqrt{2} \pi\, _2F_1\left(\frac{1}{2}, \frac{1}{2}, 1; u\right)\quad,\quad \omega^D_0(u)=  i \sqrt{2} \pi\, \,_2F_1\left(\frac{1}{2}, \frac{1}{2}, 1; 1-u\right)
\end{eqnarray}
The next two WKB orders for the action and dual action are:
\begin{eqnarray}
\label{eq:mathieu-higher-actions}
a_1(u)&=&
\frac{2 u-1}{384 (u-1) u} a_0(u) -\frac{1}{96} \frac{d a_0(u)}{d u}
\\
a_1^D(u)&=&\frac{2 u-1}{384 (u-1) u} a_0^D(u) -\frac{1}{96} \frac{d a_0^D(u)}{d u}
 \\
 a_2(u)&=&\left(\frac{4 (1-2 u)^4-153 (1-2 u)^2-75}{5898240 (u-1)^3 u^3}\right) a_0(u)
 +\left(\frac{-4 u^3+6 u^2+12 u-7}{184320 (u-1)^2 u^2}\right) \frac{d a_0(u)}{d u}
\\
 a_2^D(u)&=&\left(\frac{4 (1-2 u)^4-153 (1-2 u)^2-75}{5898240 (u-1)^3 u^3}\right)a_0^D(u)
 +\left(\frac{-4 u^3+6 u^2+12 u-7}{184320 (u-1)^2 u^2}\right)\frac{d a_0^D(u)}{d u}
\end{eqnarray}
The corresponding higher-order periods can be obtained by differentiating with respect to the energy $u$. These results confirm a non-trivial check of the quantum Wronskian condition, as in (\ref{eq:ladder2}).
In the expressions above, the rational functions that constitute the coefficients of $a_0(u)$ and $a_0^\prime(u)$ (or $a_0^D(u)$ and $a_0^{D\,\prime}(u)$) can be decomposed into a sum over poles that lie on $u=0$ and $u=1$, indicating that quantization does not introduce further singular points in the moduli space. 
These coefficient functions for the next two WKB orders are listed in Appendix B in Section \ref{app:results}, permitting verification of the higher order quantum Wronskian conditions in (\ref{eq:laddern}). 

\subsection{Symmetric degenerate triple well:  $V(x)=T_3^2(x)=x^2(3-4x^2)^2$}
\label{sec:q-tw}

The normalized classical actions and periods are:
\begin{eqnarray}
a_0(u)=  {\pi\over 3\sqrt{2}}\, u\,_2F_1\left(\frac{1}{3}, \frac{2}{3}, 2; u\right)\quad (\text{outer well}) 
\,\,&,&\,\, 
a^D_0(u)= -i {\pi\over \sqrt{6}}\, (1-u)\,_2F_1\left(\frac{1}{3}, \frac{2}{3}, 1; 1-u\right)\qquad
\\
\omega_0(u)= {\pi\over 3\sqrt{2}}\, _2F_1\left(\frac{1}{3}, \frac{2}{3}, 1; u\right) \quad (\text{outer well}) 
\,\,&,&\,\, 
\omega^D_0(u)=  i  {\pi\over \sqrt{6}}\, \,_2F_1\left(\frac{1}{3}, \frac{2}{3}, 1; 1-u\right)
\end{eqnarray}
Note the factor of $\frac{1}{\sqrt{3}}$ difference between the normalizations of the action and dual action (and period and dual period). This corresponds to the ${\sqrt{r}}$ factor difference between  (\ref{eq:hypermagic-w}) and (\ref{eq:hypermagic-wd}), with $r=3$ for the symmetric degenerate triple well potential.

The next two WKB orders for the action and dual action are:
\begin{eqnarray}
a_1(u)&=&\frac{3-5 u}{9 (u-1) u}a_0(u)+\frac{5}{6}\frac{d a_0(u)}{d u}
\\
a_1^D(u)&=&\frac{3-5 u}{9 (u-1) u}a_0^D(u)+\frac{5}{6} \frac{d a_0^D(u)}{d u}
 \\
 a_2(u)&=&\left(\frac{1280 u^3-5105 u^2+5022 u-1701}{1620 (u-1)^3 u^3}\right) a_0(u)
+\left(\frac{-145 u^2+390 u-189}{180 (u-1)^2 u^2}\right) \frac{d a_0(u)}{d u}
\\
 a_2^D(u)&=&\left(\frac{1280 u^3-5105 u^2+5022 u-1701}{1620 (u-1)^3 u^3}\right) a_0^D(u)
+\left(\frac{-145 u^2+390 u-189}{180 (u-1)^2 u^2}\right)
\frac{d a_0^D(u)}{d u}\quad
\end{eqnarray}
The corresponding higher-order periods can be obtained by differentiating with respect to the energy $u$. These results confirm  a non-trivial check of the quantum Wronskian condition, as in (\ref{eq:ladder2}).
In the expressions above, the rational functions that constitute the coefficients of $a_0(u)$ and $a_0^\prime(u)$ (or $a_0^D(u)$ and $a_0^{D\,\prime}(u)$) can be decomposed into a sum over poles that lie on $u=0$ and $u=1$, indicating that quantization does not introduce further singular points in the moduli space. 
These coefficient functions for the next two WKB orders are listed in Appendix B in Section \ref{app:results}, permitting verification of the higher order quantum Wronskian conditions in (\ref{eq:laddern}).

\subsection{Symmetric double well: $V(x)=T_2^2(x)=(1-2x^2)^2$}
\label{sec:q-dw}

The normalized classical actions and periods are:
\begin{eqnarray}
a_0(u)=  {\pi\over 2}\, u\,_2F_1\left(\frac{1}{4}, \frac{3}{4}, 2; u\right)\ 
\quad&,&\quad 
a^D_0(u)=- i {\pi\over \sqrt{2}}\, (1-u)\,_2F_1\left(\frac{1}{4}, \frac{3}{4}, 1; 1-u\right)\\
\omega_0(u)= {\pi\over 2}\, _2F_1\left(\frac{1}{4}, \frac{3}{4}, 1; u\right)
\quad&,&\quad
 \omega^D_0(u)=  i {\pi\over \sqrt{2}}\, \,_2F_1\left(\frac{1}{4}, \frac{3}{4}, 1; 1-u\right)
\end{eqnarray}
Note the $\frac{1}{\sqrt{2}}$ factor difference between the action and dual action (and period and dual period).  This corresponds to the ${\sqrt{r}}$ factor difference between  (\ref{eq:hypermagic-w}) and (\ref{eq:hypermagic-wd}), with $r=2$ for the symmetric double well potential.

The next two WKB orders for the action and dual action are:
\begin{eqnarray}
a_1(u)&=&\frac{3 u-2}{16(u-1) u} a_0(u)-\frac{1}{4}\frac{d a_0(u)}{d u}
\\
a_1^D(u)&=&\frac{3 u-2}{16(u-1) u} a_0^D(u)-\frac{1}{4} \frac{d a_0^D(u)}{d u}
 \\
 a_2(u)&=&\frac{705 u^3-2685 u^2+2652 u-896}{7680 (u-1)^3 u^3} a_0(u)
+\frac{-720 u^2+1840 u-896}{7680 (u-1)^2 u^2} \frac{d a_0(u)}{d u}
\\
 a_2^D(u)&=&\frac{705 u^3-2685 u^2+2652 u-896}{7680 (u-1)^3 u^3} a_0^D(u)
+\frac{-720 u^2+1840 u-896}{7680 (u-1)^2 u^2} \frac{d a_0^D(u)}{d u}
\end{eqnarray}
The corresponding higher-order periods can be obtained by differentiating with respect to the energy $u$. These results confirm a non-trivial check of the quantum Wronskian condition, as in (\ref{eq:ladder2}).
 In the expressions above, the rational functions that constitute the coefficients of $a_0(u)$ and $a_0^\prime(u)$ (or $a_0^D(u)$ and $a_0^{D\,\prime}(u)$) can be decomposed into a sum over poles that lie on $u=0$ and $u=1$, indicating that quantization does not introduce further singular points in the moduli space. 
These coefficient functions for the next two WKB orders are listed in Appendix B in Section \ref{app:results}, permitting verification of the higher order quantum Wronskian conditions in (\ref{eq:laddern}).

\subsection{Cubic oscillator: $V(x)=T_{3/2}^2(x)=\frac{1}{2}(1+x)(1-2x)^2$}
\label{sec:q-co}

The normalized classical actions and periods are:
\begin{eqnarray}
a_0(u)=  \pi \sqrt{2\over3} \, u\,_2F_1\left(\frac{1}{6}, \frac{5}{6}, 2; u\right)
\quad&,&\quad 
a^D_0(u)=- i \pi \sqrt{2\over3} \, (1-u)\,_2F_1\left(\frac{1}{6}, \frac{5}{6}, 1; 1-u\right)
\\
\omega_0(u)= \pi \sqrt{2\over3} \, _2F_1\left(\frac{1}{6}, \frac{5}{6}, 1; u\right) 
\quad&,&\quad 
\omega^D_0(u)=  i\pi \sqrt{2\over3} \, \,_2F_1\left(\frac{1}{6}, \frac{5}{6}, 1; 1-u\right)
\end{eqnarray}
Note there is no factor  difference between the normalizations of the action and dual action (and period and dual period). This corresponds to the ${\sqrt{r}}$ factor difference between  (\ref{eq:hypermagic-w}) and (\ref{eq:hypermagic-wd}), with $r=1$ for the cubic oscillator.

The next two WKB orders for the action and dual action are:
\begin{eqnarray}
a_1(u)&=&-\frac{5 (2 u-1)}{144 (u-1) u}a_0(u)+\frac{1}{12} \frac{d a_0(u)}{d u}
\\
a_1^D(u)&=&\-\frac{5 (2 u-1)}{144 (u-1) u} a_0^D(u)+\frac{1}{12} \frac{d a_0^D(u)}{d u}
 \\
 a_2(u)&=&-\frac{7 \left(211 u^2-211 u+72\right)}{41472 (u-1)^3 u^3}a_0(u)
+\frac{7 (2 u-1)}{576 (u-1)^2 u^2}\frac{d a_0(u)}{d u}
\\
 a_2^D(u)&=&-\frac{7 \left(211 u^2-211 u+72\right)}{41472 (u-1)^3 u^3} a_0^D(u)
+\frac{7 (2 u-1)}{576 (u-1)^2 u^2}\frac{d a_0^D(u)}{d u}
\end{eqnarray}
The corresponding higher-order periods can be obtained by differentiating with respect to the energy $u$. These results confirm a non-trivial check of the quantum Wronskian condition, as in (\ref{eq:ladder2}).
In the expressions above, the rational functions that constitute the coefficients of $a_0(u)$ and $a_0^\prime(u)$ (or $a_0^D(u)$ and $a_0^{D\,\prime}(u)$) can be decomposed into a sum over poles that lie on $u=0$ and $u=1$, indicating that quantization does not introduce further singular points in the moduli space. 
These coefficient functions for the next two WKB orders are listed in Appendix B in Section \ref{app:results}, permitting verification of the higher order quantum Wronskian conditions in (\ref{eq:laddern}).

\section{More General Genus 1 Cases}
\label{sec:more-general}

In this Section we consider more general genus 1 cases: those for which the third-order classical Picard-Fuchs equation does not reduce to a second-order equation.
We illustrate with the example of the Lam\'e equation, as it is of  physical interest, since it is directly related to the ${\mathcal N}=2^*$ SUSY QFT \cite{Nekrasov:2002qd,Nekrasov:2009rc,Nekrasov:2003rj,agt,fateev,Huang:2011qx,KashaniPoor:2012wb,Krefl:2013bsa,Gorsky:2014lia,piatek,Basar:2015xna,kpt,Ashok:2016yxz}, and also because any genus 1 example can be brought into this standard Weierstrass form by appropriate transformations \cite{bateman,byrd}. The Lam\'e equation is a Schr\"odinger equation with a doubly periodic potential which we express in terms of the Weierstrass elliptic function with lattice invariants $g_2$ and $g_3$:
\begin{eqnarray}
V(x)=\PP\left(x; g_2,  g_3 \right)\,.
\label{eq:lame}
\end{eqnarray}
Using the modular transformation properties of the Weierstrass function we  set one of the lattice periods to $1$, and we parameterize the other as $i \K(1-\nu)/\K(\nu)$. When $\nu=0$ and $1$ the Lam\'e equation reduces to the Mathieu and modified Mathieu equations, respectively. With this parametrization the zeroes of the normal cubic $4z^3-g_2z-g_3$ are
\be
e_1={2-\nu\over3}\K^2(\nu)\quad,\quad e_2= {2\nu-1\over3}\K^2(\nu)\quad,\quad e_3=-{\nu+1\over3}\K^2(\nu)
\label{eq:Lame_eis}
\ee
which are related to the lattice invariants as
\be
g_2=2(e_1^2+e_2^2+e_3^2)\quad,\quad g_3=4e_1e_2e_3.
\label{eq:Lame_gis}
\ee
It is also useful to define a modular parameter $\ft$ as the ratio of the two periods:
\be
\ft=i {\K(1-\nu)\over\K(\nu)}\,.
\label{eq:Lame_modular}
\ee
This modular parameter $\ft$ characterizes the torus that is defined by the periods of the Weierstrass function, and it is not to be confused with the modular parameter $\tau_0$ associated with the complexified phase space.
 
The corresponding spectral curve can be obtained by starting from the conservation of energy $p^2=2(u-V(x))$ and changing variables as $x\rightarrow v=V(x)$:
 \be
 y^2=8(u-v)(v-e_1)(v-e_2)(v-e_3)\,.
 \label{eq:lame_curve}
 \ee
Here we redefined the kinetic term as $y^2=p^2 V^{\prime\,2}(x)$ and used the identity  $\PP^{\prime\,2}(x)= 4\PP^3(x)-g_2\PP(x)-g_3=4\prod_{i=1}^3(\PP(x)-e_i)$. It is worth noting that any genus-1  spectral curve can be put in this Weierstrass form by appropriately choosing the lattice invariants \cite{bateman,byrd}. The classical actions and periods are defined as period integrals over two independent cycles of the phase space torus, which we choose here as
\beqa
a_0(u)&=&\oint_\alpha \sqrt{2(u-V)} \,dx=4\int_{e_3}^{e_2} {u-v\over y}dv
\,\,\,,\,\,\,
\omega_0(u)=a_0^\prime(u)=\oint_\alpha{dx \over \sqrt{2(u-V)}}=2\int_{e_3}^{e_2} {dv\over y}
\\
a^D_0(u)&=&\oint_\beta \sqrt{2(u-V)} \,dx=4\int_{e_1}^{e_2} {u-v\over y}dv
\,\,\,,\,\,\,
\omega^D_0(u)=a_0^{D\,\prime}(u)=\oint_\beta {dx \over \sqrt{2(u-V)}} =2\int_{e_1}^{e_2} {dv\over y}\qquad\,\,
\enqa
The integrals above for the classical actions and periods can be computed in terms of the standard elliptic functions as \cite{byrd}
\begin{eqnarray}
a_0(u)&=&{2\sqrt{2}\over \K(\nu)}{u-e_3\over\sqrt{u-e_2}}{\Pi}\Big({e_3-e_2\over u-e_2},\nu{u-e_1\over u-e_2}\Big)\quad,\quad
\omega_0={\sqrt{2}\over\K(\nu)\sqrt{u-e_2}}\K\Big(\nu{u-e_1\over u-e_2}\Big)
\\
a^D_0(u)&=&{i\,2\sqrt{2}\over\K(\nu)}{u-e_1\over\sqrt{u-e_2}}{\Pi}\Big({e_1-e_2\over u-e_2},(1-\nu){u-e_3\over u-e_2}\Big)
\quad,\quad
\omega^D_0={i\,\sqrt{2}\over\K(\nu)\sqrt{u-e_2}}\K\Big((1-\nu){u-e_3\over u-e_2}\Big)\qquad
\end{eqnarray}
As a general property of the geometry of the Riemann surfaces, the classical \textit{periods} satisfy a second order Picard-Fuchs equation, and the classical \textit{actions}, being related to the period through a $u$ integration, satisfy a third-order Picard-Fuchs equation. This is in contrast with the special cases discussed in the previous section where the  third order Picard-Fuchs equation reduced to a second order one. In the next section we present a simple derivation of the classical Picard-Fuchs equation before we move on to the quantum deformation of the elliptic curve.
  
\subsection{From The Elliptic Curve  to The Classical Picard-Fuchs Equation}

The classical Picard-Fuchs equation can be derived from the elliptic curve by the following procedure. 
In  the classical energy-momentum relation (\ref{eq:curve1}) change variable from $x$ to $y(x)$ where
\begin{eqnarray}
p(x,u)=\sqrt{2(u-V(x))}
\label{eq:p}
\end{eqnarray}
Define the derivative as 
\begin{eqnarray}
g(x,u)=\frac{dp}{dx}=-{1\over p}\,\frac{dV}{dx}
\label{eq:g}
\end{eqnarray}
and regard $g$ as a function of $p$. In general, inversion of $p(x,u)$ cannot be made in an explicit way. However, suppose we can invert to write $\left(\frac{dV}{dx}\right)^2$ as a third order polynomial in $V$
\begin{eqnarray}
&&\left(\frac{dV}{dx}\right)^2=P(V)\,.
\label{eq:p3}
\end{eqnarray}
This is indeed the case for the Lam\'e and Mathieu potentials:
\begin{eqnarray}
&&{\text{Lam\'e}}: \quad P(V)= 4V^3-g_2 V-g_3=4\prod_{i=1}^3(V-e_i)\quad,\quad {\rm Mathieu}: \quad P(V)=4V(1-V)\hskip0.7cm
\label{eq:lame_mathieuP}
\end{eqnarray}
With this identity, it is now possible to express $g(x,u)$ as a function of $p$, since $V=u-p^2/2$. We can then write $g$ as a function of $p$ as:
\be
g(p)=-{1\over p}\sqrt{P(u) -{1\over 2}P'(u)\,p^2+{1\over 8} P^{\prime\prime}(u)\,p^4-{1\over 48}P^{\prime\prime\prime}(u)\,p^6}
\label{eq:gp}
\ee  
where we fixed the sign of the branch.\footnote{This choice is identical with choosing the orientation of the phase space integrals for the classical action and periods and does not affect the results we will discuss in the rest of the paper as long as it is implemented consistently.} The next step is to consider the identity
\begin{eqnarray}
g(p)\frac{d}{dp}\left(\frac{g(p)}{p^2}\right)&=& -P(u)\frac{3}{p^5}+P^\prime(u) \frac{1}{p^3}-P^{\prime\prime}(u) \frac{1}{8 p}\,.
\label{eq:gpp}
\end{eqnarray}
Using \eqref{eq:p}, we can express the inverse odd powers of $p$ as $u$ derivatives acting on $p$:
\begin{equation}
{1\over p^{2n-1}}={(-1)^{n+1}\over (2n-3)!!} {d^n p\over du^n}
\label{eq:inv_p}
\end{equation}
where the double factorial is defined as $n!!=n(n-2)(n-4)\dots$ with $(-1)!!=1$. Therefore we can write the identity (\ref{eq:gpp})  as
\begin{eqnarray}
g(p)\frac{d}{dp}\left(\frac{g(p)}{p^2}\right)&=&- \left[P(u) \frac{d^3}{du^3}+P^\prime(u) \frac{d^2}{du^2}+\frac{P^{\prime\prime}(u)}{8} \frac{d}{du}\right] p
\label{eq:id}
\end{eqnarray}
On the other hand we also have the vanishing closed contour integral
\begin{eqnarray}
0=\oint \frac{d}{dx}\left(\frac{g(p(x,u))}{p^2(x,u)}\right)=\oint g(p)\frac{d}{dp}\left(\frac{g(p)}{p^2}\right)
\label{eq:zero}
\end{eqnarray}
Integrating around the turning points, we obtain the action and dual action as integrals around the two contours on the torus: $a_0(u)=\oint_\alpha p$, and $a_0^D(u)=\oint_\beta p$. Therefore, using the identity (\ref{eq:id}) we find that both  $a_0(u)$ and $a_0^D(u)$ satisfy the third-order classical Picard-Fuchs equation:
\begin{eqnarray}
P(u) \frac{d^3a_0}{du^3}+P^\prime(u) \frac{d^2 a_0}{du^2}+\frac{P^{\prime\prime}(u)}{8} \frac{d a_0}{du}&=&0 
\label{eq:lame-pf}
\\
P(u) \frac{d^3a_0^D}{du^3}+P^\prime(u) \frac{d^2 a_0^D}{du^2}+\frac{P^{\prime\prime}(u)}{8} \frac{d a_0^D}{du}&=&0
\label{eq:lame-pfd}
\end{eqnarray}
Notice the absence of  $a_0(u)$ and $a_0^D(u)$   terms in (\ref{eq:lame-pf}, \ref{eq:lame-pfd}), which means that one of the solutions is the trivial constant solution. The two independent nontrivial solutions, $a_0(u)$ and $a^D_0(u)$, are associated with the integrals over two independent cycles of the phase space torus. It also means that the periods, $\omega_0=a_0^\prime(u)$ and $\omega^D_0=a^{D\,\prime}_0(u)$, satisfy a second order equation:
\begin{eqnarray}
 \frac{d^2 \omega_0}{du^2}+{P^\prime(u)\over P(u)} \frac{d \omega_0}{du}+\frac{P^{\prime\prime}(u)}{8 P(u)} \omega_0&=&0 
\label{eq:lame-pf-omega}
\end{eqnarray}
Notice that the coefficients have first order poles at $u=e_1,e_2,e_3, \infty$, therefore all the singularities are regular. For the Mathieu system, the polynomial $P(u)=4u(1-u)$ is quadratic, so the Picard-Fuchs equation can be integrated once to a second-order equation
\begin{eqnarray}
4u(1-u) \frac{d^2a_0}{du^2}-a_0&=&0 
\label{eq:mathieu-pf}
\\
4u(1-u) \frac{d^2a_0^D}{du^2}-a_0^D&=&0 
\label{eq:mathieu-pfd}
\end{eqnarray}
which is just (\ref{eq:pfp}) with $m=\infty$, as noted before.

For Lam\'e, the classical Picard-Fuchs equation is third order:
\begin{eqnarray}
(4u^3-g_2 u-g_3) \frac{d^3 a_0}{du^3}+ (12u^2-g_2) \frac{d^2 a_0}{du^2} +3u \frac{d a_0}{du}&=&0 
\label{eq:lame-pfex}
\\
(4u^3-g_2 u-g_3) \frac{d^3 a_0^D}{du^3}+ (12u^2-g_2) \frac{d^2 a_0^D}{du^2} +3u \frac{d a_0^D}{du}&=&0 
\label{eq:lame-pfdex}
\end{eqnarray}

\subsection{From The Schr\"odinger Equation to the Quantum Picard-Fuchs Equation}
\label{sec:qpf}

The previous subsection derived the {\it classical}  Picard-Fuchs equation directly from the elliptic curve expression. Now we consider quantization. We convert the the Schr\"odinger equation (\ref{eq:schrodinger}), a differential equation with respect to the coordinate $x$,  into a tower of Picard-Fuchs equations, which are differential equations with respect to the energy variable $u$. The computational strategy is the same as the previous section. 
We first change variables from $x$ to $(u-V(x))$, and then trade derivatives with respect to $V(x)$ for derivatives with respect to $u$. We illustrate this procedure for the Lam\'e potential, which is genus-1, and has a 3rd order classical Picard-Fuchs equation for the actions. 

The quantum corrections to the classical actions are encoded in the wavefunction which we write as a WKB ansatz $\psi=\exp\left[\frac{i}{\hbar} \int^x \phi(x^\prime,u;\hbar)dx^\prime\right]$. This ansatz converts the Schr\"odinger equation into a Riccati form
\begin{eqnarray}
\phi^2(x,u;\hbar)-p^2(x,u)-i \hbar \frac{d\phi}{dx}=0
\end{eqnarray}
which can be solved recursively with the formal expansion $\phi(x,u;\hbar)=\sum_n\phi_n(x,u)\hbar^n$. The quantum actions can be written formally as
\be
a(u, \hbar)=\oint_\alpha \phi(x,u;\hbar)=\sum_{n=0}^\infty\hbar^{2n} \oint_\alpha \phi_{2n}(x,u)\,dx \quad,\quad 
a^D(u,\hbar)=\oint_\beta \phi(x,u;\hbar)=\sum_{n=0}^\infty\hbar^{2n} \oint_\beta \phi_{2n}(x,u)\,dx\,.
\ee
The first few terms of this expansion are given in (\ref{eq:dunham1}, \ref{eq:dunham2}). The expansion for the action is in $\hbar^2$ since all the odd terms $\phi_{2n+1}(x,u)$ are total derivatives and vanish when integrated along closed contours.\footnote{In fact an improved WKB ansatz $\psi=\exp\left[\frac{i}{\hbar} \int^x \phi(x^\prime,u;\hbar)dx^\prime\right]/{\sqrt{\phi}}$ automatically resums these odd terms and generates a Riccati equation that depends on $\hbar^2$. The recursion relations of this improved Riccati equation directly give the even terms in \eqref{eq:s}. Even though the modified Riccati equation is advantageous computationally, its form is a little more complicated and for pedagogical reasons we chose to present the simpler and more familiar version here.}
The next step is to change variables from $x$ to $p(x)$:
\be
\phi^2(p,u;\hbar)-p^2-i \hbar \,g(p)\, \frac{d\phi}{dp}=0\,
\ee
Formally expanding $\phi(p,u;\hbar)$ as a series in $\hbar$ and repeating the steps above leads to the expansion
\begin{eqnarray}
\phi(p,u;\hbar)=\sum_{n=0}^\infty \phi_n(p; u) \hbar^n \qquad, \quad \phi_0=p\quad , \quad \phi_1=\frac{g(p)}{2p}
\label{eq:s}
\end{eqnarray}
where we chose the upper branch for $\phi_0$. The higher order terms can be computed recursively:
\begin{eqnarray}
\phi_n(p)=\frac{1}{2p}\left( i\, g(p)\, {d\phi_{n-1}(p)\over dp} -\sum_{k=1}^{n-1} \phi_k(p) \phi_{n-k}(p)\right)\qquad, \quad n\geq 2
\label{eq:sn}
\end{eqnarray}
By using the expression for $g(p)$ in \eqref{eq:gp}, $\phi_{2n\geq2}(p)$ may be expressed as a polynomial in inverse powers of $p$, with coefficients given by derivatives of $P(u)$. For example:
\begin{eqnarray}
\phi_2&=& \frac{5}{8} P(u)\frac{1}{p^5} -\frac{3}{16} P^\prime(u) \frac{1}{p^3} +\frac{1}{64} P^{\prime\prime}(u) \frac{1}{p} +\frac{1}{384} P^{\prime\prime\prime}(u) p
\label{eq:s2}
\end{eqnarray}
Now, we can use the fact, \eqref{eq:inv_p}, that odd powers of $p$ can be written as derivatives of $p$ with respect to $u$.
Therefore, each $\phi_n$ can be written as a differential operator acting on $\phi_0=p$. In our example of $\phi_2$, this leads to:
\begin{eqnarray}
\phi_2&=& \left[\frac{5}{24} P(u)\frac{d^3}{du^3} +\frac{3}{16} P^\prime(u) \frac{d^2}{du^2} +\frac{1}{64} P^{\prime\prime}(u) \frac{d}{du} +\frac{1}{384} P^{\prime\prime\prime}(u)\right] \phi_0
\label{eq:s2u}
\end{eqnarray}
But since $\phi_n$ is the integrand for the WKB expansion coefficient functions $a_{n/2}(u)$ and $a^D_{n/2}(u)$ in (\ref{eq:fullactions1}, \ref{eq:fullactions2}), we see that these differential operators are precisely the differential operators mentioned in (\ref{eq:same1}, \ref{eq:same2}). Thus we learn, for example, that for the Lam\'e or Mathieu system the next-to-leading order WKB action $a_1(u)$ and dual action $a_1^D(u)$ are given by the following differential operator acting on $a_0(u)$ or $a_0^D(u)$, respectively, with the appropriate choice of the polynomial $P(u)$ from (\ref{eq:lame_mathieuP}):
\begin{eqnarray}
a_1(u)&=& \left[\frac{5}{24} P(u)\frac{d^3}{du^3} +\frac{3}{16} P^\prime(u) \frac{d^2}{du^2} +\frac{1}{64} P^{\prime\prime}(u) \frac{d}{du} +\frac{1}{384} P^{\prime\prime\prime}(u)\right] a_0(u) 
\label{eq:s2a} \\
a_1^D(u)&=& \left[\frac{5}{24} P(u)\frac{d^3}{du^3} +\frac{3}{16} P^\prime(u) \frac{d^2}{du^2} +\frac{1}{64} P^{\prime\prime}(u) \frac{d}{du} +\frac{1}{384} P^{\prime\prime\prime}(u)\right] a_0^D(u) 
\label{eq:s2ad}
\end{eqnarray}
Notice that it is the same differential operator acting on both $a_0(u)$ and $a_0^D(u)$, as argued in Section \ref{sec:proof}. Furthermore, using the classical Picard-Fuchs equation (\ref{eq:lame-pf}, \ref{eq:lame-pfd}), we can reduce the third order derivative term to lower order derivatives, thereby  obtaining the representation in (\ref{eq:ans1}, \ref{eq:ans2}):
\begin{eqnarray}
a_1(u)&=&- \frac{1}{48} P^\prime(u)\frac{d^2 a_0}{du^2} -\frac{1}{96} P^{\prime\prime}(u) \frac{d a_0}{du} +\frac{1}{384} P^{\prime\prime\prime}(u) a_0
\label{eq:s2areduced} \\
a_1^D(u)&=&- \frac{1}{48} P^\prime(u)\frac{d^2 a^D_0}{du^2} -\frac{1}{96} P^{\prime\prime}(u) \frac{d a_0^D}{du} +\frac{1}{384} P^{\prime\prime\prime}(u) a_0^D
\label{eq:s2adreduced}
\end{eqnarray}
For the Mathieu system, where $P(u)$ is quadratic, the last term $P^{\prime\prime\prime}(u)$ vanishes, and since the classical Picard-Fuchs equation takes the simpler 2nd order form, we can further reduce to arrive at the expressions in (\ref{eq:mathieu-higher-actions}), involving just $a_0$ and $a_0^\prime$.

This procedure is simple to implement recursively at higher order, permitting the computation of higher $a_n(u)$ and $a_n^D(u)$ for both the Lam\'e and Mathieu system in a straightforward fashion. For example, the order $\hbar^4$ actions of the Lam\'e system are found as:
\begin{eqnarray}
a_2(u)&=&
{1\over3870720\, P(u)^2}\Big(2352 P^{\prime}(u)^4-6363 P(u) P^{\prime\prime}(u) P^{\prime}(u)^2+2310 P(u)^2 P^{\prime\prime\prime}(u) P^{\prime}(u)
\nonumber\\
&&\hskip2.5cm+3150 P(u)^2 P^{\prime\prime}(u)^2\Big)a_0^{\prime\prime}(u)
\nonumber\\
&&-{1\over3870720 \,P(u)^2}\Big(-294 P^{\prime\prime}(u)P^{\prime}(u)^3+147 P(u)P^{\prime\prime\prime}(u) P^{\prime}(u)^2+630 P(u)P^{\prime\prime}(u)^2P^{\prime}(u)\nonumber\\
&&\hskip2.5cm
-840 P(u)^2 P^{\prime\prime}(u)P^{\prime\prime\prime}(u)\Big)a_0^\prime(u)-{1\over294912}P^{(3)}(u)a_0(u)
\label{eq:Lame_higherorder1}
\\
a_2^D(u)&=&
{1\over3870720 \,P(u)^2}\Big(2352 P^{\prime}(u)^4-6363 P(u) P^{\prime\prime}(u) P^{\prime}(u)^2+2310 P(u)^2 P^{\prime\prime\prime}(u) P^{\prime}(u)
\nonumber\\
&&\hskip2.5cm+3150 P(u)^2 P^{\prime\prime}(u)^2\Big)a_0^{D\,\prime\prime}(u)
\nonumber\\
&&-{1\over3870720\, P(u)^2}\Big(-294 P^{\prime\prime}(u)P^{\prime}(u)^3+147 P(u)P^{\prime\prime\prime}(u) P^{\prime}(u)^2+630 P(u)P^{\prime\prime}(u)^2P^{\prime}(u)\nonumber\\
&&\hskip2.5cm
-840 P(u)^2 P^{\prime\prime}(u)P^{\prime\prime\prime}(u)\Big)a_0^{D\,\prime}(u)-{1\over294912}P^{(3)}(u)a_0^D(u)
\label{eq:Lame_higherorder2}
\end{eqnarray}
This general structure makes it clear that each $a_n(u)$ satisfies a third-order differential equation, with coefficients being rational functions of the polynomial $P(u)$ and its derivatives. Furthermore, $a_n^D(u)$ satisfies the same third-order differential equation. Thus the all-orders quantum action and dual action satisfy a quantum Picard-Fuchs equation of the form:
\begin{eqnarray}
f^{(3)}(u, \hbar) \frac{\partial^3a(u, \hbar)}{\partial u^3}+f^{(2)}(u, \hbar) \frac{\partial^2 a(u, \hbar)}{\partial u^2}+f^{(1)}(u, \hbar) \frac{\partial a(u, \hbar)}{\partial u}&=&0 
\label{eq:lame-qpf}
\\
f^{(3)}(u, \hbar) \frac{\partial^3a^D(u, \hbar)}{\partial u^3}+f^{(2)}(u, \hbar) \frac{\partial^2 a^D(u, \hbar)}{\partial u^2}+f^{(1)}(u, \hbar) \frac{\partial a^D(u, \hbar)}{\partial u}&=&0 
\label{eq:lame-qpfd}
\end{eqnarray}
It can be shown that the quantum corrections do not introduce new singularities to the differential equation. In fact all the singularities are in the form of poles. However, the order of these poles grow with the order of $\hbar$, and therefore they are not regular singularities. In other words, the quantum corrections appear as irregular singularities (higher order poles) at the original regular singular points  of the classical Picard-Fuchs equation.

\subsection{Quantum Schwarzian Equation}
\label{sec:quantumschwarzian}
The third order quantum Picard-Fuchs equation for the actions can be written as a second order equation for the periods:
\begin{eqnarray}
\frac{\partial^2\Omega(u, \hbar)}{\partial u^2}+{f^{(2)}(u, \hbar) \over f^{(3)}(u, \hbar) }\frac{\partial \Omega(u, \hbar)}{\partial u}+{f^{(1)}(u, \hbar)\over f^{(3)}(u, \hbar) } \Omega(u, \hbar)&=&0 
\label{eq:lame-qpf-periods}
\end{eqnarray}
 whose  two independent solutions for $\Omega(u, \hbar)$ are the all-orders periods $\omega(u,\hbar)$ and $\omega^D(u,\hbar)$.  The partial differential equation (\ref{eq:lame-qpf-periods})  is understood as a tower of ordinary differential equations for the coefficients of periods (that are functions of $u$) in the formal $\hbar$ expansion. Equation \eqref{eq:lame-qpf-periods}  is second order, and  can be put into a Schwarzian form:
 \begin{eqnarray}
\{\tau,u\}-2Q(u,\hbar)=0
\quad {\text{where}}\quad  Q(u,\hbar)= {f^{(1)}(u, \hbar)\over f^{(3)}(u, \hbar) } -{1\over4}\left({f^{(2)}(u, \hbar)\over  f^{(3)}(u, \hbar) }\right)^2-{1\over2}{\partial\over\partial u}\left({f^{(2)}(u, \hbar)\over  f^{(3)}(u, \hbar) }\right)  \qquad
\end{eqnarray}
where $\tau(u,\hbar)$ is the all-orders modular parameter,
\begin{equation}
\tau(u,\hbar)={\omega^D(u,\hbar)\over\omega(u,\hbar)}\,.
\end{equation}
At any order in $\hbar$, the function $Q(u,\hbar)$ is rational, in fact a sum over higher order poles located at the original singularities of the elliptic curve. 

\subsection{Quantum Perturbative/Non-perturbative Relation for Lam\'e Potential}
\label{sec:lame}
 
For the Lam\'e system, generating higher order WKB terms for the action and dual action, we have verified the simple relation satisfied by the all-orders action $a(u, \hbar; \ft)$ and dual action $a^D(u, \hbar; \ft)$:
\begin{eqnarray}
\frac{\partial a(u, \hbar; \ft)}{\partial u} \frac{\partial a^D(u, \hbar; \ft)}{\partial \ft}- \frac{\partial a^D(u, \hbar; \ft)}{\partial u} \frac{\partial a(u, \hbar; \ft)}{\partial \ft} = 4
\label{eq:lame-qw}
\end{eqnarray}
Here $\ft$ refers to the elliptic $\ft$ parameter of the original Lam\'e potential defined in \eqref{eq:Lame_modular} (do not confuse it with the modular $\tau_0$ defined in (\ref{eq:tau})). 

Note that this is not a Wronskian condition. The standard classical Wronskian combination for a second order Picard-Fuchs equation is of course not a constant:
\begin{eqnarray}
\frac{\partial a_0(u; \ft)}{\partial u} a_0^D(u; \ft)- \frac{\partial a_0^D(u; \ft)}{\partial u} a_0(u; \ft) =
\frac{-2\pi i}{{\mathbb K}(\nu) \sqrt{(u-e_2)}} {\mathbb F}\left( {\rm arccos}\left(\sqrt{\frac{(u-e_1)}{(u-e_3)}} \right), (1-\nu)\frac{(u-e_3)}{(u-e_2)}\right)
\nonumber\\
\label{eq:lame-cw}
\end{eqnarray}
The third order classical Picard-Fuchs equation for the Lam\'e system implies the classical Wronskian condition: 
\begin{eqnarray}
\frac{\partial^2 a_0(u; \ft)}{\partial u^2} \frac{\partial a_0^D(u; \ft)}{\partial u} - \frac{\partial^2 a_0^D(u; \ft)}{\partial u^2} \frac{\partial a_0(u; \ft)}{\partial u} = \frac{2\pi i}{4u^3-g_2 u-g_3}
\nonumber\\
\label{eq:lame-cw2}
\end{eqnarray}
But using special properties of the elliptic $\mathbb K$, $\mathbb E$ and $\Pi$ functions, in particular their derivatives with respect to $\ft$ \cite{byrd}, one can verify that the classical actions also satisfy 
\begin{eqnarray}
\frac{\partial a_0(u; \ft)}{\partial u} \frac{\partial a_0^D(u; \ft)}{\partial \ft}- \frac{\partial a^D_0(u; \ft)}{\partial u} \frac{\partial a_0(u; \ft)}{\partial \ft} = 4
\label{eq:lame-cmw}
\end{eqnarray}
Remarkably, this classical identity is unchanged, as in (\ref{eq:lame-qw}), at the quantum level, when the classical actions $a_0(u; \ft)$ and $a_0^D(u; \ft)$ are replaced by the all-orders quantum actions $a(u, \hbar; \ft)$ and $a^D(u, \hbar; \ft)$.

Therefore, as argued in general in Section \ref{sec:proof}, there is a simple and  explicit quantitative relationship between the all-orders action $a(u, \hbar; \ft)$ and the all-orders dual action $a^D(u, \hbar; \ft)$. If one knows the expansion of one of them to some order in $\hbar$, the other can be deduced.
In the context of gauge theory (${\cal N}=2^*$, $SU(2)$ theory) \eqref{eq:lame-qw} is the quantum Matone relation where the modular parameter $\ft$, and $\hbar$ are identified with the complex gauge coupling and the Omega deformation parameter $\epsilon_1$ (in the limit $\epsilon_2=0$), respectively \cite{Nekrasov:2002qd,Nekrasov:2009rc,Nekrasov:2003rj,KashaniPoor:2012wb,Krefl:2013bsa,Gorsky:2014lia,piatek,Basar:2015xna,kpt,Ashok:2016yxz}.

\subsection{Other Genus 1 Cases}

Other interesting genus 1 potentials include, for example, the asymmetric double-well, the symmetric but non-degenerate triple well, the double-Sine-Gordon, and the prolate/oblate spheroidal potentials. At the classical level each can be uniformized in a standard manner, resulting in a third-order classical Picard-Fuchs equation.  A remarkable feature of these systems is the fact that there is in fact still just one independent classical action and period, even though a cursory glance at the form of the potential reveals different shaped wells. Despite these different shapes, the classical periods in different wells are identical functions of the energy (up to an overall normalization factor), and the classical actions differ only by a constant, related to the offset between the asymmetric wells. These potentials contain extra parameters, associated with this offset/asymmetry. On the gauge theory side, these extra parameters are associated with masses of extra hypermultiplets. For example, the modular $J(u)$ function for the elliptic curve associated with the prolate spheroidal potential, $V(x)=\sin^2 x+b/\sin^2 x$, is
\begin{eqnarray}
J(u)=\frac{4}{27}\frac{(u^2-u+1-3b)^3}{(4b-u^2)(u-1-b)^2}
\label{eq:psj}
\end{eqnarray}
With the identifications $b=4m^2/\Lambda^2$ and $u=\frac{1}{2}+4v/\Lambda^2$ this becomes
\begin{eqnarray}
J(v)=\frac{\left(3 \Lambda^4-48 \Lambda^2 m^2+64
   v^2\right)^3}{27 \Lambda^4 \left(\Lambda^2+8 m^2-8
  v\right)^2 \left(\Lambda^4-64 \Lambda^2 m^2+16 \Lambda^2
  v+64 v^2\right)}
   \label{eq:psj2}
\end{eqnarray}
which is the modular  $J$ function for the elliptic curve, $y^2=\left(x^2-\frac{\Lambda^4}{64}\right)(x-v)+\frac{1}{4} m^2 \Lambda^2 x -\frac{1}{32} m^2 \Lambda^4$,  associated with the two flavor ($N_f=2$, with equal masses $m$) $SU(2)$ SUSY QFT. Spectral properties of the prolate spheroidal system have been studied recently in \cite{prolate}.

\section{Conclusions}

In this paper we have shown that a wide class of quantum spectral problems, associated with a  classical genus 1 elliptic curve, have the remarkable property that the perturbative data to a certain order determines the associated non-perturbative data to a similar order. This is a very explicit realization of resurgence, with the fluctuations about a perturbative saddle determining the fluctuations about a non-perturbative saddle. We have shown that this can be understood in simple geometric terms using nothing more than classical mechanics and all-orders WKB. There is a class of potentials for which the resulting $\hbar$ deformation of the classical Wronskian relation (and Matone relation) takes a particularly simple form (\ref{eq:quantum-wronskian}), which moreover is the same as the form previously found for the Mathieu system, just with a different  numerical constant. Thus, the spectral analysis of these potentials is deeply related, even though the potentials themselves are very different. This special class is associated with Ramanujan's theory of elliptic functions in alternative bases, and is related to the special vacua of certain supersymmetric and superconformal QFTs. We have also found a simple form of the quantum perturbative/non-perturbative relation for the Lam\'e system, which is associated with $\mathcal N=2^*$ SUSY QFT.

At higher genus, the classical uniformization procedure involves hyperelliptic functions and abelian integrals, which are more complicated than the elliptic functions appearing in genus 1 problems, but which share similar geometric and modular structures \cite{saito}. On physical grounds we still expect that with a suitable basis choice of integration cycles, it should be possible to find a basis of $g$ classical actions $a_{0}^{(i)}(u)$ and classical dual actions $a_{0}^{D\, (i)}(u)$, for $i=1, ..., g$, such that the $a_{0}^{D\, (i)}(u)$ are determined by the $a_{0}^{(i)}(u)$. Upon quantization, given that the higher order WKB actions are obtained by differential operators acting on the classical ones, then our general argument, that the $a_{0}^{D\, (i)}(u, \hbar)$ are determined by the $a_{0}^{(i)}(u, \hbar)$, goes through. These differential operators may also involve derivatives with respect to the extra parameters (e.g. masses in the gauge theory). The physical implication of such a result would be that for a more general potential, corresponding to a classical genus $g$ system, there are $g$ independent actions (associated, for example,  with the wells of a higher-order polynomial potential) and $g$ independent dual actions (associated, for example, with the barriers of the higher-order polynomial potential), but that for quantization only the perturbative data of the $g$ actions is necessary, as this information encodes also the non-perturbative data of the $g$ dual (barrier) actions. There are formal arguments \cite{sibuya} concerning resurgence in such higher polynomial potentials, but it would be of interest to find concrete examples where all this could be demonstrated explicitly, to provide a constructive procedure to generate fluctuations about non-perturbative saddles directly from the fluctuations about perturbative saddles, as has now been done for genus 1 systems. We expect the methods of SUSY gauge theory and topological strings to provide a natural formalism for addressing these questions, as advocated also in  \cite{Codesido:2016dld}.

\bigskip
{\noindent\bf Acknowledgments:}
We thank G. \'Alvarez, C. Bender, O. Costin, M. Mari\~no, T. Sulejmanpasic, S-T. Yau, and K. Saito  for discussions and correspondence.
This material is based upon work supported by the U.S. Department of Energy, Office of Science, Office of High Energy Physics under Award Number DE-SC0010339 (GD), and Office of Nuclear Physics under Award Numbers DE-FG02-93ER40762 (GB) and DE-SC0013036 (M\"U). M.\"U's work was partially supported by the Center for Mathematical Sciences and Applications (CMSA) at Harvard University.

\section{Appendix A: Classical Uniformization of Genus 1 Systems}
\label{app:uniformization}

\subsection{Elliptic Curve Data}

There are several related, but different, ways to perform the uniformization of the classical torus \cite{bateman,byrd}. Each relies on the fact that from the elliptic curve expression we can simply read off some basic data about the torus. Given the elliptic curve (\ref{eq:curve1}) written in the form (after a suitable change of variables)
\begin{eqnarray}
p^2= b_4\, x^4+4 b_3\, x^3+6 b_2\, x^2+4 b_1\, x+b_0
\label{eq:curve2}
\end{eqnarray}
it is a simple matter to construct
\begin{eqnarray}
g_2&=& b_4 b_0-4 b_3 b_1+3 b_2^2 \qquad; \qquad 
g_3=b_4 b_2 b_0+2 b_3 b_2 b_1-b_4 b_1^2-b_3^2 b_0-b_2^3\\
\Delta&:=& g_2^3-27g_3^2 \qquad;\qquad  J:=\frac{g_2^3}{\Delta}=\frac{4}{27}\frac{(1-\lambda+\lambda^2)^3}{\lambda^2(1-\lambda)^2}
\label{eq:data}
\end{eqnarray}
which bring the elliptic curve to a normal form.
For a given potential, with elliptic curve $p^2=2(u-V)$,  the elliptic curve data, $g_2$, $g_3$, $\Delta$, $J$, and $\lambda$ all become functions of the energy $u$. Some illustrative explicit examples are shown in Table \ref{table:uni}. For these Chebyshev potentials the normalization is chosen such that the bottom of each well is at $u=0$, and the top of each barrier is at $u=1$, and we recognize these special points as zeros and poles in $\Delta$ and $J$.
\begin{table}[htb]
\centerline{\begin{tabular}{|c|c|c|c|c|}  
\hline 
 &  & &  & \\
potential (in Chebyshev form) & $g_2(u)$ & $g_3(u)$ & $\Delta(u)$ & $J(u)$ \\ &&&& \\
\hline
&&&& \\
Mathieu: $V(x)=\cos^2(x)$ &  $\frac{16}{3}(1-u+u^2)$  & $\frac{32}{27}(1+u)(1-2u)(2-u)$  &$2^{10}u^2(1-u)^2$  & $\frac{4}{27}\,\frac{(1-u+u^2)^3}{u^2(1-u)^2}$  \\ 
&&&& \\
\hline
&&&& \\
triple-well: $V(x)=x^2(3-4x^2)^2$ & $48(9-8u)$ & $64(27-36u+8u^2)$ & $2^{18} 27 u^3(1-u)$ & $\frac{1}{64}\ \frac{(9-8 u)^3}{u^3 (1-u)}$ \\
&&&& \\
\hline
&&&& \\
double-well: $V(x)=(1-2x^2)^2$ & $\frac{16}{3}(4-3u)$ & $\frac{64}{27}(8-9u)$ & $2^{12} u^2(1-u)$ & $\frac{1}{27}\ \frac{(4-3 u)^3}{u^2 (1-u)}$ \\
&&&& \\
\hline
&&&& \\
cubic: $V(x)=\frac{1}{2}(1+x)(1-2x)^2$ & $3$ & $(1-2u)$ & $2^2\, 27 u(1- u)$ & $\frac{1}{4}\,\frac{1}{u(1- u)}$ \\
&&&& \\
\hline
\end{tabular}}
\caption{Uniformizing elliptic curve data for the four arithmetic Chebyshev potentials. The modular $J(u)$ functions are used in Table \ref{table:p2} for a comparison with associated mirror curve properties in Table \ref{table:klemm}.}
\label{table:uni}
\end{table}

\subsection{Uniformization with Modular $J$ Function}

The modular $J$ parameter provides a convenient uniformization of the torus. A classical result of Klein is that the classical period $\omega_0(u)$, viewed as a function of $J$, which is itself a function of $u$, can be written in terms of a hypergeometric function as 
\begin{eqnarray}
\omega_0(J)&=& \sqrt{\frac{g_3(J)}{g_2(J)}} \left(\frac{J}{1-J}\right)^{1/4} \, ~_2F_1\left(\frac{1}{12}, \frac{5}{12}, 1; \frac{1}{J}\right)
\label{eq:wJ}
\end{eqnarray}
Thus, $\omega_0(J)$ satisfies a second-order differential equation with respect to the variable $J$, and $\omega_0^D(J)$ is another suitably chosen independent solution of this equation.

This Kleinian form for the classical periods agrees with our results in Section \ref{sec:chebyshev} owing to some non-trivial higher-order changes of variable for the hypergeometric function $~_2F_1\left(\frac{1}{12}, \frac{5}{12}, 1; \frac{1}{J}\right)$ \cite{maier,vidunas}:

\begin{enumerate}
\item
Mathieu: $V(x)=\cos^2(x)$:
\begin{eqnarray} 
\omega_0(u)&=&\frac{1}{(1-u+u^2)^{1/4}} ~_2F_1\left(\frac{1}{12}, \frac{5}{12}, 1; \frac{27\, (1-u)^2 u^2}{4(1-u+u^2)^3}\right)=~_2F_1\left(\frac{1}{2}, \frac{1}{2}, 1; u \right)
\label{eq:hypermagic-sg}
\end{eqnarray}
\item
Double-well: $V(x)=(1-2x^2)^2$:
\begin{eqnarray} 
\omega_0(u)&=& \frac{\sqrt{2}}{(4-3 u)^{1/4}} ~_2F_1\left(\frac{1}{12}, \frac{5}{12}, 1; \frac{27\, u^2(1- u)}{(4-3 u)^3}\right)=~_2F_1\left(\frac{1}{4}, \frac{3}{4}, 1; u\right) 
\label{eq:hypermagic-dw}
\end{eqnarray}
\item
Cubic oscillator: $V(x)=\frac{1}{2}(1+x)(1-2x)^2$:
\begin{eqnarray} 
\omega_0(u)&=& ~_2F_1\left(\frac{1}{12}, \frac{5}{12}, 1; 4 u(1-u)\right)=~_2F_1\left(\frac{1}{6}, \frac{5}{6}, 1;  u\right) 
\label{eq:hypermagic-co}
\end{eqnarray}
\item
Triple well: $V(x)= x^2(3-4x^2)^2$:
\begin{eqnarray} 
\omega_0(u)&=& \frac{\sqrt{3}}{(9-8 u)^{1/4}} ~_2F_1\left(\frac{1}{12}, \frac{5}{12}, 1; \frac{64\, u^3(1-  u)}{(9-8 u)^3} \right)=~_2F_1\left(\frac{1}{3}, \frac{2}{3}, 1; u\right)
\label{eq:hypermagic-tw}
\end{eqnarray}
\end{enumerate}
Note that these expressions are valid as expansions about $u=0$. The behavior near $u=1$ must be matched with the other independent solution for the correct monodromy behavior.

The advantage of the Kleinian hypergeometric form (\ref{eq:wJ}) is that it is valid for all potentials associated with a genus 1 elliptic curve, while the simplifed transformed hypergeometric expressions,  $~_2F_1(\frac{1}{2}-\frac{1}{2m}, \frac{1}{2}+\frac{1}{2m}; 1, u)$, appear only to be valid for the special genus 1 cases in which the Picard-Fuchs equation can be integrated up to a second-order equation (instead of the generic third-order equation), as in Section \ref{sec:chebyshev}. The advantage of the transformed expressions is evident: the argument is linear in the energy $u$, so these expressions are much easier to work with, and their analytic continuation properties are corresponding simpler to analyze. In the Kleinian representation, the conversion between $J$ and $u$ is non-trivial, as the above examples show.

\subsection{Uniformization with Modular $\lambda$ Function}

The general genus 1 uniformization can also be expressed in terms of the modular  $\lambda$ parameter, related to $J(u)$ via:
\begin{eqnarray}
J=\frac{4}{27}\frac{(1-\lambda+\lambda^2)^3}{\lambda^2(1-\lambda)^2}
\label{eq:jlambda}
\end{eqnarray}
Then the classical periods can always be written in terms of $\lambda$ as:
\begin{eqnarray}
\omega_0(\lambda)&=&
f(\lambda) \, \mathbb K(\lambda):=  
\sqrt{\frac{(1+\lambda)(1-2\lambda)(2-\lambda)}{(1-\lambda+\lambda^2)} \frac{g_2(\lambda)}{g_3(\lambda)}} \, \mathbb K(\lambda) 
\label{eq:omega-lambda1}\\
\omega_0^D(\lambda)&=&
i\, f(\lambda) \, \mathbb K(1-\lambda):=
 i \sqrt{\frac{(1+\lambda)(1-2\lambda)(2-\lambda)}{(1-\lambda+\lambda^2)} \frac{g_2(\lambda)}{g_3(\lambda)}} \, \mathbb K(1-\lambda)
\label{eq:omega-lambda2}
\end{eqnarray}
where $\mathbb K(\lambda)$ is the complete elliptic integral of the first kind. The classical modular $\tau$ parameter takes its canonical uniformized form, as the complicated $f(\lambda)$ prefactors in (\ref{eq:omega-lambda1}, \ref{eq:omega-lambda2}) cancel, leaving just the ratio of elliptic $\mathbb K$ functions:
\begin{eqnarray}
\tau_0(\lambda):= \frac{\omega_0^D(\lambda)}{\omega_0(\lambda)}=\frac{i\, \mathbb K(1-\lambda)}{\mathbb K(\lambda)}
\label{eq:tau}
\end{eqnarray}

For the special subclass of systems for which the classical Picard-Fuchs equation for the classical actions reduces to a second-order equation, the actions can be written in terms of the function $f(\lambda)$ defined in (\ref{eq:omega-lambda1}, \ref{eq:omega-lambda2}):
\begin{eqnarray}
a_0(\lambda) &=&
\left[2\lambda(1-\lambda) \frac{f^\prime(\lambda)}{f^2(\lambda)}-(1-\lambda) \frac{1}{f(\lambda)} \right] \K(\lambda) +\frac{1}{f(\lambda)}\, \E(\lambda) 
\label{eq:aw0}
\\
a_0^D(\lambda) &=& i \left[2\lambda(1-\lambda) \frac{f^\prime(\lambda)}{f^2(\lambda)}+\lambda \frac{1}{f(\lambda)} \right] \K(1-\lambda) -i \frac{1}{f(\lambda)} \, \E(1-\lambda)
\label{eq:awd0}
\end{eqnarray}
The Wronskian relation for these two independent solutions simply expresses 
the Legendre relation in terms of $\lambda$:
\begin{eqnarray}
-\K (\lambda) \, \K(1-\lambda)+ \E(\lambda) \K (1-\lambda) +\E(1-\lambda) \K(\lambda) =\frac{\pi}{2}
\label{eq:legendre}
\end{eqnarray}

Once again, the implication of all this is that the dual action $a_0^D(\lambda)$ is determined by the action $a_0(\lambda)$: given  $a_0$, we simply deduce the dual action  as:
\begin{eqnarray}
a_0^D(\lambda)=\tau_0(\lambda) \, a_0(\lambda)-\frac{i  S}{\omega_0(\lambda)}
\label{eq:magic}
\end{eqnarray}
At the classical level, the only things that change between different genus 1 cases (with second order Picard-Fuchs equation) are (i) the relation between the energy $u$ and modular $\lambda$, and (ii) the constant $S$, which is the instanton action.

In summary: the advantage of the $\lambda$ uniformization approach is that it is very general, and relies on minimal information, much of which is obtained from the elliptic curve by simple algebraic manipulations. A disadvantage is that if we are ultimately interested in finding the eigenvalues $u$, the expression relating $\lambda$ to $u$ can be quite complicated, and appropriate branches must be specified for certain parts of the energy spectrum. This is relatively straightforward, but can be messy in practice.
For example, for the symmetric double-well potential, the relation between the energy $u$ and $\lambda$ gives three possibilities:
\begin{eqnarray}
u&=& \frac{4\lambda}{(1+\lambda)^2}\\
u&=&\frac{4(1-\lambda)}{(2-\lambda)^2} \\
u&=& \frac{4\lambda(\lambda-1)}{(2\lambda-1)^2}
\label{eq:dw-ulambda}
\end{eqnarray}
We choose $u=\frac{4\lambda}{(1+\lambda)^2}$, in which case the spectral region inside the well, $u\in [0, 1]$, is described by $\lambda\in [0, 1]$, which is parametrized by $\lambda(\tau_0)$, with pure imaginary $\tau_0\in i\, [0, \infty)$.
Then  the classical  periods and actions, in terms of the energy variable $u$, are transformed into the $\lambda$ form as:
\begin{eqnarray}
\omega_0(u)&=& ~_2 F_1\left(\frac{1}{4}, \frac{3}{4}, 1;  u\right)
=\frac{2}{\pi} \sqrt{1+\lambda}\, \K(\lambda)
=\frac{2}{\pi}\,\frac{1}{\sqrt{1+\sqrt{u}}} \K\left(\frac{2\sqrt{u}}{1+\sqrt{u}}\right) 
\\
\omega_0^D(u)&=& i~_2 F_1\left(\frac{1}{4}, \frac{3}{4}, 1; 1- u\right)
=i \frac{2}{\pi} \sqrt{1+\lambda}\, \K(1-\lambda)
=\frac{2 i \sqrt{2}}{\pi}\,\frac{1}{\sqrt{1+\sqrt{u}}} \K\left(\frac{1-\sqrt{u}}{1+\sqrt{u}}\right)
\label{eq:dw-periods-u}
\end{eqnarray}
where $\lambda=(2-2\sqrt{1-u})/u-1$ in this spectral region. Similarly, the classical actions are:
 \begin{eqnarray}
a_0(u)&=& u~_2 F_1\left(\frac{1}{4}, \frac{3}{4}, 2;  u\right) \\
&=&\frac{16}{3\pi}\frac{1}{(1+\lambda)^{3/2}}\left[-(1-\lambda)\, \K(\lambda) +(1+\lambda) \E(\lambda)\right]
\nonumber\\
&=& \frac{8}{3\pi \sqrt{1+\sqrt{u}} } \left[-(1-u)\K\left(\frac{2\sqrt{u}}{1+\sqrt{u}}\right) 
+(1+\sqrt{u})\,  \E\left(\frac{2\sqrt{u}}{1+\sqrt{u}}\right)\right] \\
a_0^D(u)&=&  -\sqrt{2} \, i\, (1-u)~_2 F_1\left(\frac{1}{4}, \frac{3}{4}, 2; 1- u\right) 
\\
&=&\frac{16}{3\pi} \frac{i}{(1+\lambda)^{3/2}}\left[2\lambda\, \K(\lambda) -(1+\lambda) \E(\lambda)\right]
\nonumber\\
&=&  \frac{16 i}{3\pi \sqrt{1+\sqrt{u}}}  \left[\sqrt{u}(1+\sqrt{u})\K\left(\frac{1-\sqrt{u}}{1+\sqrt{u}}\right) 
-(1+\sqrt{u})\,  \E\left(\frac{1-\sqrt{u}}{1+\sqrt{u}}\right)\right] 
\label{eq:dw-actions-u}
\end{eqnarray}
These conversions rely on the following nontrivial identities:
\begin{eqnarray}
\K\left(\frac{\pm 4\sqrt{\lambda}}{(1\pm \sqrt{\lambda})^2}\right) 
&=& \left(1\pm \sqrt{\lambda}\right) \K(\lambda) \\
\E\left(\frac{\pm 4\sqrt{\lambda}}{(1\pm \sqrt{\lambda})^2}\right) 
&=&\frac{2}{1\pm \sqrt{\lambda}} \E(\lambda)- \left(1\mp \sqrt{\lambda}\right) \K(\lambda) \\
\K\left(\left(\frac{1-\sqrt{\lambda}}{1+\sqrt{\lambda}}\right)^2 \right) &=&
 \frac{1}{2} \left(1+\sqrt{\lambda}\right) \K(1-\lambda) \\
\E\left(\left(\frac{1-\sqrt{\lambda}}{1+\sqrt{\lambda}}\right)^2 \right) &=& 
\frac{1}{1+\sqrt{\lambda}} \E(1-\lambda) +\frac{\sqrt{\lambda}}{1+\sqrt{\lambda}} \K(1-\lambda)
\label{eq:dw-identities}
\end{eqnarray}
which are relevant because $u=\frac{4\lambda}{(1+\lambda)^2}$ implies:
\begin{eqnarray}
1+\sqrt{u}=\frac{(1+\sqrt{\lambda})^2}{1+\lambda}\quad, \quad
\frac{2\sqrt{u}}{1+\sqrt{u}}=\frac{4\sqrt{\lambda}}{(1+\sqrt{\lambda})^2}\quad, \quad 
\frac{1-\sqrt{u}}{1+\sqrt{u}}=\left(\frac{1-\sqrt{\lambda}}{1+\sqrt{\lambda}}\right)^2
\label{eq:dw-ulambda-relations}
\end{eqnarray}

\section{Appendix B: Higher order actions for the special arithmetic Chebyshev genus 1 systems}
\label{app:results}

In this appendix we list the first 4 non-trivial  WKB orders of the expansions
of  the quantum action and dual action:
\begin{eqnarray}
a(u, \hbar) = \sum_{n=0}^\infty \hbar^{2n} a_n(u)\quad,\quad a^D(u)=\sum_{n=0}^\infty \hbar^{2n} a^D_n(u)
\end{eqnarray}
For the four special arithmetic Chebyshev systems, at each order $n$, the energy dependent coefficients of this expansion, $a_n(u)$ and $a_n^D(u)$, can be expressed in the form:
\begin{eqnarray}
a_n(u)= f_n^{(1)}(u) a_0(u)+f_n^{(2)}(u) \frac{d a_0(u)}{d u}\quad,\quad a_n^D(u)= f_n^{(1)}(u) a_0^D(u)+f_n^{(2)}(u) \frac{d a_0^D(u)}{d u}
\label{eq:fn}
\end{eqnarray}
Below, for each of the four special arithmetic Chebyshev systems, we list the normalized classical action and dual action (for the outermost well and barrier, respectively), and we list the first four orders of the function $f_n^{(1)}(u)$ and $f_n^{(2)}(u)$.

\subsection{Mathieu potential: $V(x)=\cos^2(x)$}
The classical actions and periods are:
\begin{eqnarray}
a_0(u)&=&\sqrt{2} \pi\, u\,_2F_1\left(\frac{1}{2}, \frac{1}{2}, 2; u\right)\ \quad,\quad a^D_0(u)=- i \sqrt{2} \pi\, (1-u)\,_2F_1\left(\frac{1}{2}, \frac{1}{2}, 2; 1-u\right)\\
\omega_0(u)&=& \sqrt{2} \pi\, _2F_1\left(\frac{1}{2}, \frac{1}{2}, 1; u\right)\quad,\quad \omega^D_0(u)=  i \sqrt{2} \pi\, \,_2F_1\left(\frac{1}{2}, \frac{1}{2}, 1; 1-u\right)
\end{eqnarray}
The coefficient functions in (\ref{eq:fn}) are:
\begin{eqnarray}
f_1^{(1)}(u)&=&\frac{1}{384 u}+\frac{1}{384 (u-1)}
\\
 f_1^{(2)}(u)&=&-\frac{1}{96}
 \\
f_2^{(1)}(u)&=&\frac{7}{184320 u^3}+\frac{23}{1474560 u^2}+\frac{23}{1474560 (u-1)^2}-\frac{7}{184320 (u-1)^3}+\frac{1}{49152 u}-\frac{1}{49152 (u-1)}
\nonumber \\
 f_2^{(2)}(u)&=&-\frac{7}{184320 u^2}+\frac{7}{184320 (u-1)^2}-\frac{1}{92160 u}-\frac{1}{92160 (u-1)}
 \\
f_3^{(1)}(u)&=& \frac{31}{10321920 u^5}-\frac{13}{55050240 u^4}+\frac{589}{1321205760 u^3}+\frac{461}{1321205760 u^2}+\frac{589}{1321205760 (u-1)^3}
 \nonumber \\
&&+\frac{13}{55050240 (u-1)^4}+\frac{31}{10321920 (u-1)^5}-\frac{461}{1321205760 (u-1)^2}
\\
f_3^{(2)}(u)&=& -\frac{31}{10321920 u^4}+\frac{101}{165150720 u^3}-\frac{19}{66060288 u^2}-\frac{19}{66060288 (u-1)^2}-\frac{101}{165150720 (u-1)^3}
 \nonumber \\
&&-\frac{31}{10321920 (u-1)^4}-\frac{31}{165150720 u}+\frac{31}{165150720 (u-1)}
\\
f_4^{(1)}(u)&=&\frac{127}{220200960 u^7}-\frac{257}{1056964608 u^6}+\frac{197}{4227858432 u^5}+\frac{989}{32212254720 u^4}+\frac{3641}{169114337280 u^3}
 \nonumber \\
&&
+\frac{461}{37580963840 u^2}+\frac{461}{37580963840 (u-1)^2}-\frac{3641}{169114337280 (u-1)^3}+\frac{989}{32212254720 (u-1)^4}
\nonumber \\
&&-\frac{197}{4227858432 (u-1)^5}-\frac{257}{1056964608 (u-1)^6}-\frac{127}{220200960 (u-1)^7}+\frac{461}{18790481920 u}
\nonumber\\
&&-\frac{461}{18790481920 (u-1)}
\\
f_4^{(2)}(u)&=&-\frac{127}{220200960 u^6}+\frac{119}{377487360 u^5}-\frac{577}{14092861440 u^4}-\frac{359}{21139292160 u^3}-\frac{167}{21139292160 u^2}
 \nonumber \\
&&-\frac{359}{21139292160 (u-1)^3}+\frac{577}{14092861440 (u-1)^4}+\frac{119}{377487360 (u-1)^5}+\frac{127}{220200960 (u-1)^6}
\nonumber\\
&& +\frac{167}{21139292160 (u-1)^2}
\end{eqnarray}

\subsection{Symmetric  degenerate triple well potential:  $V(x)=T_3^2(x)=x^2(3-4x^2)^2$}
The classical actions and periods are:
\begin{eqnarray}
a_0(u)=  {\pi\over 3\sqrt{2}}\, u\,_2F_1\left(\frac{1}{3}, \frac{2}{3}, 2; u\right)\, (\text{outer well}) 
\,&,&\,
a^D_0(u)=- i {\pi\over \sqrt{6}}\, (1-u)\,_2F_1\left(\frac{1}{3}, \frac{2}{3}, 2; 1-u\right)
\\
\omega_0(u)= {\pi\over 3\sqrt{2}}\, _2F_1\left(\frac{1}{3}, \frac{2}{3}, 1; u\right) \quad (\text{outer well}) 
\quad&,&\quad 
\omega^D_0(u)=  i  {\pi\over \sqrt{6}}\, \,_2F_1\left(\frac{1}{3}, \frac{2}{3}, 1; 1-u\right)
\end{eqnarray}
The coefficient functions in (\ref{eq:fn}) are:
\begin{eqnarray}
f_1^{(1)}(u)&=&-\frac{1}{3 u}-\frac{2}{9 (u-1)}
\\
 f_1^{(2)}(u)&=&\frac{5}{6}
 \\
f_2^{(1)}(u)&=&\frac{21}{20 u^3}+\frac{1}{20 u^2}+\frac{41}{405 (u-1)^2}-\frac{14}{45 (u-1)^3}+\frac{49}{324 u}-\frac{49}{324 (u-1)}
\\
 f_2^{(2)}(u)&=&-\frac{21}{20 u^2}+\frac{14}{45 (u-1)^2}+\frac{1}{15 u}-\frac{1}{15 (u-1)}
 \\
f_3^{(1)}(u)&=&-\frac{3069}{140 u^5}+\frac{321}{56 u^4}-\frac{1697}{3780 u^3}-\frac{18109}{29160 u^2}+\frac{5729}{29160 (u-1)^2}-\frac{6269}{17010 (u-1)^3}
 \nonumber \\
&&-\frac{41}{189 (u-1)^4}-\frac{248}{105 (u-1)^5}-\frac{619}{1458 u}+\frac{619}{1458 (u-1)}
\\
f_3^{(2)}(u)&=&\frac{3069}{140 u^4}-\frac{2287}{280 u^3}-\frac{1339}{7560 u^2}+\frac{13093}{51030 (u-1)^2}+\frac{151}{315 (u-1)^3}+\frac{248}{105 (u-1)^4}+\frac{2317}{29160 u}
\nonumber\\
&&-\frac{2317}{29160 (u-1)}
\\
f_4^{(1)}(u)&=&\frac{1327023}{1120 u^7}-\frac{135171}{224 u^6}+\frac{46309}{1680 u^5}+\frac{495533}{45360 u^4}+\frac{10218619}{816480 u^3}+\frac{53806507}{4408992 u^2}+\frac{970421}{4408992 (u-1)^2}
 \nonumber \\
&&-\frac{11722283}{22044960 (u-1)^3}+\frac{745363}{306180 (u-1)^4}-\frac{4471}{1215 (u-1)^5}-\frac{3208}{189 (u-1)^6}-\frac{1524}{35 (u-1)^7}
\nonumber\\
&&+\frac{570593}{45927 u}-\frac{570593}{45927 (u-1)}
\\
f_4^{(2)}(u)&=&-\frac{1327023}{1120 u^6}+\frac{411651}{560 u^5}-\frac{88541}{3360 u^4}+\frac{2911}{1620 u^3}-\frac{60407}{816480 u^2}-\frac{3674767}{7348320 (u-1)^2}
 \\
&&-\frac{1259149}{918540 (u-1)^3}+\frac{107}{35 (u-1)^4}+\frac{20612}{945 (u-1)^5}+\frac{1524}{35 (u-1)^6}-\frac{421843}{734832 u}+\frac{421843}{734832 (u-1)}
\nonumber
\end{eqnarray}

\subsection{Symmetric double well potential: $V(x)=T_2^2(x)=(1-2x^2)^2$}

The classical actions and periods are:
\begin{eqnarray}
a_0(u)=  {\pi\over 2}\, u\,_2F_1\left(\frac{1}{4}, \frac{3}{4}, 2; u\right)\ 
\quad&,&\quad 
a^D_0(u)=- i {\pi\over \sqrt{2}}\, (1-u)\,_2F_1\left(\frac{1}{4}, \frac{3}{4}, 2; 1-u\right)\\
\omega_0(u)= {\pi\over 2}\, _2F_1\left(\frac{1}{4}, \frac{3}{4}, 1; u\right)
\quad&,&\quad
 \omega^D_0(u)=  i {\pi\over \sqrt{2}}\, \,_2F_1\left(\frac{1}{4}, \frac{3}{4}, 1; 1-u\right)
\end{eqnarray}
The coefficient functions in (\ref{eq:fn}) are:
\begin{eqnarray}
f_1^{(1)}(u)&=&\frac{1}{8 u}+\frac{1}{16 (u-1)}
\\
 f_1^{(2)}(u)&=&-\frac{1}{4}
 \\
f_2^{(1)}(u)&=&\frac{7}{60 u^3}+\frac{3}{640 u^2}+\frac{23}{2560 (u-1)^2}-\frac{7}{240 (u-1)^3}+\frac{7}{512 u}-\frac{7}{512 (u-1)}
\\
 f_2^{(2)}(u)&=&-\frac{7}{60 u^2}+\frac{7}{240 (u-1)^2}+\frac{1}{160 u}-\frac{1}{160 (u-1)}
 \\
f_3^{(1)}(u)&=&\frac{62}{105 u^5}-\frac{137}{1120 u^4}+\frac{1453}{107520 u^3}+\frac{2649}{143360 u^2}-\frac{1973}{286720 (u-1)^2}+\frac{10189}{860160 (u-1)^3}
 \nonumber \\
&&+\frac{39}{8960 (u-1)^4}+\frac{31}{420 (u-1)^5}+\frac{95}{8192 u}-\frac{95}{8192 (u-1)}
\\
f_3^{(2)}(u)&=&-\frac{62}{105 u^4}+\frac{199}{1120 u^3}+\frac{269}{53760 u^2}-\frac{1829}{215040 (u-1)^2}-\frac{101}{8960 (u-1)^3}
\nonumber \\
&&
-\frac{31}{420 (u-1)^4}-\frac{251}{71680 u}+\frac{251}{71680 (u-1)}
\\
f_4^{(1)}(u)&=&\frac{254}{35 u^7}-\frac{1019}{336 u^6}+\frac{1657}{26880 u^5}+\frac{97073}{1146880 u^4}+\frac{217803}{2293760 u^3}+\frac{45531}{524288 u^2}+\frac{1569}{524288 (u-1)^2}
 \nonumber \\
&&
-\frac{101877}{9175040 (u-1)^3}+\frac{77407}{2621440 (u-1)^4}-\frac{2167}{61440 (u-1)^5}-\frac{257}{1792 (u-1)^6}-\frac{127}{280 (u-1)^7}
 \nonumber \\
&&+\frac{11775}{131072 u}-\frac{11775}{131072 (u-1)}
\\
f_4^{(2)}(u)&=& -\frac{254}{35 u^6}+\frac{3119}{840 u^5}+\frac{1213}{53760 u^4}+\frac{5147}{430080 u^3}-\frac{2627}{430080 u^2}+\frac{109}{860160 (u-1)^2}-\frac{5219}{286720 (u-1)^3}
\nonumber \\
&&+\frac{22091}{860160 (u-1)^4}+\frac{119}{640 (u-1)^5}+\frac{127}{280 (u-1)^6}-\frac{49}{8192 u}+\frac{49}{8192 (u-1)}
\end{eqnarray}

\subsection{Cubic oscillator: $V(x)=T_{3/2}^2(x)=\frac{1}{2}(1+x)(1-2x)^2$}
The classical actions and periods are:
\begin{eqnarray}
a_0(u)=  \pi \sqrt{2\over3} \, u\,_2F_1\left(\frac{1}{6}, \frac{5}{6}, 2; u\right)
\quad&,&\quad 
a^D_0(u)=- i \pi \sqrt{2\over3} \, (1-u)\,_2F_1\left(\frac{1}{6}, \frac{5}{6}, 2; 1-u\right)
\\
\omega_0(u)= \pi \sqrt{2\over3} \, _2F_1\left(\frac{1}{6}, \frac{5}{6}, 1; u\right) 
\quad&,&\quad 
\omega^D_0(u)=  i\pi \sqrt{2\over3} \, \,_2F_1\left(\frac{1}{6}, \frac{5}{6}, 1; 1-u\right)
\end{eqnarray}
The coefficient functions in (\ref{eq:fn}) are:
\begin{eqnarray}
f_1^{(1)}(u)&=& -\frac{5}{144 u}-\frac{5}{144 (u-1)}
\\
 f_1^{(2)}(u)&=& \frac{1}{12}
 \\
f_2^{(1)}(u)&=& \frac{7}{576 u^3}+\frac{35}{41472 u^2}+\frac{35}{41472 (u-1)^2}-\frac{7}{576 (u-1)^3}+\frac{35}{20736 u}-\frac{35}{20736 (u-1)}
\\
 f_2^{(2)}(u)&=& \frac{7}{576 (u-1)^2}-\frac{7}{576 u^2}
 \\
f_3^{(1)}(u)&=& -\frac{31}{1344 u^5}+\frac{587}{193536 u^4}-\frac{11845}{13934592 u^3}-\frac{11845}{13934592 u^2}-\frac{11845}{13934592 (u-1)^3}
\\ \nonumber
&&-\frac{587}{193536 (u-1)^4}-\frac{31}{1344 (u-1)^5}+\frac{11845}{13934592 (u-1)^2}
\\
f_3^{(2)}(u)&=& \frac{31}{1344 u^4}-\frac{299}{64512 u^3}+\frac{1475}{10450944 u^2}+\frac{1475}{10450944 (u-1)^2}+\frac{299}{64512 (u-1)^3}
\\ \nonumber
&&+\frac{31}{1344 (u-1)^4}+\frac{1475}{5225472 u}-\frac{1475}{5225472 (u-1)}
\\
f_4^{(1)}(u)&=& \frac{381}{3584 u^7}-\frac{24235}{774144 u^6}+\frac{1301}{3981312 u^5}+\frac{2195473}{1146617856 u^4}+\frac{4401817}{2579890176 u^3}+\frac{2217215}{1719926784 u^2}
 \nonumber \\
&&+\frac{2217215}{1719926784 (u-1)^2}-\frac{4401817}{2579890176 (u-1)^3}+\frac{2195473}{1146617856 (u-1)^4}-\frac{1301}{3981312 (u-1)^5} 
\nonumber \\
&&-\frac{24235}{774144 (u-1)^6}-\frac{381}{3584 (u-1)^7}
+\frac{2217215}{859963392 u}-\frac{2217215}{859963392 (u-1)}
\\
f_4^{(2)}(u)&=& -\frac{381}{3584 u^6}+\frac{14975}{387072 u^5}+\frac{1537}{884736 u^4}-\frac{3199}{13436928 u^3}-\frac{3199}{13436928 u^2}-\frac{3199}{13436928 (u-1)^3}
 \nonumber \\
&&-\frac{1537}{884736 (u-1)^4}+\frac{14975}{387072 (u-1)^5}+\frac{381}{3584 (u-1)^6}+\frac{3199}{13436928 (u-1)^2}
\end{eqnarray}

\end{document}